\documentclass[showpacs,showkeys, groupedaddress,twocolumn, floatfix,aps, prb,reprint]{revtex4-1}
\usepackage[english]{babel}
\usepackage{fontenc}
\usepackage{graphicx}
\usepackage{amsmath}
\usepackage{epstopdf}
\usepackage{amssymb}
\usepackage{amsbsy}
\usepackage{amscd}
\usepackage{float}
\usepackage{color}
\usepackage{bbm}
\usepackage{tikz}
\usepackage{braket}
\usepackage{standalone}
\usepackage{bm}
\usepackage{siunitx}
\usepackage[normalem]{ulem} 
\usepackage{verbatim}
\usepackage{url}
\usetikzlibrary{positioning,calc}
\tikzset{>=latex}
\usepackage[verbose,hypertexnames=false,bookmarksopenlevel=1,filecolor=blue,
linkcolor=blue,citecolor=blue,urlcolor=blue,pdfstartview=FitH,bookmarksopen,bookmarksnumbered,
colorlinks,plainpages=false,linktocpage]{hyperref}
\DeclareSymbolFont{mathbold}{OML}{cmm}{b}{it}
\DeclareMathSymbol{\bsigma}{\mathord}{mathbold}{27}

\newcommand{\dd}{\mathrm{d}}
\begin{document}
\title{Driven Hofstadter Butterflies and Related Topological Invariants}
\author{Martin Wackerl}
\email{martin.wackerl@ur.de}
\author{Paul Wenk}
\email{paul.wenk@ur.de}
\author{John Schliemann}
\email{john.schliemann@ur.de}
\affiliation{Institute for Theoretical
  Physics, University of Regensburg, 93040 Regensburg, Germany}
\date{\today }
\begin{abstract}
  The properties of the Hofstadter butterfly, a fractal, self similar
  spectrum of a two dimensional electron gas, are studied in the case
  where the system is additionally illuminated with monochromatic
  light.  This is accomplished by applying Floquet theory to a tight
  binding model on the honeycomb lattice subjected to a perpendicular
  magnetic field and either linearly or circularly polarized light. It
  is shown how the deformation of the fractal structure of the
  spectrum depends on intensity and polarization.  Thereby, the
  topological properties of the Hofstadter butterfly in presence of
  the oscillating electric field are investigated. A thorough
  numerical analysis of not only the Chern numbers but also the
  $W_{3}$-invariants gives the appropriate insight into the topology
  of this driven system. This includes a comparison of a direct
  $W_3$-calculation to the method based on summing up Chern numbers of
  the truncated Floquet Hamiltonian.
\end{abstract}

\maketitle
\allowdisplaybreaks
\section{Introduction}
The integer quantum Hall effect \cite{Klitzing,KlitzingQHE} marks, in
hindsight, the inception of the field of topological insulators
\cite{Hasan10,Qi11}. This discovery was preceded by a few years by
Hofstadter\rq s seminal work on hopping models on a two-dimensional
square lattice in a perpendicular magnetic field
\cite{Hofstadter76}. The celebrated Hofstadter butterfly contains the
Landau level structure underlying the quantum Hall effect in the limit
of small fluxes per unit cell. The relation of the band structure to
the Hall conductance at general flux was clarified shortly later
\cite{TKNN} in terms of Chern numbers \cite{FukuiChern}.

Moreover, an important recent direction of work in the area of
topological insulators are systems under external driving, mainly by
electromagnetic radiation, and the formation of nontrivial topological
phases dubbed Floquet topological insulators
\cite{Oka09,Kitagawa10,Lindner11,Gu11,Cayssol13,Rudner13,Mikami16,Holthaus16,Klinovaja16}.
In fact, the study of light-matter interaction is one of the fastest
growing research areas in physics. Here, two-dimensional systems with
underlying honeycomb lattice structure have attracted particular
interest including graphene
\cite{Oka09,Karch10,Calvo11,Zhou11,Gu11,Scholz13,Usaj14,Sentef15,Lopez15a,Wang16},
silicene \cite{Lopez15b,Mohan16}, germanene \cite{Mohan16,Tahir16},
and transition metal dichalcogenides \cite{Claasen16}.  To access
e.g. in graphene the feasibility of ac-driven fields to generate a
finite spin polarization of carriers the effect of periodically driven
spin-orbit coupling was studied in Refs.~\onlinecite{Lopez12,Lopez13}.

Furthermore, as seen from the quantum Hall effect
  \cite{KlitzingQHE}, the topological properties of two-dimensional
systems are also drastically altered by applying a perpendicular
magnetic field also leading to fractal structures as the Hof\-stadter
butterfly \cite{Hofstadter76, Rammal85, Hasegawa06, Wang09, Rhim12,
  Yilmaz15, Yilmaz17, Asboth17}. The question arises in which way an external
periodic driving can modify or destroy the fractal
structure. Moreover, following the seminal paper by Rudner \textit{et
  al.}, Ref.~\onlinecite{Rudner13}, it becomes clear that the topology
analysis of driven systems needs a different approach compared to the
static case which goes beyond the Chern number calculation. We are
going to address these problems in the present paper.

Concerning the experimental realizability of the theory
developed in this paper we first emphasize the pioneering work of
measuring the Hofstadter butterfly in m$\acute{\mathrm{o}}$ire
superlattices \cite{Dean13} showing the possibility of measuring the
Hofstadter butterfly as well on a hexagonal lattice
structure. Utilizing superlattice structures the necessary magnetic
field can be lowered to easily accessible field strengths of about
tens of Tesla. Furthermore, the formation of Floquet bands exist not
only on paper. Using ARPES methods the periodic band structure was
resolved in momentum space and even the gap opening of driven
topological insulators was realized and measured\cite{Wang13}.
Thus, the path to experimental accessibility is already paved by
modern techniques and the study presented in this paper aims at
giving a better understanding of the fundamental building blocks by
focusing on a single graphene sheet subjected to a strong
perpendicular magnetic field and externally driven by polarized
light.

This paper is organized as follows. First, we treat in section
\ref{hofstadterproblem} the Hofstadter butterfly problem
\cite{Hofstadter76} on the honeycomb lattice \cite{Rammal85,
  Hasegawa06, Yilmaz15, Owerre2018} in a rigorous manner. Then we
generalize it in section \ref{floquet-hofstadter} to the case with
periodic driving, realized by linearly and circularly polarized
light. We show some representative numerical results for different
frequencies, intensities and polarizations. Finally, the topological
properties of the Floquet-Hofstadter problem are characterized with
Chern numbers and $W_{3}$-invariants in section
\ref{chernnumbers}. Thereby we compare this invariant with the often
used summation over Chern numbers in the truncated Floquet space for
different frequencies and intensities.  We combine an analytical as
well as a numerical approach to the above quantities, and close with a
summary in section \ref{summary}.\\

Shortly after a previous version of this work other works on the same subject appeared
which stress the role of different regimes of the driving frequency
\cite{Kooi18} and other lattice types \cite{Du18}.
\section{Hofstadter butterfly for the honeycomb lattice}
\label{hofstadterproblem}
\subsection{Derivation of the Hamiltonian}
\begin{figure}[t]
	\includegraphics[width=0.9\columnwidth]{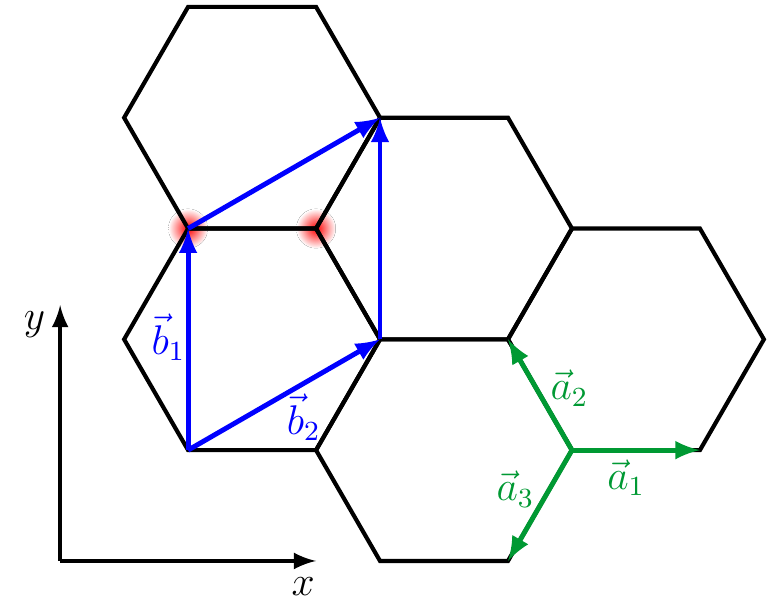}
	\caption{Coordinate geometry used on the honeycomb lattice: The
      green arrows represent the different nearest neighbor vectors $\vec{a}_i$
      and the blue ones the lattice vectors $\vec{b}_i$.}\label{Coord}
\end{figure}

To model graphene we use a tight-binding model where only nearest neighbor
hopping can take place. We choose the lattice vectors as
\begin{align}
\vec{b}_{1} = a \begin{pmatrix} 0\\ \sqrt{3} \end{pmatrix}
\quad,\quad \vec{b}_{2}
= a \begin{pmatrix}\frac{3}{2}\\ \frac{\sqrt{3}}{2} \end{pmatrix}
\label{latticeVecs}
\end{align}
with $a$ being the distance between the carbon atoms.
The nearest neighbor vectors are
\begin{align}
\vec{a}_{1} = a \begin{pmatrix} 1\\ 0 \end{pmatrix} \ , \ \vec{a}_{2} = \frac{a}{2} \begin{pmatrix} -1\\ \sqrt{3} \end{pmatrix} \  , \  
\vec{a}_{3} = \frac{a}{2} \begin{pmatrix} -1\\ -\sqrt{3} \end{pmatrix}
\end{align}
as depicted in Fig.~\ref{Coord}.
The position of an arbitrary unit cell is 
\begin{align}
\vec{R}(m,n) = m \vec{b}_{1} + n\vec{b}_{2},\quad m,n\in\mathbb{Z}\;.
\end{align}
In presence of a vector potential the hopping 
parameter $t$ gets modified by the Peierls phase,
\begin{align}
t\mapsto t_{m,n}e^{i\phi_{m,n}^{(j)}}\,,
\label{PeierlsPhase}
\end{align}
where the phase is the integral over the vector potential along the 
hopping path
\begin{align}
  \phi_{m,n}^{(j)} = \frac{e}{\hbar}\int\limits_{\vec{R}(m,n)}^{\vec{R}(m,n)+\vec{a}_{j}} 
  \vec{A}(\vec{r}\,)\cdot\dd\vec{r} \;,\quad j = 1,2,3 \ .
\end{align}
The magnetic field is applied in $z$ direction, $\vec{B} = B
\vec{e}_z$. For Landau gauge $\vec{A}(\vec{r}\,) = (0,Bx,0)^T$ the Peierls phase becomes independent of the index $m$,
\begin{align}
   \int\limits_{\vec{R}(m,n)}^{\vec{R}(m,n)+\vec{a}_{2,3}} \vec{A}(\vec{r}\,) 
   \cdot \dd\vec{r} = \pm \frac{3\sqrt{3}}{4}Ba^2\bigl(n-\tfrac{1}{6}\bigr)
   \label{peierlsexpl}
\end{align}
and zero for the hopping in $\vec{a}_{1}$ direction. 
Note that the prefactor in the above expression is related to the
area of the elementary unit cell $A_{\mathrm{cell}}$
by $3\sqrt{3}a^2/4=A_{\mathrm{cell}}/2$.
As usual, we restrict the  flux per 
unit cell in units of the elementary charge over Planck's constant to a 
rational value
\begin{align}
   \phi \equiv \frac{e}{h}\,BA_{\mathrm{cell}} = \frac{p}{q}\,.
\end{align}
Thus, the Peierls phase can be written as
\begin{align}
   \frac{e}{\hbar} \frac{3\sqrt{3}}{4}Ba^2\bigl(n-\tfrac{1}{6}\bigr) = \pi \phi \bigl(n-\tfrac{1}{6}\bigr)
   \label{Peierls}
\end{align}
which leads then to the explicit form of of the Hamiltonian
\begin{align}
\begin{aligned}
   H = -t \sum_{mn} \Bigl[& a^{\dagger}_{m,n}\bigl( b_{m,n}  + e^{i\pi\phi(n-\frac{1}{6})} b_{m+1,n-1}\\
   & +   e^{-i\pi\phi(n-\frac{1}{6})}b_{m,n-1}   \bigr) + \text{h.c.} \Bigr]\;,
\end{aligned} \label{ButterflyH}
\end{align}
where the sum is over all unit cell positions.
The solutions of the stationary Schr\"odinger equation are plane-wave type states of the
general form
\begin{align}
\begin{aligned}
   |\vec{k}\rangle = \sum_{mn} & e^{i\vec{k}\cdot\vec{R}(m,n)}
   \bigl(\alpha_{n} a_{m,n}^{\dagger} + \beta_{n}
   b_{m,n}^{\dagger}\bigr)|0\rangle\;, \label{KinSecondQ}
\end{aligned}
\end{align}
where the creation operators $a^{\dagger}_{m,n},b^{\dagger}_{m,n}$ for the different 
sublattice sites are acting on the fermionic vacuum $|0\rangle$. 
$\alpha_{n},\beta_{n}$ are complex amplitudes depending only on $n$ since the 
Peierls phase does so, see Eq.~\eqref{Peierls}.
Making a projection on a state $\langle0|a_{m',n'}$ or $\langle0|b_{m',n'}$ 
leads to a system of coupled equations for the amplitudes
\begin{align}
   -\frac{\varepsilon}{t} \alpha_{n} & = \beta_{n} + z_{n}(\vec{k})\beta_{n-1}\;,
   \label{ButterflyAlpha}  \\
   -\frac{\varepsilon}{t} \beta_{n} & = \alpha_{n} + z^{*}_{n+1}(\vec{k})\alpha_{n+1}\;,
\label{ButterflyBeta}
\end{align}
with
\begin{align}
   z_{n}(\vec{k}) = e^{-i\pi\phi(n-\frac{1}{6})-i\vec{k}\cdot\vec{b_{2}}} 
   + e^{i\pi\phi(n-\frac{1}{6})}e^{i\vec{k}\cdot(\vec{b}_{1}-\vec{b}_{2})} \ .
\end{align}
\begin{figure}[t]
	\includegraphics[width=1.0\columnwidth]{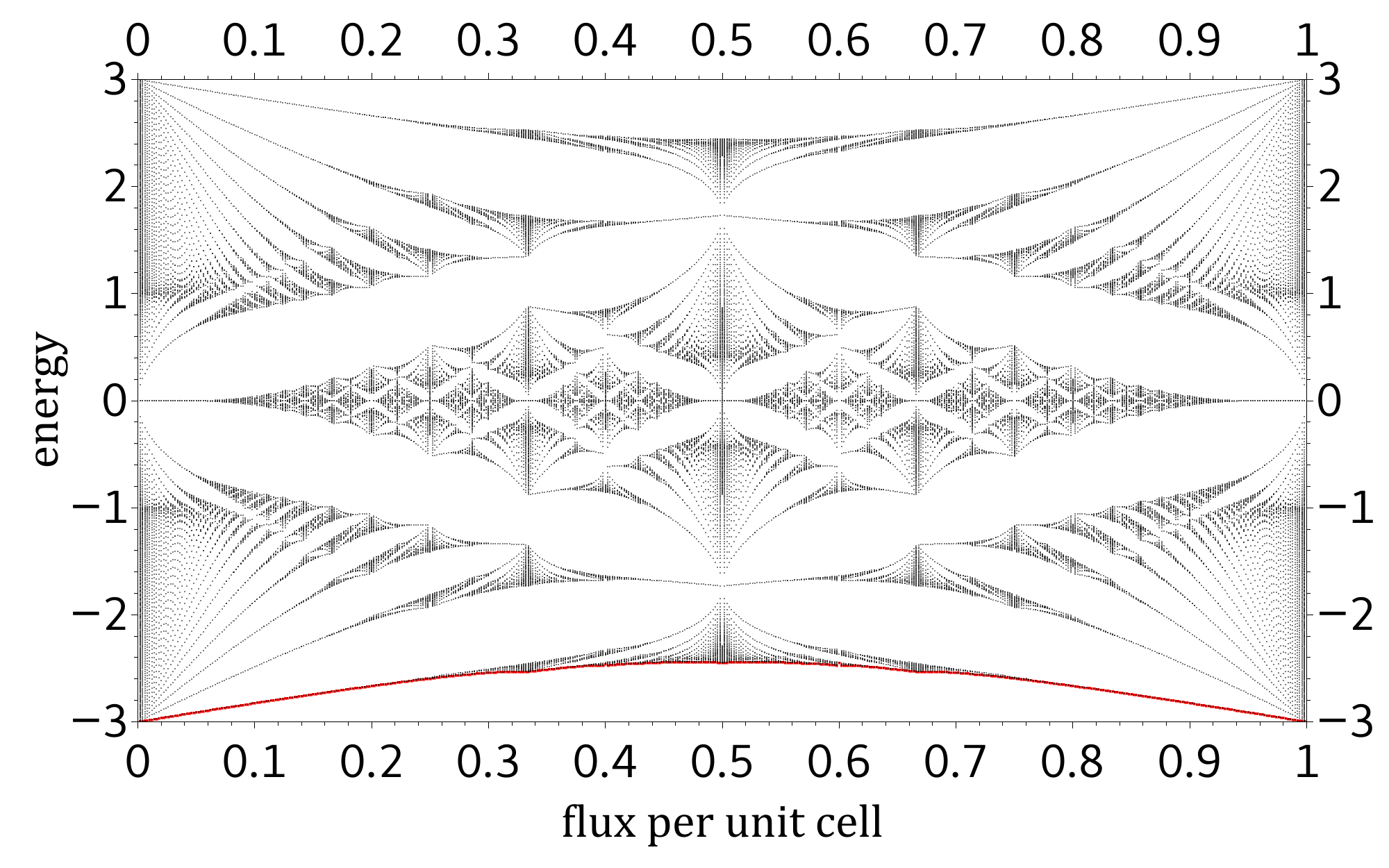}
	\caption{Hofstadter butterfly for the honeycomb lattice.
          The energy is given in units of the hopping parameter $t$.The ground state of the Hofstadter spectrum is defined as the state with lowest energy represented by the red line.}
        \label{Butterfly}
\end{figure}
\subsection{Periodicity of the Hofstadter Problem}
The Eqs.~(\ref{ButterflyAlpha}), (\ref{ButterflyBeta}) define a prima vista
infinite system of linear equation, which, however, closes to a finite one
due to periodicity properties of the amplitudes involved.
First, we define the operators
\begin{align}
   T_{r} \begin{pmatrix} a_{m,n} \\ b_{m,n} \end{pmatrix} T^{\dagger}_{r} 
   = \begin{pmatrix} a_{m,n+r} \\ b_{m,n+r} \end{pmatrix} \\
   u \begin{pmatrix} a_{m,n} \\ b_{m,n} \end{pmatrix} u^{\dagger} 
   =  (-1)^{n} \begin{pmatrix} a_{m,n} \\ b_{m,n} \end{pmatrix}
\end{align}
such that for
\begin{align}
  \text{$p$ even:}&\quad T_{q} H  T_{q}^{\dagger} = H\;,\\
  \text{$p$ odd :}&\quad uT_{q}H T_{q}^{\dagger}u^{\dagger} =  H\;.
\end{align}
For even $p$, the translation operator $T_{q}$ acts on the state ansatz as
\begin{align}
  \ket{\vec{k}} = e^{i\vec{k}\cdot\vec{b}_{2}q}\, T_{q}\ket{\vec{k}}
\end{align}
and consequently the amplitudes have the periodicity
\begin{align}
   \alpha_{n+q} = \alpha_{n}\;,\quad  \beta_{n+q} = \beta_{n}\ . \label{qeven}
\end{align}
In the other case where $p$ is odd
\begin{align}
   \ket{\vec{k}} = e^{i\vec{k}\cdot\vec{b}_{2}q}\,  uT_{q}\ket{\vec{k}} \label{kTtilde}
\end{align}
and the amplitudes have to fulfill
\begin{align}
   \alpha_{n+q} = (-1)^{n+q}\alpha_{n} 
   \;, \quad \beta_{n+q} = (-1)^{n+q}\beta_{n} \;. \label{qodd}
\end{align}
The relations \eqref{qeven}, \eqref{qodd} can be summarized as
\begin{align}
   \alpha_{q} = (-1)^{pq} \alpha_{0} \;,\quad \beta_{q} = (-1)^{pq} \beta_{0}\; .
\end{align}
Thus, Eqs.~(\ref{ButterflyAlpha}), (\ref{ButterflyBeta}) define a finite
linear system of equation for, say, $\alpha_0\dots\alpha_{q-1}$ and
$\beta_0\dots\beta_{q-1}$, and if both $p$ and $q$ are odd the relation 
between the missing amplitudes $\alpha_q$, $\beta_q$ and
$\alpha_0$, $\beta_0$, resp., contains an additional minus sign.
This sign can be compensated by shifting the wave vectors by half of a reciprocal
lattice vector as $k_{x}\to k_{x}+\frac{2\pi}{3q}$
leading to
\begin{align}
   &\alpha_{n+q} = (-1)^{n+1+q} \alpha_{n}\ , \\ 
   &\beta_{n+q} = (-1)^{n+1+q} \beta_{n} \ . \label{kshiftb}
\end{align}
This allows us to use Eq.~\eqref{qeven} for all flux values in the calculation 
of the Hofstadter spectrum and Chern numbers but one should keep in mind that 
one gets a shifted band structure for odd flux values according to
Eqs.~\eqref{kTtilde}-\eqref{kshiftb}.
As a result, in order to calculate the Hofstadter butterfly a $2q\times 2q$ 
matrix is sufficient to obtain the full Hofstadter spectrum.
%
%
\section{Floquet-Hofstadter spectrum}
\label{floquet-hofstadter}
In this section we generalize the Hofstadter butterfly to the case of an 
additional oscillating electric field.  We will focus on linear and circular 
polarization and show how the two polarization states affect the 
Hofstadter spectrum.
\subsection{Circularly polarized light}
\begin{figure}[t]
	\includegraphics[width=1.00\columnwidth]{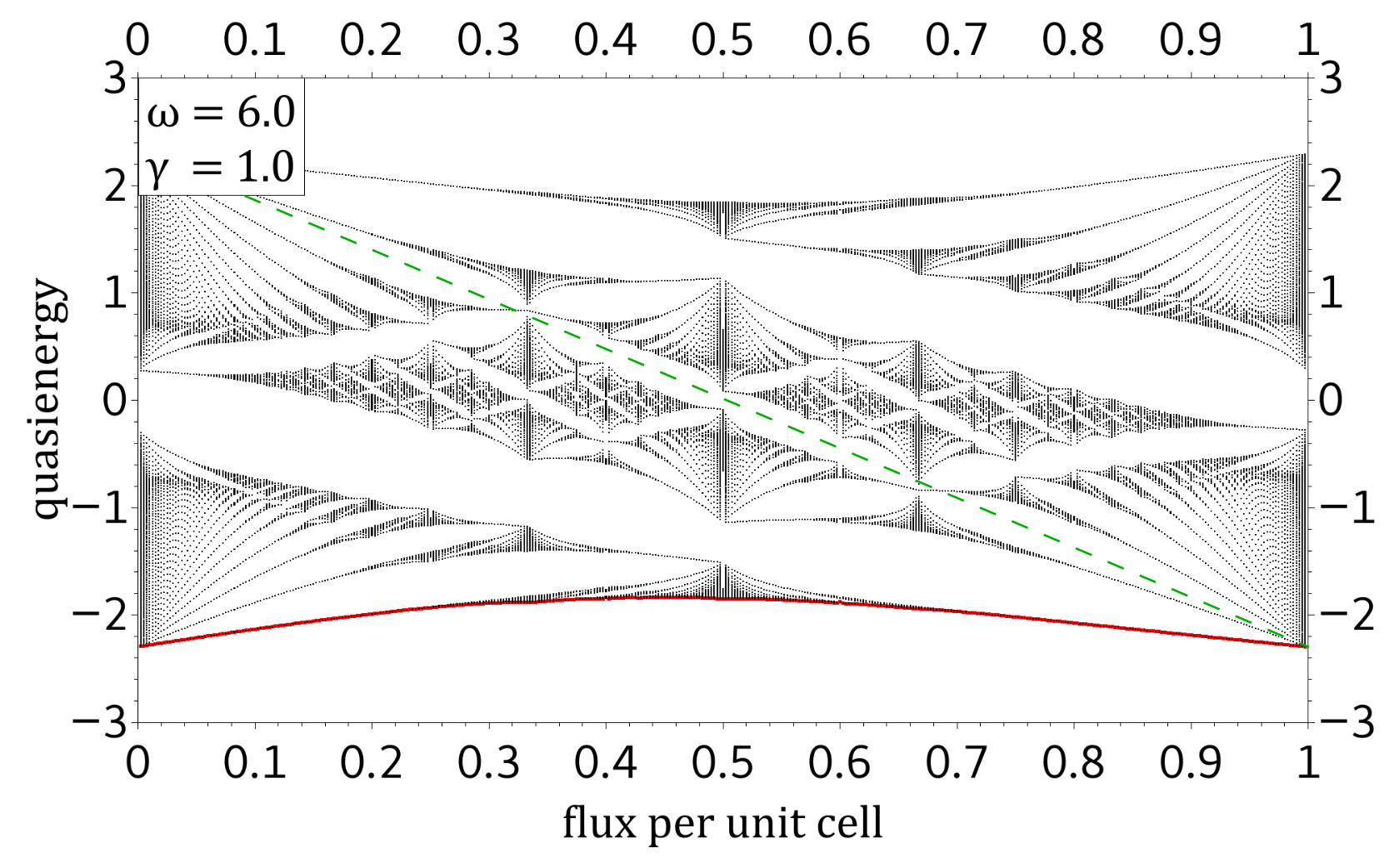}
	\caption{The Hofstadter butterfly gets deformed in presence of
      circularly polarized light. The frequency $\omega$ of the
      periodic driving was set to $6.0\,g/\hbar$ and the intensity
      $\gamma$ to
      $1.0\,eAa/\hbar$. With the present choice of frequency the
      different butterflies of the different Floquet modes do not
      overlap. The red line shows the state with lowest quasienergy of
      the central Floquet mode.}\label{Butterfly61}
\end{figure}

The following vector potential $\vec{A}$ is representing a in $xy$-plane circularly polarized 
light of frequency $\omega$ and amplitude $A$, and the perpendicular
magnetic field $B$,
\begin{align}
   \vec{A}(\vec{r},t) = \begin{pmatrix}
   A \sin(\omega t)\qquad\ \ \ \\ A\cos(\omega t) + Bx
   \end{pmatrix} \label{VectorPotential} \ . 
\end{align}
The vector potential is included in the Hamiltonian via Peierls substitution. 
In what follows, the hopping parameter is renamed to $g$, and $\vec A(t)$ 
is representing only the time-dependent part of
Eq.~\eqref{VectorPotential}. The resulting Hamiltonian reads
\begin{align}
\begin{aligned}
   H ={}&  -g \sum_{mn} \Bigl[ a^{\dagger}_{m,n} \bigl( \
   e^{i\frac{e}{\hbar} \vec{A}(t)\cdot\vec{a}_{1}}b_{m,n} \\
   & + e^{i\pi\phi(n-\frac{1}{6})+i\frac{e}{\hbar} \vec{A}(t)\cdot\vec{a}_{2}}b_{m+1,n-1}\\
   & + e^{-i\pi\phi(n-\frac{1}{6})+i\frac{e}{\hbar} \vec{A}(t)\cdot\vec{a}_{3}}b_{m,n-1} \bigr) + \text{h.c.} \Bigr]\;.
\end{aligned} \label{FloqetHofHamiltonian}
\end{align}
The time-dependent Schr\"odinger equation can be expressed in the
Floquet form
\begin{align}\label{FloquetEq}
   H_F\ket{\vec{k},t}:= (H - i\hbar\partial_{t} ) \ket{\vec{k},t} ={}& 
  \varepsilon \ket{\vec{k},t}\;,
\end{align}
where $\varepsilon$ is the quasienergy which is only defined modulo
integer multiples of $\hbar \omega$. The state $\ket{\vec{k},t}$ is
periodic in time with a period $T=2\pi/\omega$ which allows for a
discrete Fourier transformation. According to Eq.~\eqref{KinSecondQ},
the general solution of $H_F$ can be written in the form
\begin{align}
\begin{aligned}
   |\vec{k},t\rangle = \sum_{mn} & e^{i\vec{k}\cdot\vec{R}(m,n)} \Bigl(\alpha_{n}(t) a_{m,n}^{\dagger} + \beta_{n}(t) b_{m,n}^{\dagger}\Bigr)|0\rangle \label{kt}\;.
\end{aligned}
\end{align}
Due to the periodicity of $\ket{\vec{k},t}$, one can expand the terms $\alpha_{n}(t),\beta_{n}(t)$ using the Fourier series
\begin{align}
  \alpha_{n}(t) = \sum_{l} \alpha_{n,l}e^{il\omega t}\;,\label{alphaFourier}
\end{align}
where the index $l$ is the quantum number of the Floquet mode (also
called Floquet replica). The equivalent to Brillouin zones (BZ) for the real space are the Floquet modes for the time space.
Additionally use the Jacobi-Anger expansion\cite{Wang16}
\begin{align}
   e^{iz\cos(\omega t)} = \sum_{n=-\infty}^{\infty} J_{n}(z)e^{in\bigl( \omega t +\frac{\pi}{2}\bigr) } \;,
\end{align}
where $J_{n}$ denotes the $n$-th order Bessel function of the first kind. 
The Floquet equation \eqref{FloquetEq}
leads to the following coupled expressions for the amplitudes 
\begin{align}
\begin{aligned}
   l\hbar\omega\alpha_{n,l}-g\sum_{l'}& J_{l'}(\gamma)\biggl[\beta_{n,l-l'} + \\ 
   & f_{n,l'}(\vec{k})\beta_{n-1,l-l'} \biggr] =\varepsilon \alpha_{n,l}\;,\label{FloHofalpha}
\end{aligned}
\end{align}
\begin{align}
\begin{aligned}
   l\hbar\omega \beta_{n,l}-g\sum_{l'}& J_{l'}(\gamma)\biggl[\alpha_{n,l+l'} + \\ 
   & f_{n+1,l'}^{*}(\vec{k})\alpha_{n+1,l+l'} \biggr] =\varepsilon \beta_{n,l}\;, \label{FloHofbeta}
\end{aligned}
\end{align}
with
\begin{align}
\begin{aligned}
   f_{n,l'}(\vec{k}) = e^{i\pi\phi(n-\frac{1}{6})-il'\frac{4\pi}{3}}
   e^{i\vec{k}\cdot(\vec{b}_{1}-\vec{b}_{2})}
   \\ + e^{-i\pi\phi(n-\frac{1}{6})-il'\frac{2\pi}{3}} e^{-i\vec{k}\cdot\vec{b}_{2}}\;,
\end{aligned}
\end{align}
where $\gamma\equiv eAa/\hbar$, termed light parameter. An exemplary numerical result can be seen in Fig.~\ref{Butterfly61}. The bending direction represented by the green dashed line depends on the sign of the driving frequency $\omega$. 
\begin{figure}[t]
	\includegraphics[width=1.00\columnwidth]{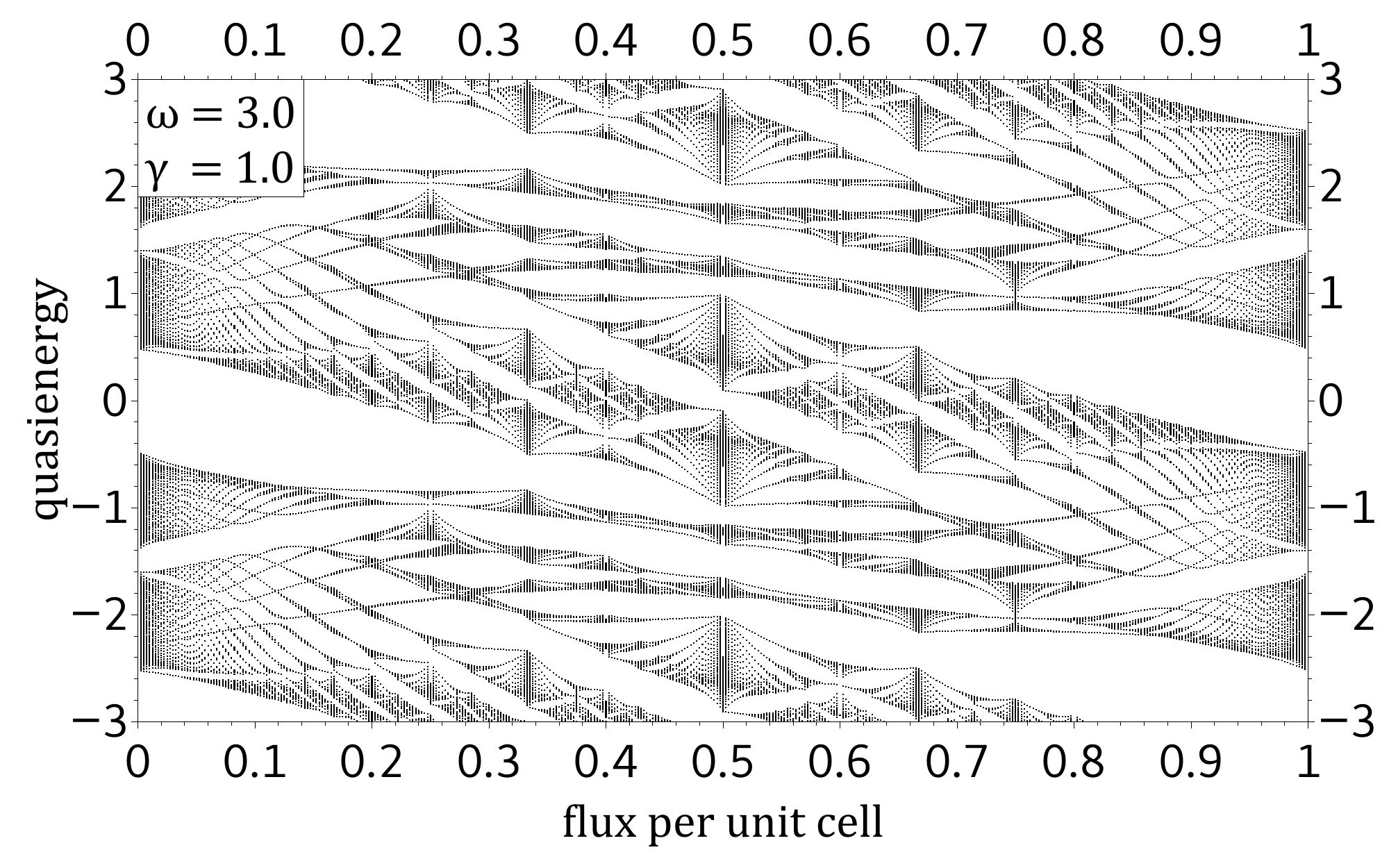}
	\caption{The frequency $\omega$ of the circularly polarized
      radiation was set to $3.0\,g/\hbar$ and the intensity $\gamma$
      to $1.0\,eAa/\hbar$. The spectra of the different Floquet modes
      overlap. }\label{Butterfly31}
\end{figure}
\subsection{Linearly polarized light}
We investigate now the case of linear polarization of the light represented by
\begin{align}
   \vec{A}(\vec{r},t) = \begin{pmatrix}
   A_{x} \cos(\omega t) \qquad\ \ \  \\ A_{y}\cos(\omega t) + Bx
   \end{pmatrix} \ .
\end{align}
The orientation of the linear polarization can be tuned by varying $A_{x}$ and $A_{y}$. The effective amplitude for the three different hopping paths is then governed by
\begin{align}
   \vec{A}(t)\cdot \vec{a}_{i}= A_{i}\cos(\omega t) \quad \text{with}\quad i = 1,2,3 \ . \label{EffAmp}
\end{align}
In contrast to the case of circularly polarized light, where the
transitions between the different Floquet modes are for all hopping
directions equally suppressed, they are for linear polarization
not. This can be seen from the fact that the argument of the Bessel
function is different for each hopping direction. The equivalent
equations to Eqs.~\eqref{FloHofalpha} and \eqref{FloHofbeta} for
linearly polarized light read
\begin{align}
\begin{aligned}
   & l\hbar\omega\alpha_{n,l} - g\sum_{l'}\biggl[\Bigl(J_{l'}(\gamma_{1}) \beta_{n,l-l'} \\
   & + J_{l'}(\gamma_{2})e^{i\pi\phi(n-\frac{1}{6}) +i\vec{k}\cdot(\vec{b}_{1}-\vec{b}_{2})} \\
   & + J_{l'}(\gamma_{3}) e^{-i\pi\phi(n-\frac{1}{6})-i\vec{k}\cdot\vec{b}_{2}}\Bigr)\beta_{n-1,l-l'} \biggr] =\varepsilon	\alpha_{n,l} \;,
\end{aligned}
\end{align}
\begin{align}
\begin{aligned}
  & l\hbar\omega\beta_{n,l} - g\sum_{l'}\biggl[\Bigl(J_{l'}(\gamma_{1}) \alpha_{n,l+l'} \\
  & + J_{l'}(\gamma_{2})e^{-i\pi\phi(n+\frac{5}{6}) -i\vec{k}\cdot(\vec{b}_{1}-\vec{b}_{2})} \\
  & +  J_{l'}(\gamma_{3}) e^{i\pi\phi(n+\frac{5}{6})+i\vec{k}\cdot\vec{b}_{2}}\Bigr)\alpha_{n+1,l+l'} \biggr] =\varepsilon	\beta_{n,l}\;.
\end{aligned}
\end{align}
Here, we have introduced three different light parameters
\begin{align*}
   \gamma_{i} = \frac{eA_{i}a}{\hbar} \ .
\end{align*}
\begin{figure}[t]
	\includegraphics[width=1.00\columnwidth]{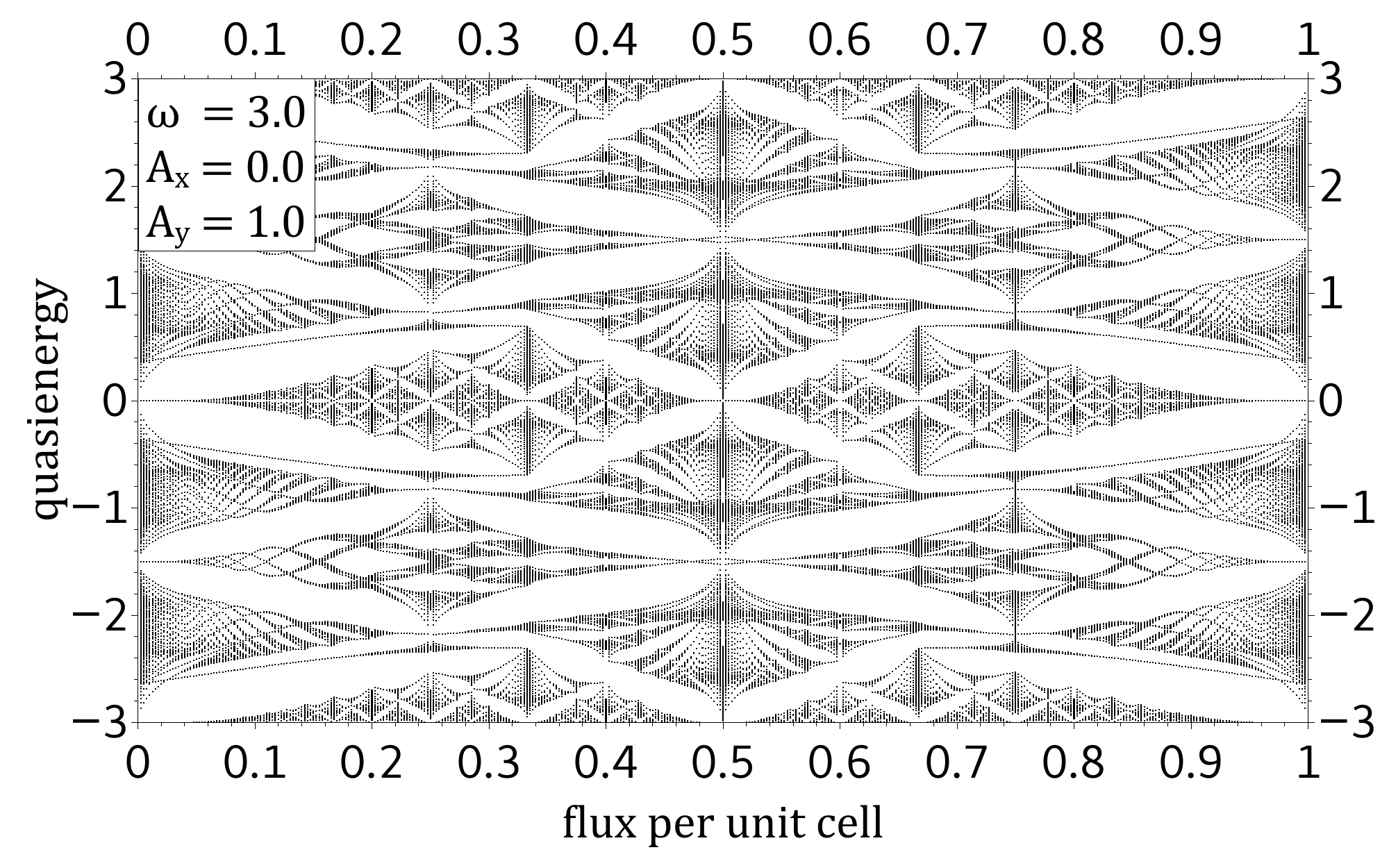}
	\caption{Hofstadter butterfly in presence of linearly polarized
      light where $A_{x}=0.0\,,\,A_{y}=1.0$ at a frequency $\omega$ of
      $3.0\,g/\hbar$. The effective amplitude for each
      hopping direction is governed by Eq.~\eqref{EffAmp}.}
\end{figure}
One should note that particle hole symmetry is conserved for linear light polarization, whereas it is not for circular polarization.
\subsection{Gap size}
To prepare for the following section, where we analyze the Chern
numbers of the static Hofstadter and the Floquet-Hofstadter problem,
we investigate the gap size occurring between the different
Floquet-Butterfly modes. To do so, we first clarify what is meant by
the gap between the different butterflies. We always calculate the gap size
numerically between the lowest band of the central Floquet
mode, being in the interval $[-\hbar\omega/2 , \hbar\omega/2 )$, and
the highest band of the minus one Floquet mode, lying in
$[-3\hbar\omega/2 , -\hbar\omega/2 )$ . Due to the periodicity of the
Floquet-Hofstadter spectrum on the quasienergy axis the gap between
neighboring modes is always the same. It is obvious that the
quasienergetic gap is not equal for all flux values, e.g., in
Fig.~\ref{Butterfly61} the lowest band of the central Floquet mode is
not constant as a function of the flux per unit cell.

\begin{figure}[t]
	\includegraphics[width=1.00\columnwidth]{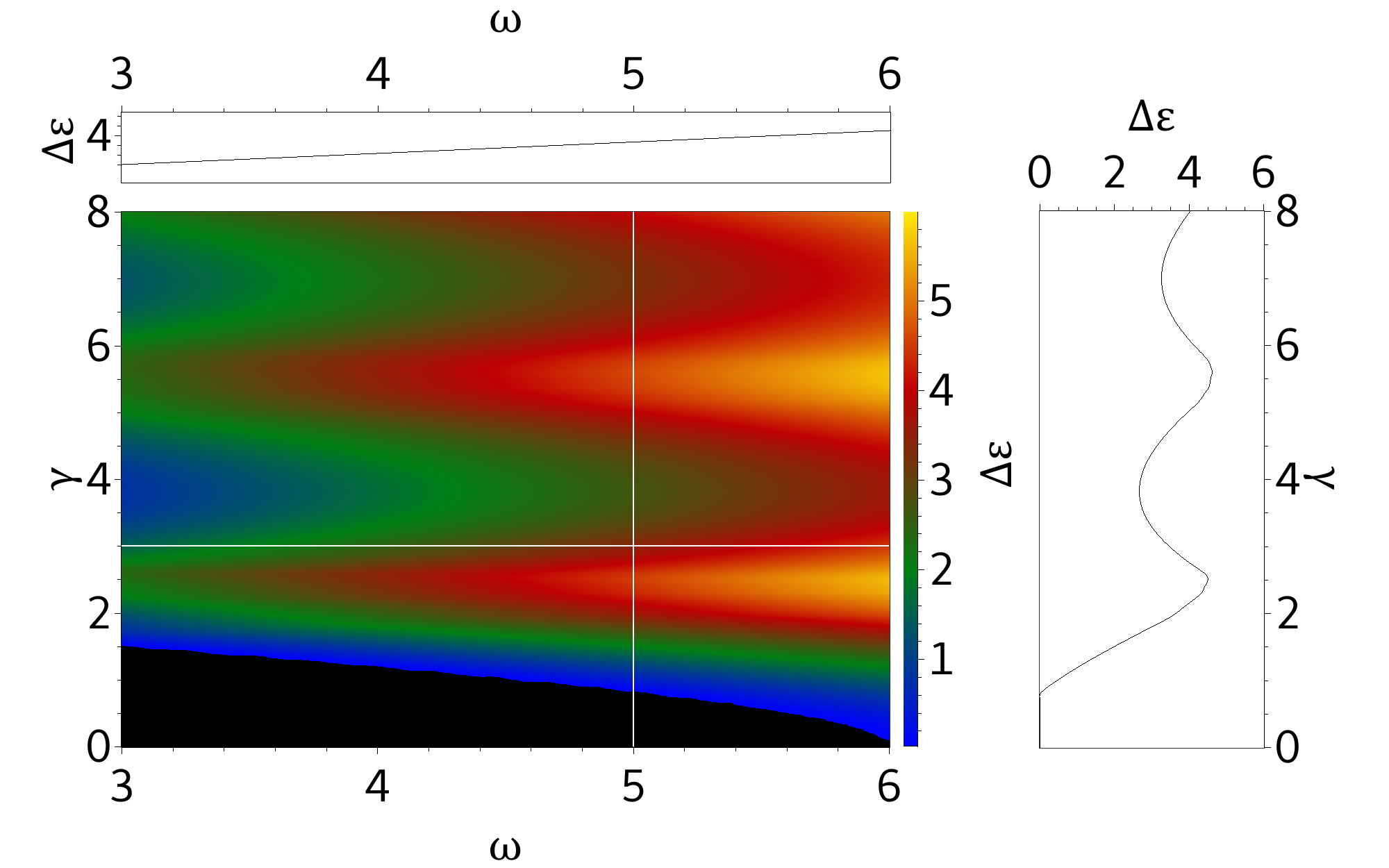}
	\caption{The gap $\Delta\epsilon$ of the Floquet-Hofstadter spectrum as a function
      of frequency $\omega$ and intensity $\gamma$ for circularly polarized light. In
      the investigated intensity range two maxima occur rising
      linearly with the driving frequency.}\label{GapCirc}
\end{figure}

\begin{figure}[t]
	\includegraphics[width=1.00\columnwidth]{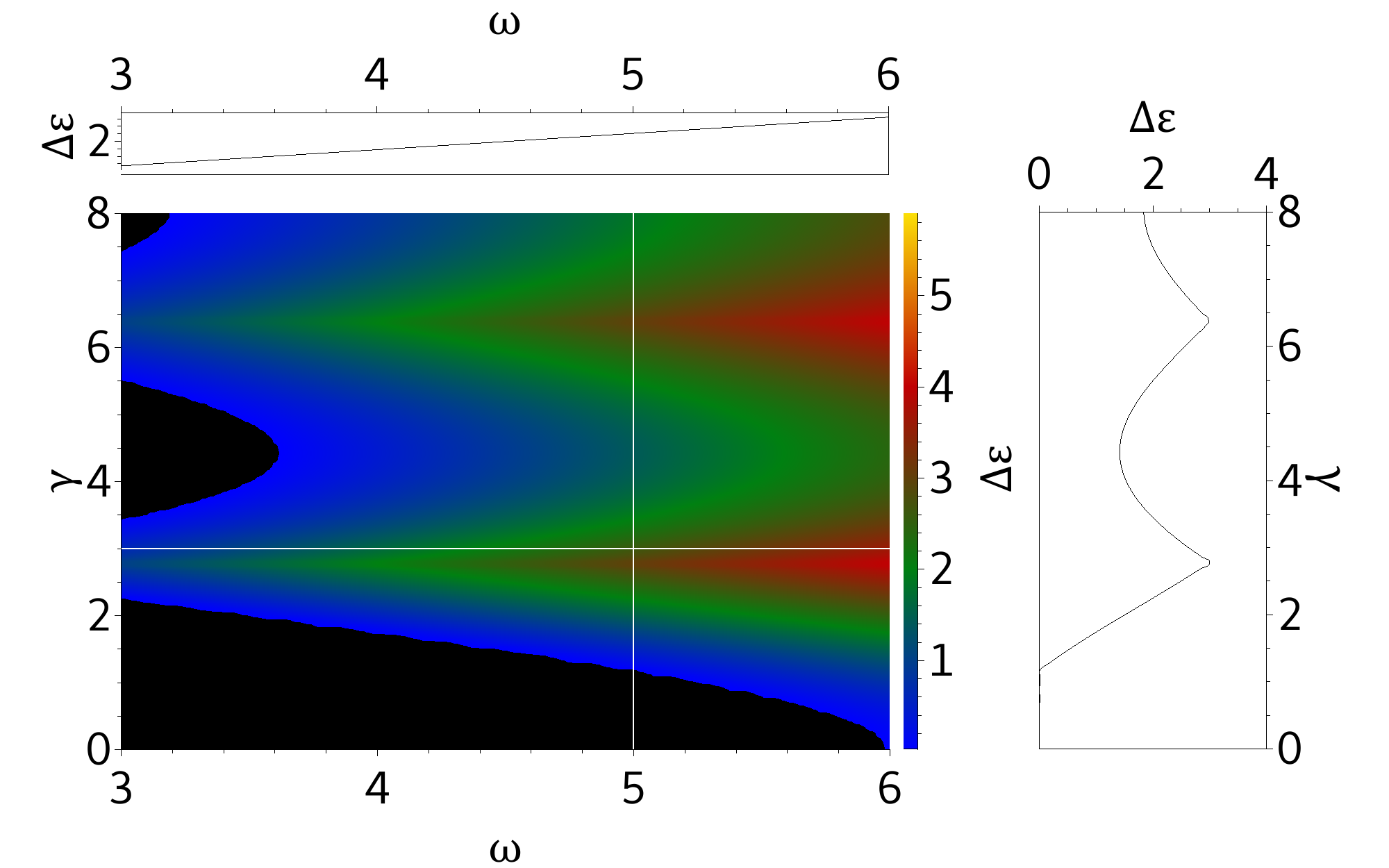}
	\caption{The polarization shows in $y$-direction. The gap shows
      qualitatively a similar behavior as for circularly polarized
      light but the gap is overall smaller.}\label{Gap01}
\end{figure}

\begin{figure}[t]
	\includegraphics[width=1.00\columnwidth]{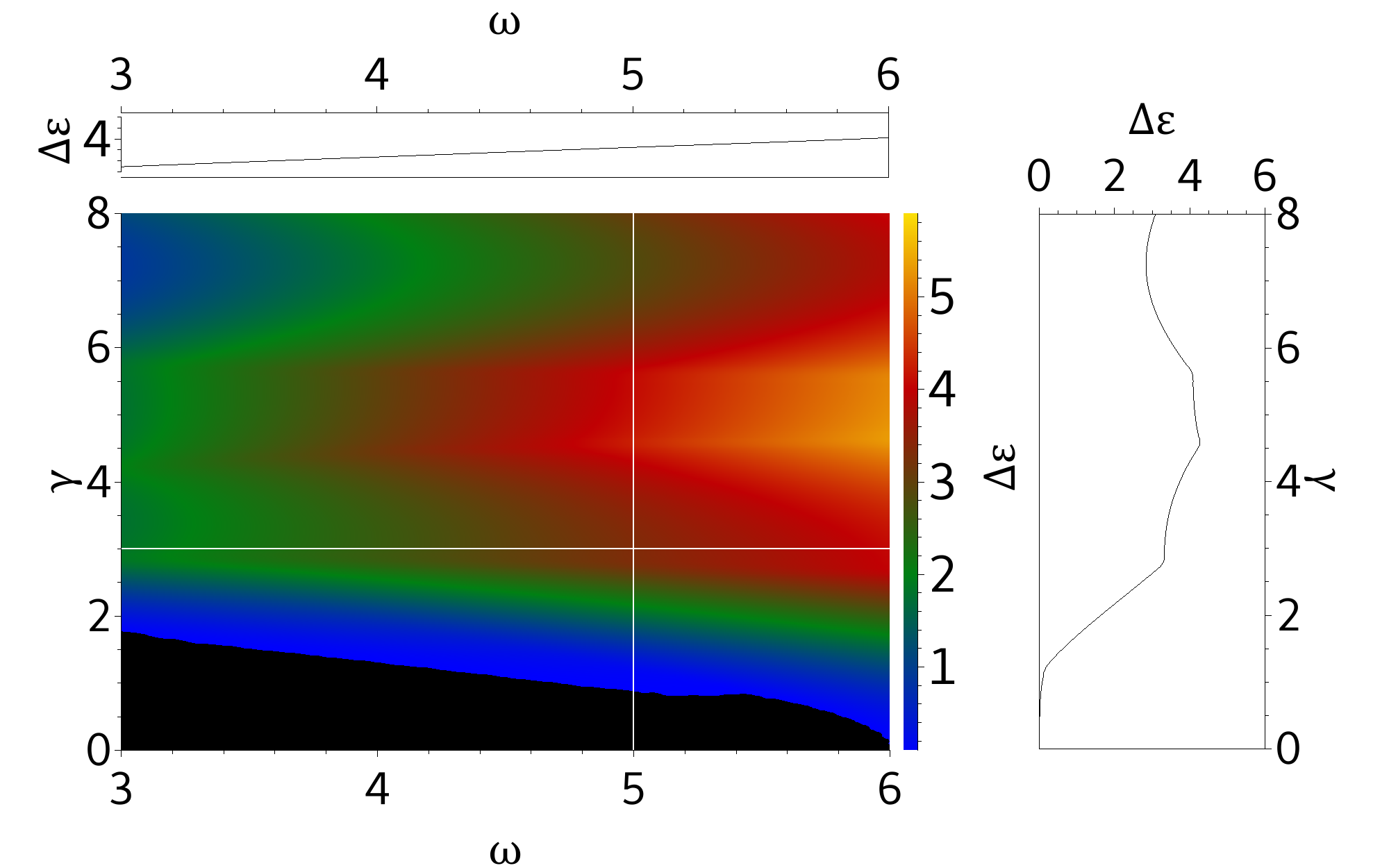}
	\caption{The polarization was set in $x$-direction, parallel to
      the $\vec{a}_{1}$ bonding. The two gap maxima at
      $\gamma=2.75\, eAa/\hbar$ and $\gamma=6.4\,eAa/\hbar$ at
      $\omega=5.0\,g/\hbar$ for $y$-polarization move together when
      changing the polarization into the $x$-direction.}\label{Gap10}
\end{figure}

As already mentioned, we focus on Chern numbers in the following
section. A change of the Chern number is always related to a band
touching. Hence, we are interested in the minimal gap as a function of
flux, denoted as $\Delta\varepsilon$ in Fig.~\ref{GapCirc},
\ref{Gap01}, and \ref{Gap10}. We refer to a gap between the
butterflies if there is no flux value where the lowest band of the
central Floquet mode and the highest band of the $n=-1$ mode touch.
The right plots of Fig.~\ref{GapCirc}, \ref{Gap01}, and \ref{Gap10}
show cuts through the contour plot at a frequency of
$5.0\, g/\hbar$. The upper plots show cuts at an intensity of
$3.0\, eAa/\hbar$. We can see that the gap size rises linearly with
the frequency. In anticipation to the following section, we can state
that the change of Chern numbers for $\omega=6.0\, g/\hbar$ is
for all polarizations only induced by band touchings of butterfly
bands lying in the same Floquet zone and not by touching of bands from
different Floquet modes.
\section{Topological Characterization}
\label{chernnumbers}
\subsection{Chern numbers} \label{Chern numbers}
Now, we turn to the topological characterization of the Hofstadter
bands \cite{TKNN,Hockendorf18,Hockendorf17,Nathan15,Wang16}, focusing
first on Chern numbers.  This topological invariants can be defined
for quantum states with two periodic parameters. They are calculated
by an integral of the Berry curvature $\vec{F}$ over a two-dimensional
compact surface $\mathcal{T}^2$, in this case the BZ in the quasienergy space of
$H_F$: Since the eigenstate $\ket{\alpha,\vec{k},t}$ with
$H_F\ket{\alpha,\vec{k},t} = \varepsilon_\alpha
\ket{\alpha,\vec{k},t}$ is periodic in time we can, according to
Eq.~\eqref{alphaFourier}, also formally write
\begin{align}
  \ket{\alpha,\vec{k},t} ={}& \sum_{n}^{}e^{i n \omega t}\ket{u_{\vec{k}\alpha}^n}\;,\label{FourierExpansion}
\end{align}
where $\alpha$ refers to a band index within one Floquet replica $n$.
The Chern number associated to a Floquet band
$\alpha$ with a Floquet state $\ket{u_{\vec{k}\alpha}^n}$ and
quasienergy $\varepsilon_{\vec{k}\alpha}$ is given by
\begin{align}\label{Cherndef}
  C_\alpha ={}& \frac{1}{2\pi}\int_{\text{BZ}}\dd^2k\,\vec{F}_\alpha(\vec{k})\cdot{\hat{z}}\;,
\end{align}
with the Berry curvature\cite{TKNN,Berry1984,Simon1983} given by
\begin{align}
  \vec{F}_\alpha(\vec{k}) ={}& \sum_{\beta \neq \alpha}\mathrm{Im}
                          \frac{\braket{u_{\vec{k}\alpha}^n|\vec{\nabla}_{\vec{k}}H_F|u_{\vec{k}\beta}^n}
                          \times
                          \braket{u_{\vec{k}\beta}^n|\vec{\nabla}_{\vec{k}}H_F|u_{\vec{k}\alpha}^n}}{(\varepsilon_{\vec{k}\alpha}
                          - \varepsilon_{\vec{k}\beta})^2}\;.
\end{align}
As long as the Floquet space is not truncated
$\vec{F}_\alpha(\vec{k})$ does not depend on the Floquet mode $n$. The
effect of a truncation of the Floquet space will be discussed in
Sec.~\ref{W3invariant}.

\begin{figure}[t]
	\includegraphics[width=1.00\columnwidth]{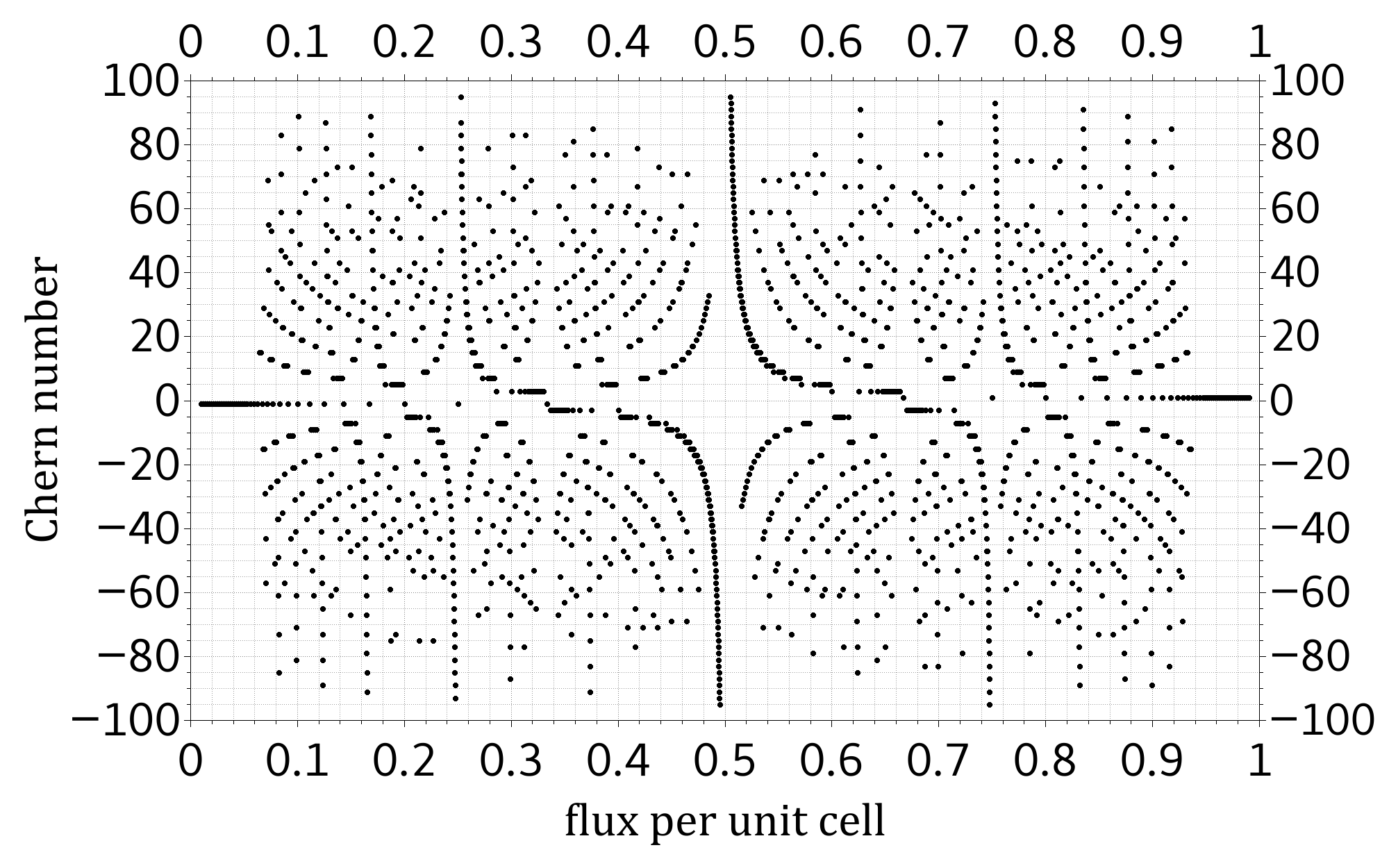}
	\caption{The Chern number of the state of lowest energy of the
      Hofstadter butterfly in dependence of the magnetic flux per unit
      cell. The flux values are all $p/q$ with $p$ co-prime to
      $q<101$.}\label{Chernnumber}
\end{figure}

Following Goldman \cite{Goldman}, we concentrate on the state of lowest
energy in one Floquet mode at given flux per unit cell as indicated in
Fig.~\ref{Butterfly}.
The Chern number is calculated numerically by the method proposed by Fukui \textit{et
al.}, Ref.~\onlinecite{FukuiChern}. Fig.~\ref{Chernnumber} reproduces the data of
Ref.~\onlinecite{Goldman} and extends it to a larger number of different flux values
$\phi=p/q$.
The computation is effectively limited by the fact that with growing
$q$ (being coprime to $p$) energy bands move closer to each other and
are increasingly difficult to resolve, an effect which is most
pronounced at fluxes near zero and unity. From a numerical perspective
the bands are degenerate impeding the use of the computation scheme by
Fukui \textit{et al.} constructed for non-degenerate band
structures.

Next, we analyze how polarized light affects the Chern numbers of the
Hofstadter butterfly. First, let us consider the case of circularly
polarized light.  The basis for this analysis are the
Eqs.~\eqref{FloHofalpha} and \eqref{FloHofbeta}. At high frequencies
the Hofstadter butterflies of the different Floquet modes are
quasienergetically separated since the distance of the Floquet modes
is governed by the photon energy. Hence, the change of Chern numbers
is induced by band touchings within the Floquet zone, as can be seen
in Fig.~\ref{GapCirc}, \ref{Gap01}, and \ref{Gap10}. At frequencies
large compared to the hopping energy the butterfly spectrum has an
overall gap in a broad intensity range. For intensities considered
in this section the topological phase transitions are all due to band
touchings within the same Floquet mode. Again, we concentrate on the
state of lowest quasienergy in the central Floquet mode, see
Fig.~\ref{Butterfly61}.  With the
Eqs.~\eqref{FloHofalpha}, \eqref{FloHofbeta} we were able to reproduce
several results of Mikami \textit{et al.}, Ref.~\onlinecite{Mikami16}, in
the limit of vanishing magnetic field strength.

As already stressed in
several works\cite{Mikami16,Seetharam15,Desbuquois17} the distribution
function in a driven system is in general not an equilibrium
distribution function. Despite that the Chern number maintains its
significance\cite{Mikami16} keeping in mind that one needs another
topological invariant to fully characterize a driven system
\cite{Rudner13}. We use the term \textit{ground state} as the state with lowest
quasienergy of the central Floquet mode, emphasizing that we do not
touch the question of the occupation of the Floquet modes in
general. However, we assume that the ground state depends adiabatically
on the intensity at least in the high frequency regime. As long as the
driving is far from resonances the driving does not significantly
change the ground state and with that the distribution function. This
also requires that the driving must not induce a heating of the
system. Hence, if we only occupy the ground state of the static system
we also assume that in the off resonantly driven system only the
ground state is occupied.

\begin{figure}[t]
	\includegraphics[width=1.0\columnwidth]{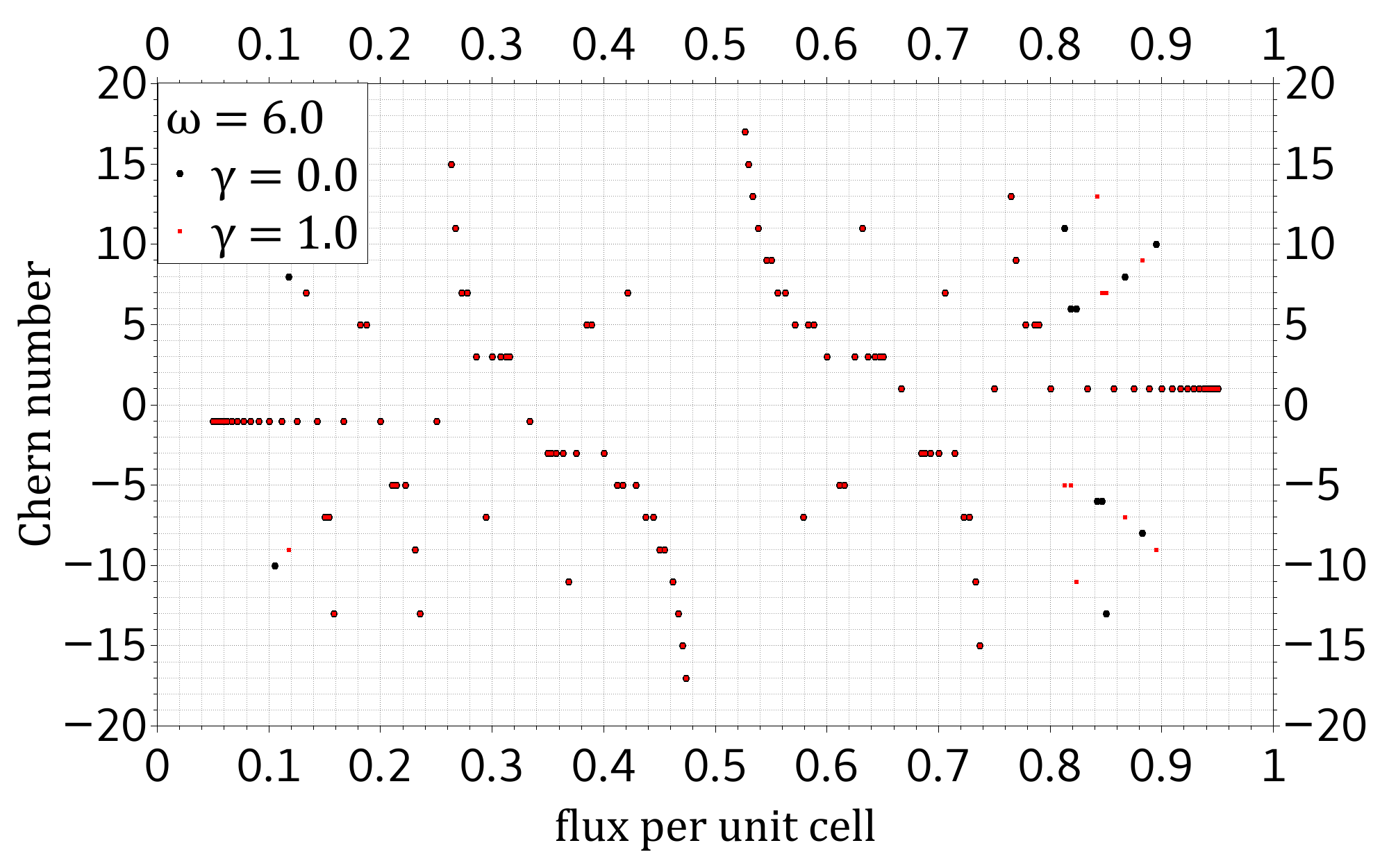}
	\caption{The ground state Chern number of the Floquet-Hofstadter
      spectrum: The frequency $\omega$ of the circularly polarized light was
      set to $6.0\,g/\hbar$. The black values show the case of
      vanishing light intensity $\gamma$ and the red values are calculated for
      an intensity of $1.0\, eAa/\hbar$.}\label{FloquetChern1}
\end{figure}

\begin{figure}[t]
	\includegraphics[width=1.0\columnwidth]{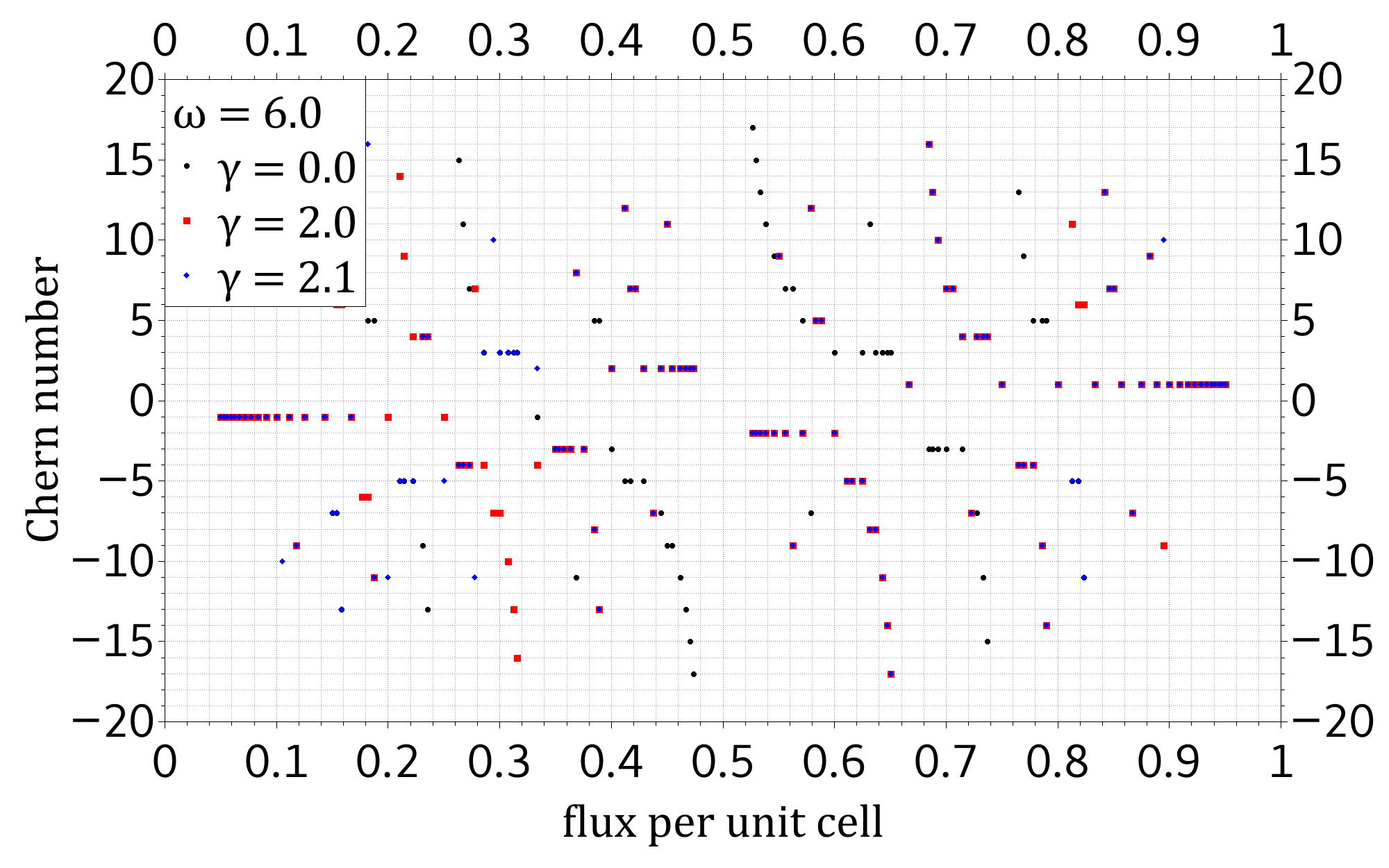}
	\caption{The distribution of the ground state Chern number in
      presence of circularly polarized light exhibits for intensities $\gamma$
      of $2.0\, eAa/\hbar$ and $2.1\, eAa/\hbar$ a rather different
      behavior as for vanishing intensity. The plot shows flux values
      for $q<21$.}\label{FloquetChern3}
\end{figure}

\begin{figure}[t!]
	\includegraphics[width=1.0\columnwidth]{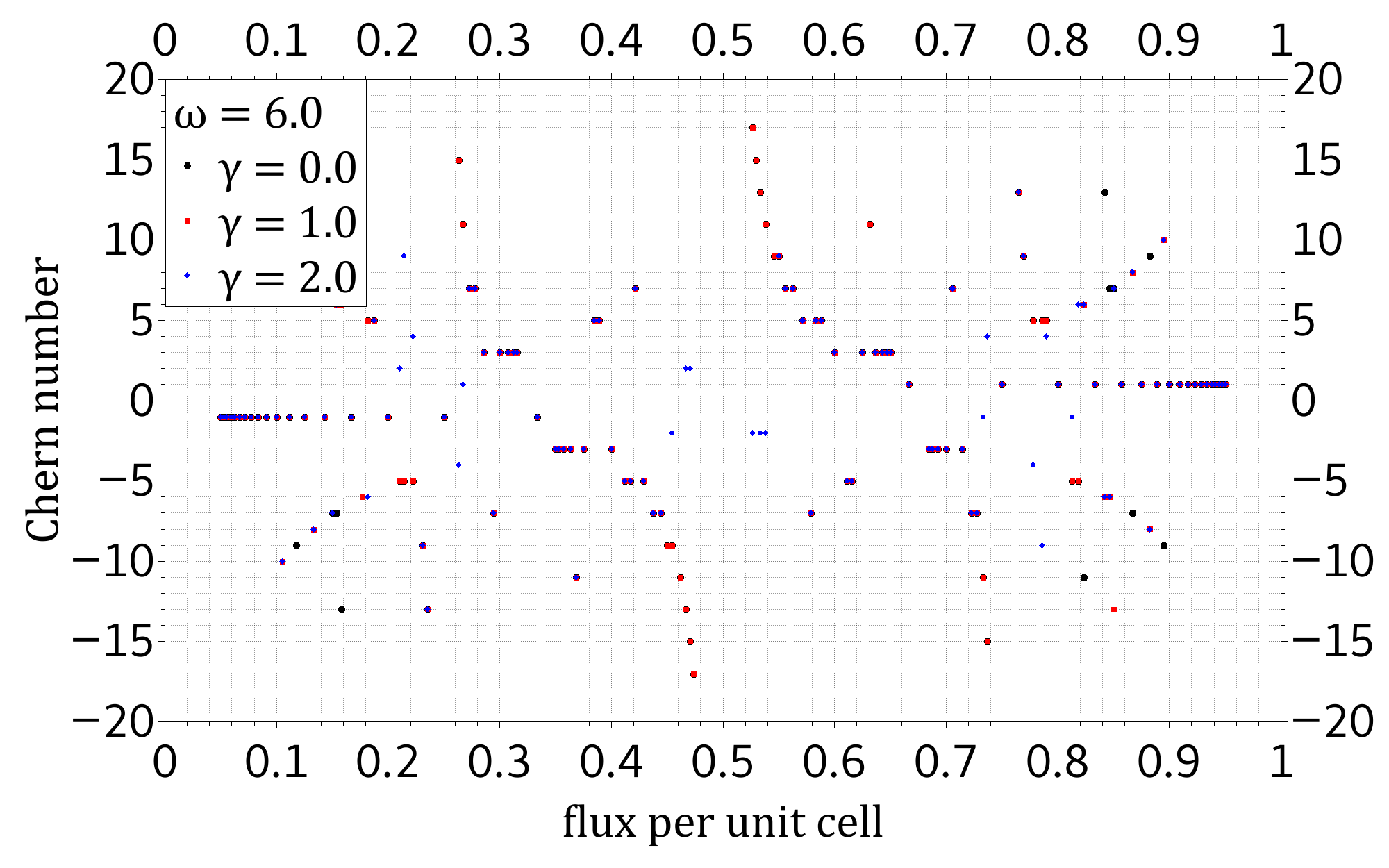}
	\caption{In the case of linear polarization with
      $A_{x}=0.0 , A_{y} = \gamma$ the distribution of the Chern
      number is for linearly polarized light similar affected as for
      circularly polarized light. The frequency $\omega$ was again fixed at
      $6.0\, g/\hbar$ and the intensity is governed by
      Eq.~\eqref{EffAmp}.}\label{LinPolChern}
\end{figure}

Our Chern number computations are done in
the off-resonant frequency regime. Hence, the ground state of
the driven system undergoes the topological phase transitions
presented in Fig.~\ref{FloquetChern1}, \ref{FloquetChern3}, and
\ref{LinPolChern}. For a vanishing light parameter $\gamma$ and high frequencies,
the ground state Chern numbers are the same as in the
undriven case, see Fig.~\ref{Chernnumber}.
In Fig.~\ref{FloquetChern1} most Chern numbers coincide with the case
of a vanishing intensity. When the intensity is further increased the
ground state Chern number exhibits a rather different behavior. Even
small intensity changes can have a vast influence on the Chern number
\cite{Mikami16}, see Fig.~\ref{FloquetChern3}. Since the
Floquet-Hofstadter spectrum gets twisted in presence of circularly
polarized light and keeps particle-hole symmetry for linearly polarized
light it is obvious that the band structure of graphene is differently
affected for the two polarization states. The deformation of the band
structure and the associated gap closing and opening is related to the
change of Chern numbers. Hence, we investigate as well the influence
of linearly polarized light on the distribution of Chern
numbers. Similar to the case of circularly polarized light, for rather
small intensities only few Chern numbers deviate from the static Chern
number distribution. An increase of the intensity leads to a
significantly different behavior, as shown in Fig.~\ref{LinPolChern}.
For circularly polarized light the ground state is uniquely
defined. Whereas, for linearly polarized light this is not the case
for all flux values. At flux values of, e.g., $6/11$, $6/13$ or
$3/17$ a band crossing of the ground state occurs. This effect can be
seen at eight different flux values for $q<21$. The occurrence of the
band crossing of the ground state seems not to follow a simple rule.
\subsection{$W_{3}$-invariants}\label{W3invariant}
The topological invariant $\nu_3$ associated with the third homotopy group of the
periodic unitary maps $\{U_{\vec{k}}(T)\}$ is given in $\mathbb{R}^3$ by
\begin{align}
  \nu_3[U_{\vec{k}}] ={}&  \frac{1}{24\pi^{2}}\int_{\text{BZ}}\dd^{3}k\,
  \varepsilon_{\alpha\beta\gamma}\, \mathrm{tr}\big[
  (U^{-1}_{\vec{k}}(T)\partial^{\alpha}U_{\vec{k}}(T))\nonumber\\
 {}&  \cdot(U^{-1}_{\vec{k}}(T)\partial^{\beta}U_{\vec{k}}(T))(U^{-1}_{\vec{k}}(T)\partial^{\gamma}U_{\vec{k}}(T))
  \big]\;.\label{nu3}
\end{align}
Rudner \textit{et al.}, Ref.~\onlinecite{Rudner13}, have devised an
invariant specifically designed for the characterization of
periodically driven systems. The idea is to replace in Eq.~\eqref{nu3}
one $k$-dimension with the time and choose a unitary matrix
$\tilde U_{\vec{\mu}}$ which is periodic in time and topologically
equivalent to a time evolution operator $U_{\vec{\mu}}$\footnote{More precisely, the following
  conditions have to be fulfilled\cite{Rudner13}: 1)
  ${\tilde U}(\vec{k},T) = \mathbb{I}$. 2) There should exist a
  one-parameter family of evolution operators $\{ U_s:s\in[0,1] \}$
  which interpolates between $U$ and $\tilde U$ as follows:
  $U_{s=0}(\vec{k},t) = U(\vec{k},t)$ and
  $U_{s=1}(\vec{k},t) = \tilde U(\vec{k},t)$. 3)
  $\tilde U(\vec{k}, T)$ has to maintain a gap around $\varepsilon_s$
  with $\varepsilon_{s=0} = \varepsilon$, $\varepsilon_{s=1} = \pi/T$
  and a smooth interpolation from $s=0$ to $s=1$.},
\begin{align}
  W_{3}[\tilde U_{\vec{\mu}}] ={}& 
  \frac{1}{24\pi^{2}}\int_{[0,1)^{3}}\dd^{3}\mu\,
  \varepsilon_{\alpha\beta\gamma}\, \mathrm{tr}\big[ (\tilde
  U_{\vec{\mu}}^{-1}\partial^{\alpha}\tilde U_{\vec{\mu}})\nonumber\\
  {}& \cdot(\tilde U_{\vec{\mu}}^{-1}\partial^{\beta}\tilde
  U_{\vec{\mu}})(\tilde U_{\vec{\mu}}^{-1}\partial^{\gamma}\tilde
  U_{\vec{\mu}}) \big]\;,\label{W3Utilde}
\end{align}
where the cube $[0,1)^3$ is spanned by two normalized in-plane wave
vectors and the time $t/T$ with $\vec{\mu} \in [0,1)^3$. The indices
$\alpha$, $\beta$, $\gamma$ are given modulo 3 and
$\partial^{\alpha} \equiv \partial^{\mu_\alpha}$.  This new invariant
is related to the lowest quasienergy gap in the central Floquet
mode. The relation between the $W_3$ invariants of different gaps $\xi_{n}$ with $\exp(i\xi_{n})\in\mathbb{S}^{1}$
around quasienergies $\varepsilon_{\nu}$ is closely
related to Chern numbers $C^{\nu}$ of appropriate
bands $\nu$. It is given
by\cite{Hockendorf17}
\begin{align}
  W_3[\tilde U,\xi_b] ={}& W_3[\tilde U,\xi_a] - \sum_{\nu =
                           \nu_1,\ldots,\nu_k} C_\nu^{(3)}
                           \Big\vert_{\mu_3\equiv \frac{t}{T}=1}\;,
\end{align}
where the bands $\nu_1,\ldots,\nu_k$ are the bands one passes through
when the value $\xi$ changes from gap at $\xi = \xi_a$ to the gap at
$\xi = \xi_b$. The Chern number is calculated by\cite{Hockendorf17}
\begin{align}
  C_\nu^{(\alpha)} ={}&
                    \frac{1}{2\pi
                    i}\int_0^1\int_0^1\dd\mu_{\alpha-1}\dd\mu_{\alpha+1}\,
                    [\epsilon_{\alpha\beta\gamma}(\partial^\beta(S^\dagger\partial^\gamma
                    S))]_{\nu\nu}\;,
\end{align}
where $C_{\nu}^{(3)}|_{\mu_{3}=1}$ is equivalent to Eq.~\eqref{Cherndef}.
The columns of the matrix $S$ contain the eigenvectors of $U_{\vec{\mu}}$.
Clearly, the full computation of the invariant
constructed in Ref.~\onlinecite{Rudner13} is more complicated
\cite{Hockendorf17} than for Chern numbers \cite{FukuiChern}.

The calculation scheme suggested by Rudner \textit{et al.},
Ref.~\onlinecite{Rudner13}, in frequency space is described in the following. In
order to calculate the generalized topological invariant for driven
systems one first computes the Chern number of all bands below
the investigated gap of a truncated Floquet matrix. The generalized invariant is then given by the
sum off all Chern numbers below this gap. In Fig.~5 in
Ref.~\onlinecite{Rudner13} the lowest band of the truncated Floquet matrix has a Chern number
$C_{0}$ different from $C_{F}$. The reason why that Chern number is
not $C_{F}$ is due to the truncation. As already
shown by Shirley\cite{Shirley63,Shirley65}, from the Fourier expansion
in Eq.~\eqref{FourierExpansion} it follows that the corresponding
eigenvector to a quasienergy $\varepsilon_{\lambda}$ differs from the
eigenvector of the quasienergy $\varepsilon_{\lambda}+\hbar\omega$
only by an index shift of the entries and a phase $\phi$ which one is
free to choose\cite{Shirley63}
\begin{align} \label{Shiftprop}
   \varepsilon_{\lambda} \leftrightarrow \begin{pmatrix}
   \vdots \\ u_{\lambda}^{-2} \\ u_{\lambda}^{-1} \\ u_{\lambda}^{0} \\ u_{\lambda}^{1} \\ u_{\lambda}^{2} \\ \vdots
   \end{pmatrix} \Longleftrightarrow \varepsilon_{\lambda} + \hbar\omega \leftrightarrow e^{i\phi}\begin{pmatrix}
   \vdots \\ u_{\lambda}^{-3} \\ u_{\lambda}^{-2} \\ u_{\lambda}^{-1} \\ u_{\lambda}^{0} \\ u_{\lambda}^{1} \\ \vdots
\end{pmatrix}\;,
\end{align}
where $\lambda$ labels a discrete set of quantum numbers, e.g., spin or
sublattice degrees. This holds equivalently for arbitrary shifts
$n\hbar\omega$, with $n\in \mathbb{Z}$, of the quasienergy. It shows
that the Chern number $C_{\varepsilon_{\lambda}}$ of a band described
by $\varepsilon_{\lambda}$ has to be equal to the Chern number of the
shifted band
\begin{align}
   C_{\varepsilon_{\lambda}} = C_{\varepsilon_{\lambda}+n\hbar\omega}\, .
\end{align}
This means for the numerics that if we assume that only a finite
number of eigenvector entries are different from zero we have to
choose the truncation of the Floquet modes large enough in order to
achieve convergence of these. Let us assume that we have to limit the
number of Floquet modes to $m$ in order to achieve convergence of the
central quasienergy $\varepsilon_{\lambda}$ up to a needed
precision. If the eigenvector corresponding to
$\varepsilon_{\lambda}\pm m\hbar\omega$ is computed this eigenvalues and
eigenvectors are in general not converged leading to different results
in the quasienergy spectrum as well as Chern numbers. To sum up, these
non converged Chern numbers might lead to an incorrect topological
characterization. Indeed, H\"ockendorf \textit{et al.} give a counterexample in
Ref.~\onlinecite{Hockendorf17} where the summation over Chern
numbers suggested by Rudner \textit{et al.} \cite{Rudner13} fails to give the correct
$W_{3}$-invariant. The authors consider a spin-$1/2$ rotation
described by the Hamiltonian
\begin{align} \label{H}
   H_{w} = 2\pi w\,\vec{f}(\mu_1, \mu_2)\cdot\vec{\sigma}
\end{align}
together with the corresponding time evolution operator
\begin{align}\label{U1}
   U(\vec{\mu}) = e^{-iH \mu_3}\;,
\end{align}
where the $\mu_i$ are chosen as in Eq.~\eqref{W3Utilde},
$w\in\mathbb{Z}$ and the function $\vec{f}$ is a map from the square
to the unit sphere $\vec{f}:[0,1]^{2} \rightarrow \mathbb{S}^2$. For
further details we refer to Ref.~\onlinecite{Hockendorf17}. The
corresponding two bands have Chern number $\pm 1$, whereas
$W_{3}=2w$. Despite the fact that the Hamiltonian $H$ is
time-independent, the system exhibits a nontrivial topology when
investigating its time evolution. We are now in the position to
clarify why the summation over Chern numbers proposed by Rudner \textit{et al.} fails
for this example. If we apply Floquet theory to the Hamiltonian
\eqref{H} with vanishing driving amplitude and frequency
$\omega = 2\pi/T$, we create Floquet copies identical to the undriven
system. This implies that the Chern numbers of the two bands in each
Floquet zone are equal to the Chern numbers of the undriven system,
i.e., they are $\pm 1$. Therefore, summing over all Floquet copies
yields a topological invariant of zero in contrast to the correct
$W_{3}$-invariant of $2w$. The above mapping
$\vec{f}(\mu_1,\mu_2)$ can be easily constructed by concatenating
three different mappings. The first one is
shifting and stretching the square
\begin{align}
   \vec{s}(\vec{\mu}) &: [0,1]^{2} \to [-1,1]^{2}\\
   \vec{s}(\vec{\mu}) &: \begin{pmatrix} \mu_1 \\ \mu_2 \end{pmatrix} \mapsto \begin{pmatrix} 2\mu_1-1 \\ 2\mu_2-1 \end{pmatrix}\ .
\end{align}
The second one is a map from a square to a circle
\begin{align}
   \vec{c}(\vec{\mu}) &: [-1,1]^{2} \to \{ |\vec{\mu}|\leq 1 : \vec{\mu}\in \mathbb{R}^{2} \} \\
   \vec{c}(\vec{\mu}) &: \begin{pmatrix}\mu_1\\ \mu_2 \end{pmatrix} \mapsto \begin{pmatrix} \mu_1\sqrt{1-\tfrac{\mu_2^{2}}{2} }\\ \mu_2\sqrt{1-\tfrac{\mu_1^{2}}{2} } \\  \end{pmatrix}
\end{align}
and the third one maps a circle to a sphere
\begin{align}
   \vec{b}(\vec{\mu}) &: \{ |\vec{\mu}|\leq 1 : \vec{\mu}\in \mathbb{R}^{2} \} \to \{ |\vec{\mu}|= 1 : \vec{\mu}\in \mathbb{R}^{3} \} \\
   \vec{b}(\vec{\mu}) &: \begin{pmatrix}\mu_1\\\mu_2 \end{pmatrix} \mapsto \begin{pmatrix}\frac{\mu_1}{n}\sin(\pi n)\\ \frac{\mu_2}{n}\sin(\pi n) \\ \cos(\pi n) \end{pmatrix} 
\end{align}
with $n = \sqrt{\mu_1^2+\mu_2^2}$. This finally yields the sought mapping $f$,
\begin{align}
   \vec{f}(\mu_1,\mu_2) = \vec{b}\big( \vec{c}\big( \vec{s}(\vec{\mu}) \big)\big)\ .
\end{align}
Let us now consider the case $w=1$. The operator in Eq.~\eqref{U1} can
be interpreted as a time evolution operator of a time-independent Hamiltonian
\begin{align} \label{H1}
   H_{w=1} = \frac{2\pi}{T} \vec{f}(\mu_1,\mu_2)\cdot\vec{\sigma}\;,
\end{align}
which has however a trivial but periodic time evolution with a period
$T=1$. Note that the eigenvector matrix $\Lambda$ of $H_{w=1}$ allows
for the transformation
\begin{align}
   \Lambda\vec{f}(\mu_1,\mu_2)\cdot\vec{\sigma}\Lambda^{\dagger} = \sigma^{z}\;.
\end{align}
Rudner \textit{et al.}, Appendix C in Ref.~\onlinecite{Rudner13}, made the attempt to map all time-independent flat band Hamiltonians onto
\begin{align}\label{HP}
   H_{P}(\vec{\mu}) = \frac{2\pi}{T} P(\vec{\mu})\;,
\end{align}
with $P(\vec{\mu})$ being a projection operator. The authors were able
to show that for these class of Hamiltonians the $W_{3}$-invariant
$W_3[U]$ is equal to the Chern number of the bands with quasienergy
$\varepsilon = -2\pi/T$. One should stress that the quasienergies of a
Hamiltonian of the form \eqref{HP} are degenerate everywhere whereas
the Chern numbers are still defined. But there is a class of flat band
Hamiltonians which cannot be mapped onto $H_{P}$. One example is
$H_{w=1}$ since the spectra differ. Here, the mentioned relation between
the $W_{3}$-invariant and the Chern number fails. Furthermore, very
much as in Appendix \ref{SecondExample}, one can show that the
quasienergies of the Floquet Hamiltonian corresponding to Eq.~\eqref{H1}
are both zero and thus degenerate everywhere. Nevertheless, the Chern
numbers are $\pm 1$ and summation over these will never lead to the
same number of edge modes as predicted by $W_{3}=2$. This shows that
the summation over Chern numbers of the truncated Floquet Hamiltonian
is not justified for every system.  Another example is discussed in
Appendix \ref{SecondExample}. Despite this counterexamples, the
summation over Chern numbers over the truncated Floquet matrix and the
calculation of the $W_{3}$-invariant for graphene without magnetic
field show a striking accordance, see Appendix \ref{W3graphene}.

In order to assure the correctness of the topological invariant we
applied the algorithm proposed by H\"ockendorf \textit{et al.},
Ref.~\onlinecite{Hockendorf17}, to compute numerically the
$W_{3}$-invariant for the Floquet-Hofstadter spectrum at $p/q =
1/3$. The result is plotted in Fig.~\ref{W3_2}.
\begin{figure}[t]
	\includegraphics[width=1.0\columnwidth]{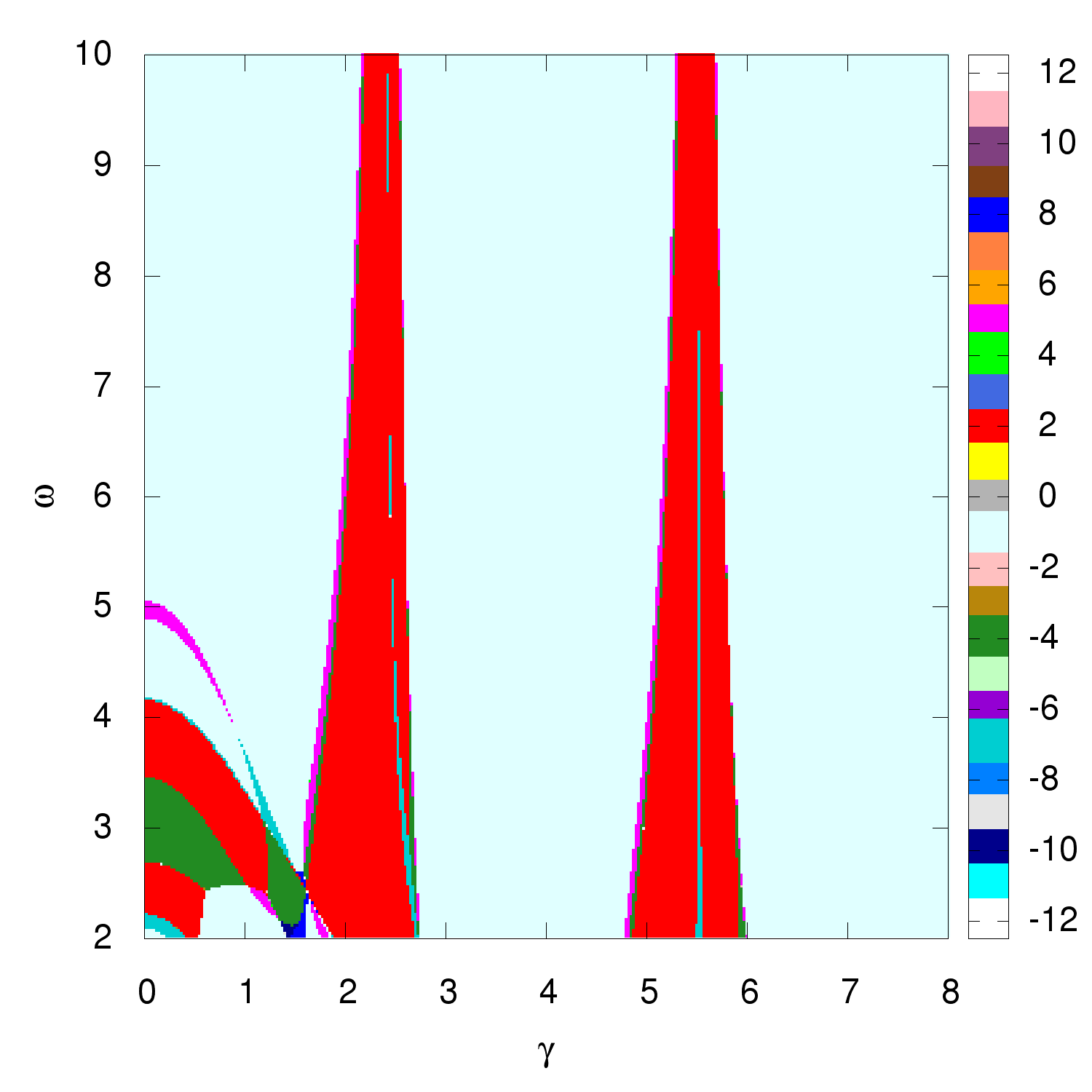}
	\caption{The Chern number of the state with lowest energy of the
      central Floquet zone for a flux per unit cell of $p/q = 1/3$ and
      circularly polarized driving.}\label{C2_F}
\end{figure}
To have a comparison to the static topological invariants, we first
compute the Chern number of the state with lowest energy of the
central Floquet zone for a flux per unit cell of $p/q = 1/3$ and
circularly polarized driving. The three dimensional momentum-time BZ is
discretized by 200$\times$200$\times$200 points together with 30
Floquet replicas. The resulting Chern numbers are plotted in
Fig.~\ref{C2_F} for different amplitudes $\gamma$ and frequencies
$\omega$ of the driving field. In the left lower region of
Fig.~\ref{C2_F}, inside the arc from
$(\gamma,\omega)=(0.0\, eAa/\hbar,5.1\, g/\hbar)$ to
$(\gamma,\omega)=(1.9\, eAa/\hbar,2.0\, g/\hbar)$, we can not trust
the numerical values. The reason can be understood by investigating
the band structure. In the parameter space where
$\hbar\omega < 6.0\,g$ the bands of the Floquet-Hofstadter spectrum
overlap and the Chern numbers are not well defined. With rising
intensity the degeneracies are lifted and anticrossings
occur. Moreover, there are $(\gamma,\omega)$ regions where no gap
between the lowest and the second lowest exists but the bands are
nowhere degenerate, see Appendix \ref{AppNoGap}.

\begin{figure}[t]
	\includegraphics[width=1.0\columnwidth]{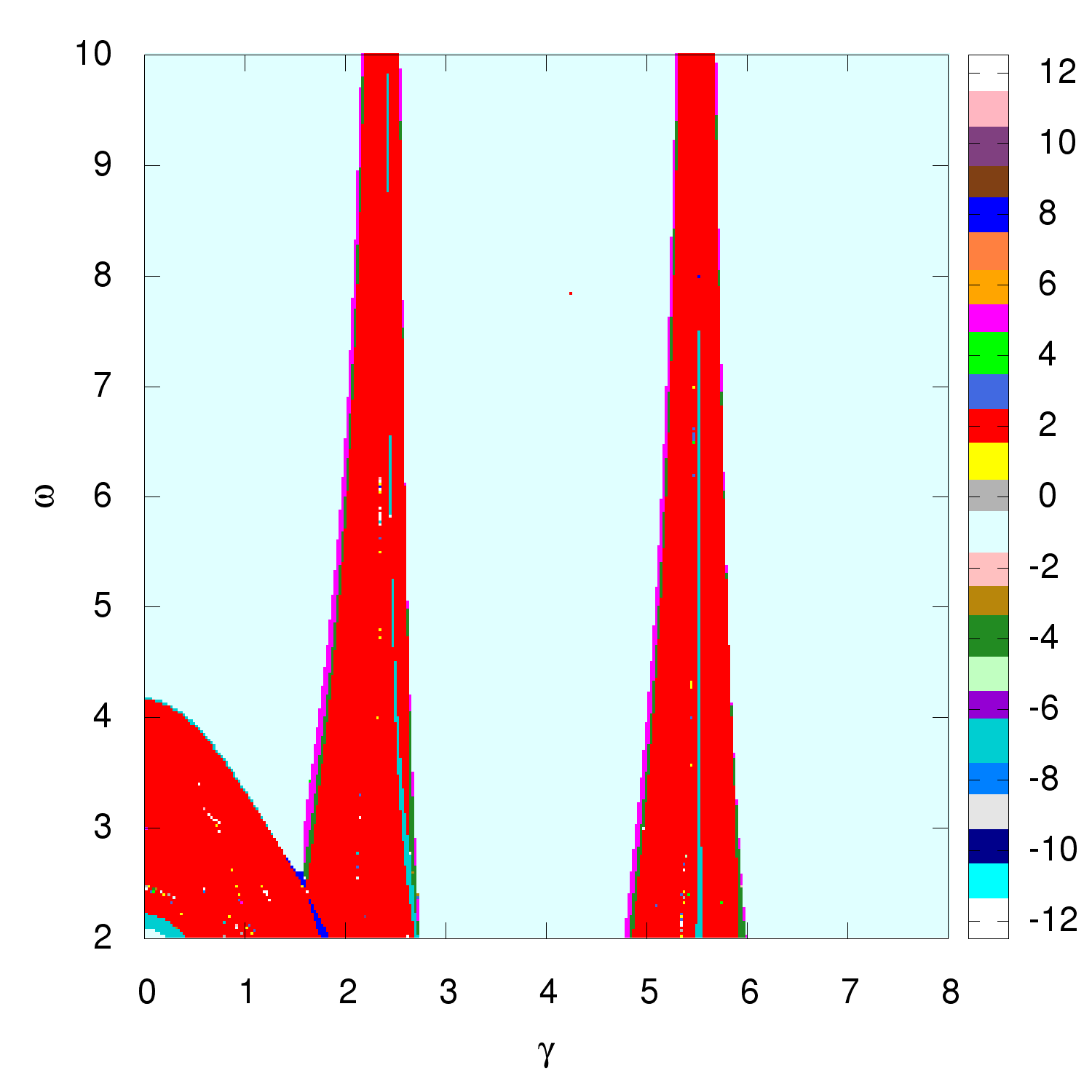}
	\caption{The sum over all Chern numbers below $\varepsilon=0$
      computed from the truncated Floquet Hamiltonian with the same
      flux and polarization as in Fig.~\ref{C2_F} and
      Fig.~\ref{W3_2}.}\label{C2_sum}
\end{figure}

\begin{figure}[t]
	\includegraphics[width=1.0\columnwidth]{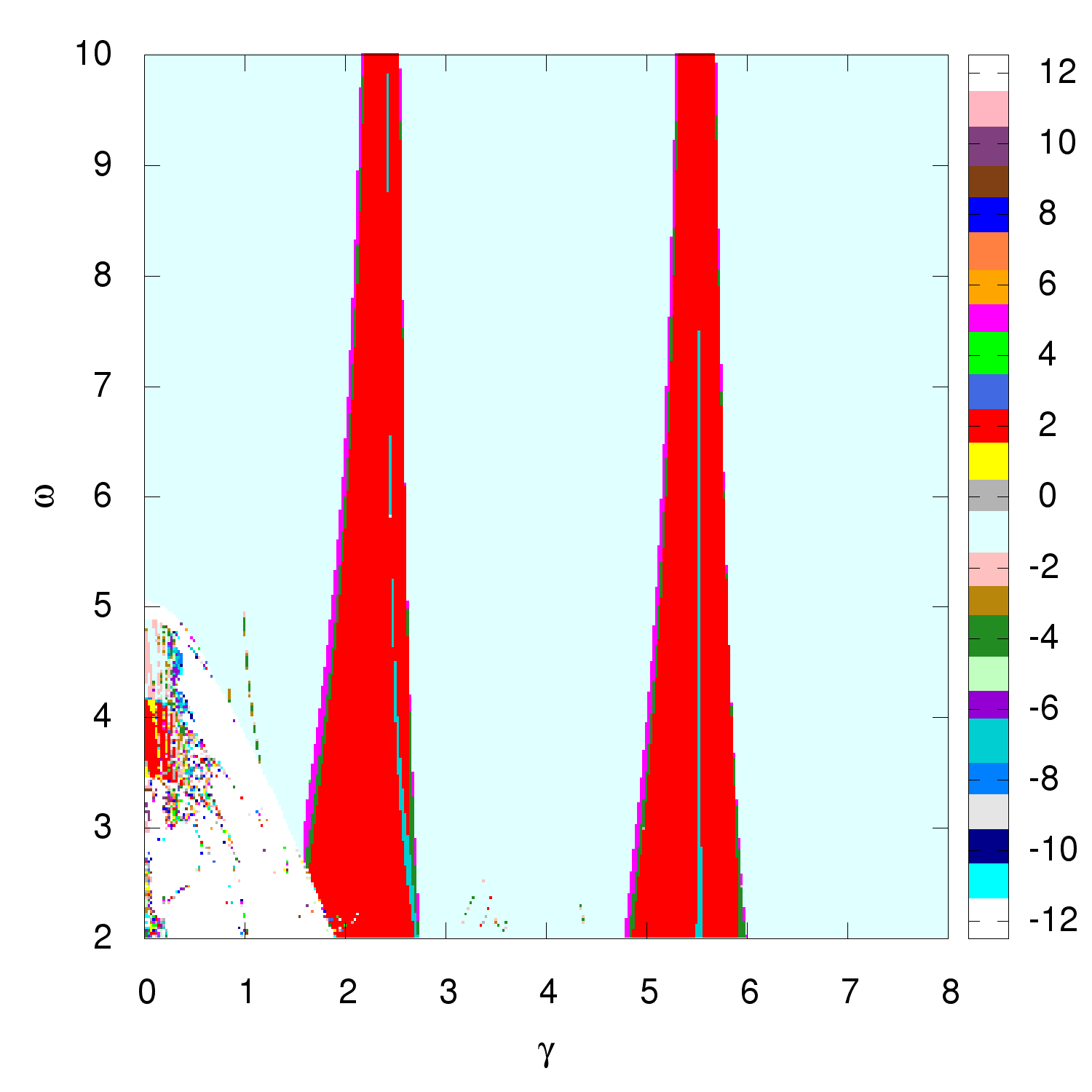}
	\caption{The $W_{3}$-invariant computed with the algorithm by
      H\"ockendorf  \textit{et al.}\cite{Hockendorf17} for the Floquet-Hofstadter
      spectrum at $p/q = 1/3$. The driving was circularly
      polarized.}\label{W3_2}
\end{figure}

In the last step we apply the $W_3$ calculation scheme following
Ref.~\onlinecite{Rudner13} as mentioned before. The same flux and polarization is used as for Fig.~\ref{C2_F}. The result is plotted in Fig.~\ref{C2_sum}. In the following, we compare the results
of both $W_3$ calculations and contrast them against the corresponding Chern numbers.

\begin{figure}[t]
	\includegraphics[width=1.0\columnwidth]{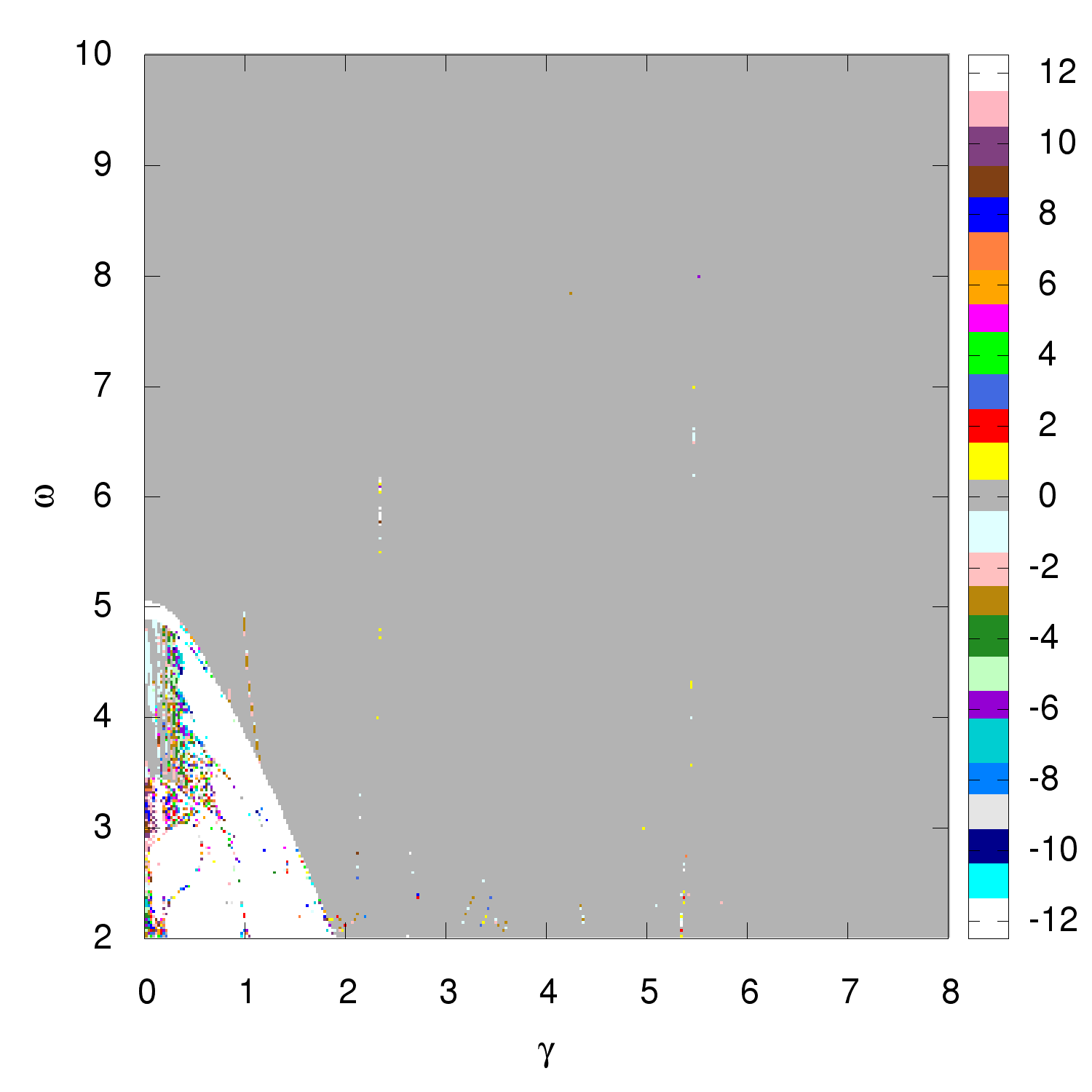}
	\caption{The difference between the sum over Chern numbers and the
      $W_{3}$-invariant. Grey regions show parameter configurations
      $(\gamma,\omega)$ where the Chern number sum and the
      $W_{3}$-invariant coincide.} \label{W3_Diff_2}
\end{figure}

The difference between both results for the $W_3$-invariant is depicted in
Fig.~\ref{W3_Diff_2}. The comparison shows that apart from zones close to
topological phase transitions the results coincide.
Interestingly, the Chern number itself show as well a great agreement
with both the sum over the Chern numbers and $W_{3}$. This justifies
once more the topological characterization presented in
Sec.~\ref{Chern numbers}. Using the connection between edge modes and
the $W_{3}$-invariant which has been proven in
Ref.~\onlinecite{Rudner13}, this result allows for the prediction of
the number of edge modes in this driven system.

Furthermore, we would like to stress that although the here presented
topological characterization is different from the one presented in
Ref.~\onlinecite{Kooi18} by Kooi \textit{et al.} the Chern numbers for
a flux per unit cell of $p/q = 1/3$ agree with our results up to the
sign of the $W_{3}$-invariants due to a different sign choice of the
driving frequency.
%
%
%
\section{Summary}
\label{summary}
In this paper we presented an explicit and rigorous treatment of the
Hofstadter problem on the hexagonal lattice. One important result is the
explicit proof of the periodicity of the Hofstadter
butterfly: Depending on whether
the numerator of the flux per unit cell is even or odd the periodicity
of the fractal spectrum is different.
To understand how illumination of graphene with both circularly and
linearly polarized light in presence of a magnetic field will effect
the fractal spectrum we unified the Hofstadter butterfly with the Floquet theory. These two polarization modes lead to clearly
different scenarios. Circularly polarized
light in combination with a magnetic field is able to lift the
symmetry of the quasienergy spectrum around zero energy, whereas linearly
polarized light is not, as shown by representative data. Furthermore,
we investigated the gap size between different Floquet modes of the
Floquet-Hofstadter spectrum.

To investigate the topological properties of this dynamical system, we
studied the Chern number of the state with lowest quasienergy in the
central Floquet mode for different flux values. Limiting the
computations to the high frequency regime we were able to identify
that the topological phase transitions induced by the external
radiation field are only caused by gap closings and openings of
butterfly bands and not by touching of different Floquet modes. For
vanishing intensity the computed Chern numbers coincide with the ones
of the undriven system.  Furthermore, we found that the system
undergoes several topological phase transitions when tuning the flux
per unit cell or the intensity. Thereby the distribution of the Chern
numbers changes in presence of an oscillating electric field for both
linearly and circularly polarized light similarly. For moderate
intensities only few Chern numbers are different from the Chern
numbers of the static case whereas for higher intensities the
distribution is substantially altered.

Yet, the appropriate invariant to look at in case of a periodically
driven system is the $W_{3}$-invariant.  We computed this topological
indicator for the Floquet-Hofstadter spectrum to give a comparison
with the results on Chern numbers. In the high frequency limit both
the Chern number and the $W_{3}$-invariant coincide, yielding the
correct number of edge modes appearing in a system of finite size. The
latter allows for an experimental access. Finally, we were able to
show agreement with other topology studies on the Floquet-Hofstadter
spectrum in the off resonant regime. Whereas, our topology analysis of
the system is valid in all driving regimes, resonant and off resonant.
\acknowledgments
The authors thank Vanessa Junk, Philipp Reck, Klaus Richter, Bastian H\"ockendorf, and Andreas Alvermann for
various useful discussions. This work was supported by the Deutsche
Forschungsgemeinschaft via GRK 1570 and project 336985961.
\appendix
\section{Gapless non-degenerate states}\label{AppNoGap}
\begin{figure}[t]
   \includegraphics[width=1.0\columnwidth]{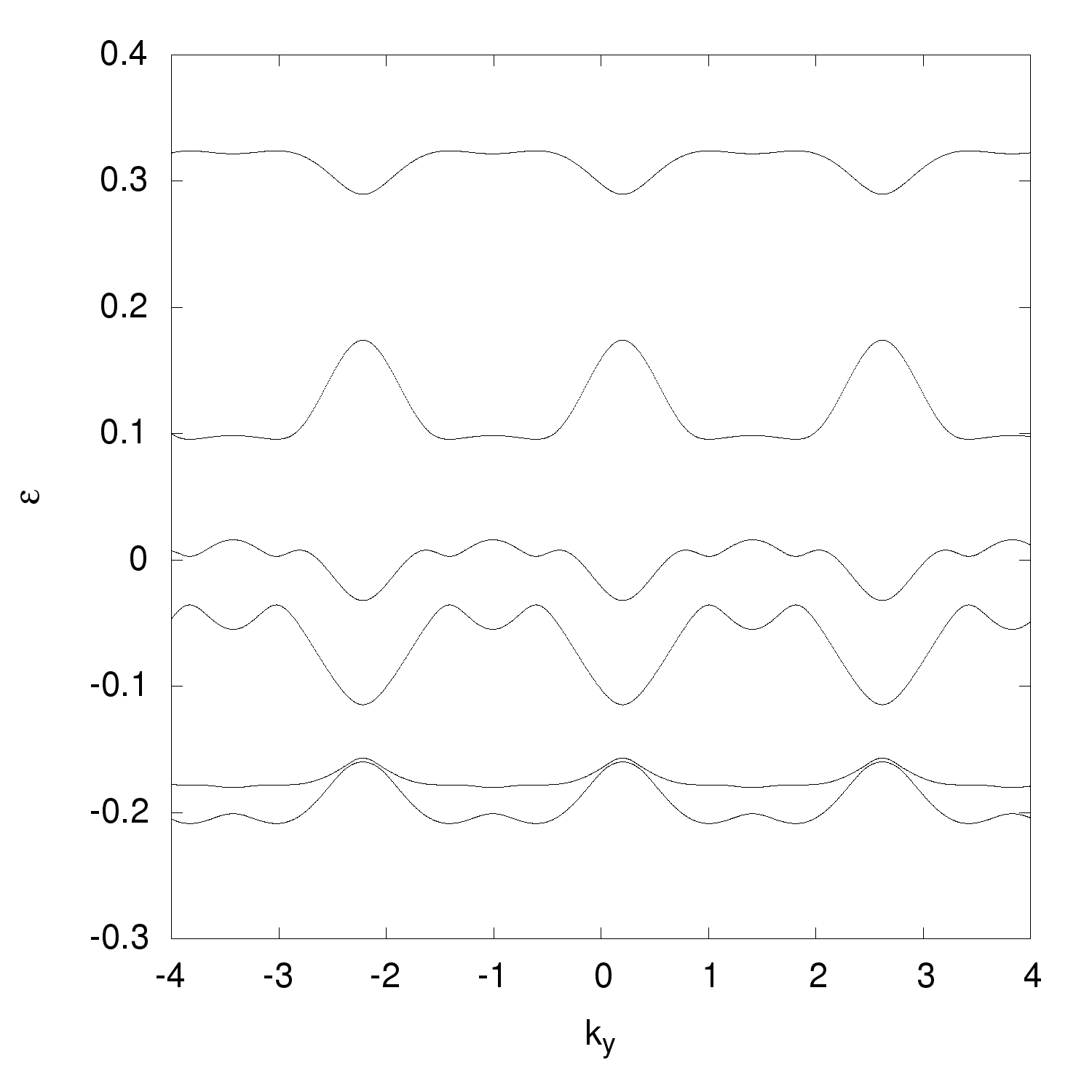}
   \caption{The quasienergy band structure for $p/q=1/3$,
     $(\gamma,\omega)=(2.65\,eAa/\hbar,3.0\,g/\hbar)$ and
     $k_{x}=0$. The lowest two bands are not
     degenerate but they do not have a gap in the sense that the
     minimum of the second lowest band is always greater than the
     maximum of the lowest band.}\label{Energy}
\end{figure}

\begin{figure}[t]
	\includegraphics[width=1.0\columnwidth]{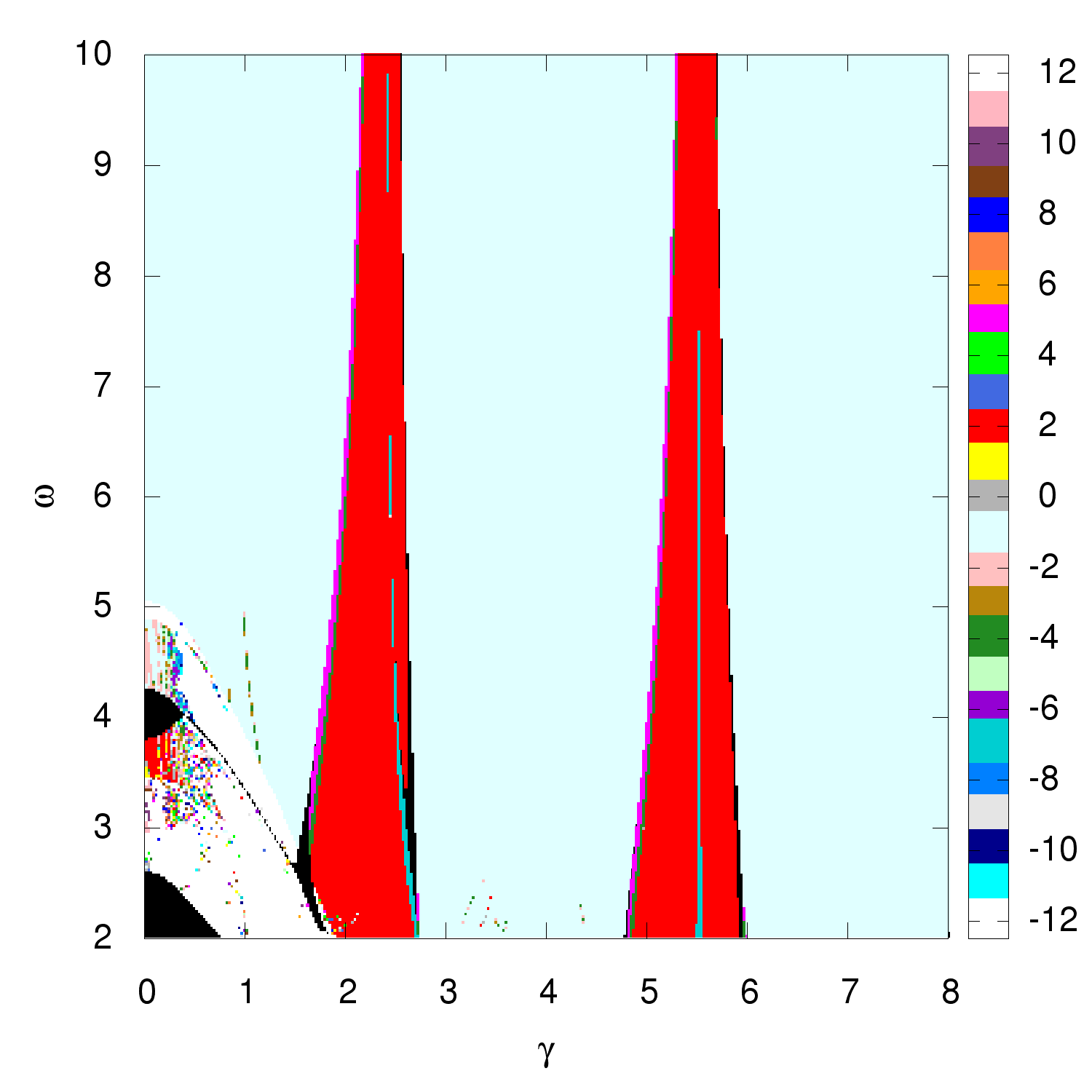}
	\caption{The $W_{3}$-invariant computed with the algorithm of
		H\"ockendorf  \textit{et al.}\cite{Hockendorf17} for the Floquet-Hofstadter
		spectrum at $p/q = 1/3$. The driving was circularly
		polarized. Parameter spaces $(\gamma,\omega)$ without a gap are
		marked black. }\label{W3}
\end{figure}

There is a global gap between two bands if the minimum of the upper
band is always greater than the maximum of the lower band. Consider
the case of two bands without a global energy gap. It does not imply
that there is a degeneracy of the two bands. This scenario occurs for
specific $(\gamma,\omega)$ configurations of the Floquet Hofstadter
spectrum between the lowest and second lowest band marked as black
stripes in Fig.~\ref{W3}. An exemplary quasienergy band structure is
shown in Fig.~\ref{Energy}. There is no gap between the lowest two non
degenerate bands.
%
\section{$W_{3}$-invariant for graphene without magnetic field} \label{W3graphene}
Although there are examples where the summation over the Chern numbers
of the truncated Floquet Hamiltonian fails to give the correct
topological invariant, as shown, e.g, by two examples in
Ref.~\onlinecite{Hockendorf17}, the procedure gives the correct
results for several models including circularly polarized driven
graphene. In the seminal work by Mikami \textit{et al.} on Floquet
topological insulators\cite{Mikami16} the authors were able to relate
topological phase transitions to effective hopping
amplitudes. Moreover, the topological phase diagram of graphene with
circularly polarized driving has been investigated.
\begin{figure}[!htb]
	\includegraphics[width=1.0\columnwidth]{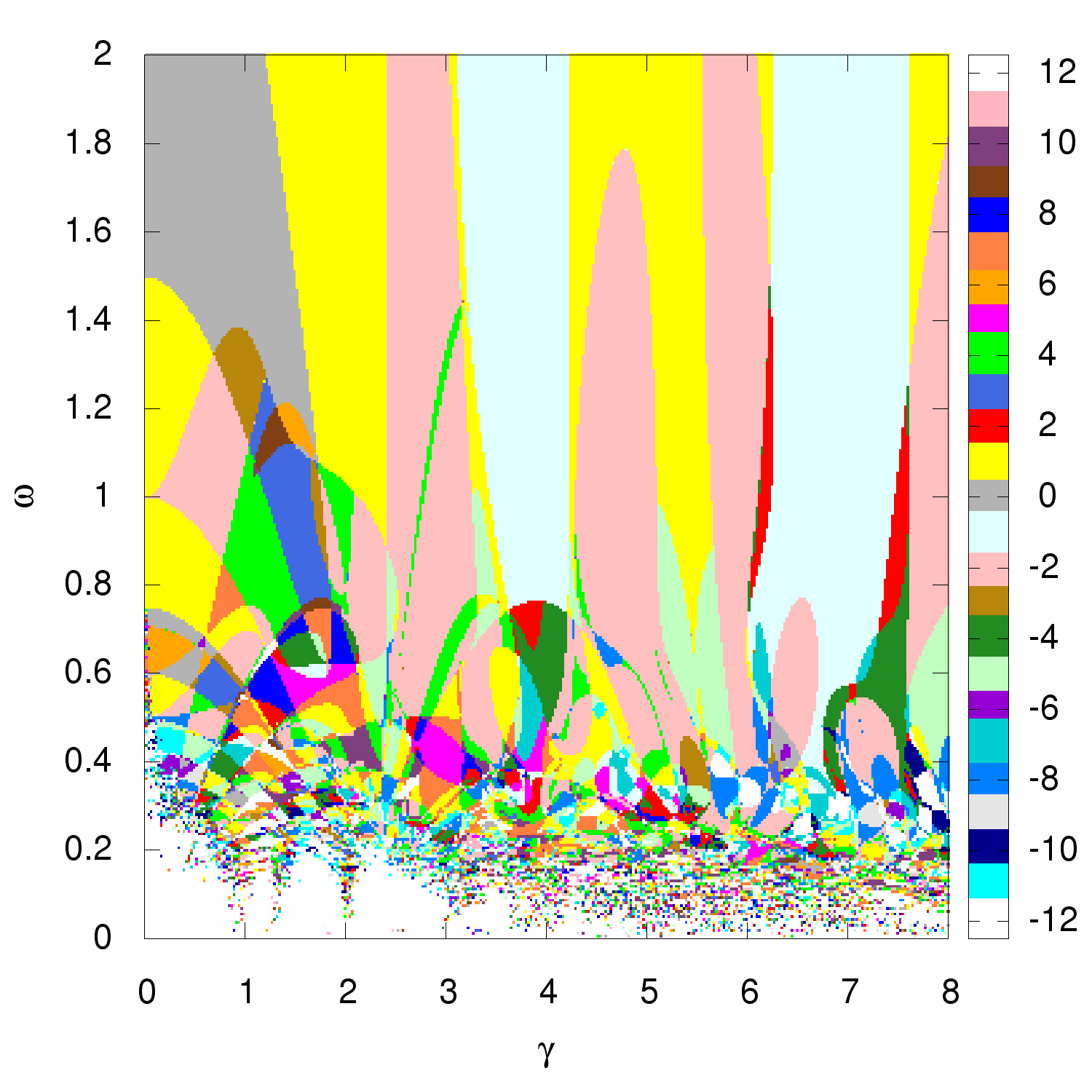}
	\caption{The sum over all Chern numbers of the truncated Floquet
      Hamiltonian below $\varepsilon=0$ for graphene with circularly polarized
      driving and without magnetic field. Our data almost perfectly
      reproduce the results from
      Ref.~\onlinecite{Mikami16}.}\label{C2_sumB_0}
\end{figure}
In order to make direct contact to the work by Mikami\cite{Mikami16}
we have set the discretization of the time-momentum BZ to
$200\times 200\times 200$ and the number of Floquet replicas to 50.
\begin{figure}[!htb]
	\includegraphics[width=1.0\columnwidth]{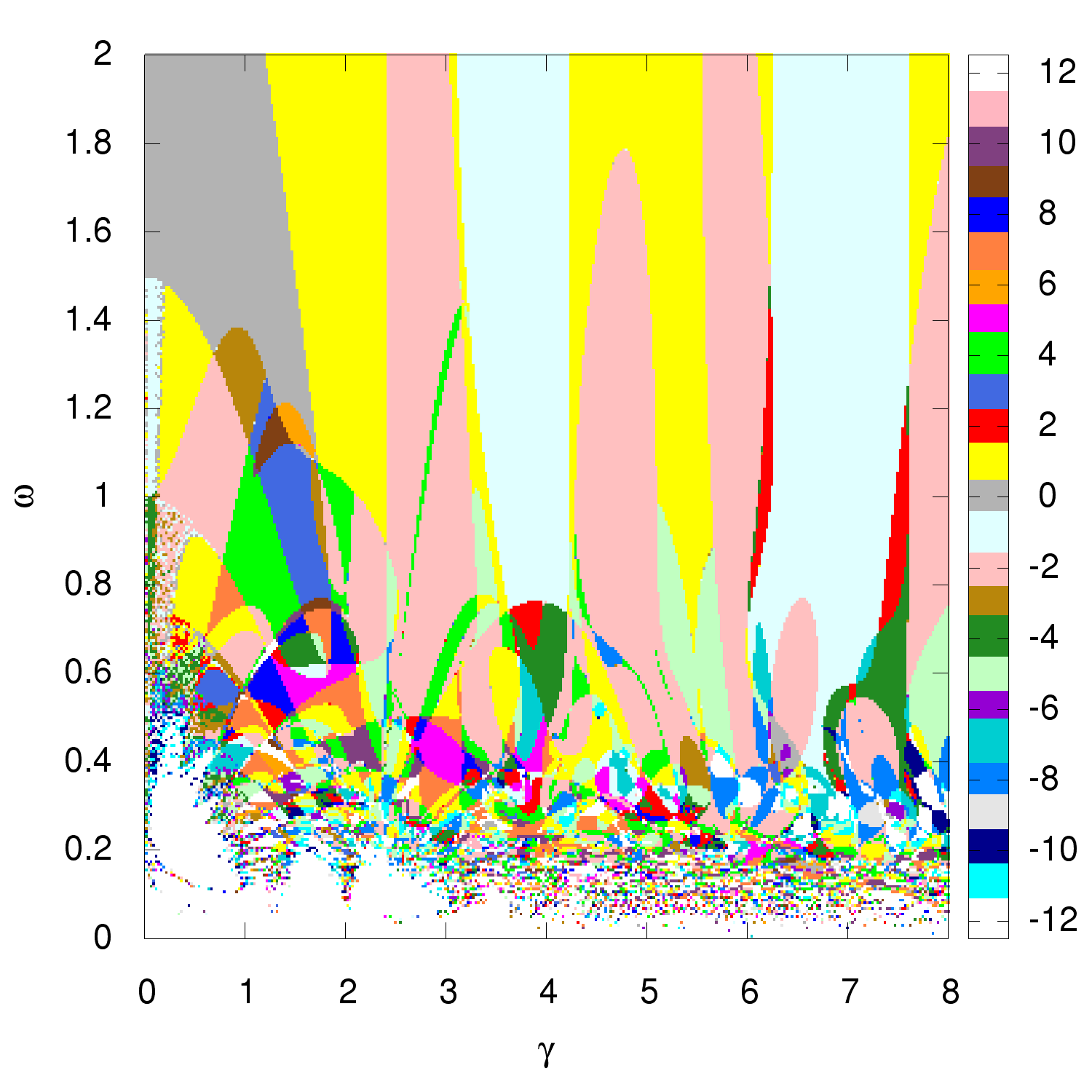}
	\caption{The $W_{3}$-invariant coincides in reliable regions with
      the sum over Chern numbers. Except for numerical unstable
      regions the Chern number sum and the $W_{3}$-invariant show a
      striking agreement.}\label{W3B_0}
\end{figure}
Although the lowest and topmost eigenvalues and eigenvectors of the
truncated Floquet Hamiltonian are not converged, i.e., they are
different from the index shifted eigenvectors with eigenvectors taken from
the central Floquet zone (compare Eq.~\eqref{Shiftprop}) they
remain relevant for the topological classification of driven
graphene. In the converged Floquet zones the sum over all bands has to
be zero inside one specific Floquet zone\cite{Hockendorf17}. For the
lowest and highest Floquet zones this is not necessarily the case. The
deviation from the converged Chern numbers contains the
information about the difference of Chern numbers and the
$W_{3}$-invariants such that the summation gives indeed the correct
topological invariant. This can be seen when comparing the sum over
all Chern numbers of the truncated Floquet Hamiltonian
Fig.~\ref{C2_sumB_0} with the $W_{3}$-invariant Fig.~\ref{W3B_0}. The
difference between the two values is plotted in
Fig.~\ref{W3_DiffB_0}. In the region of small intensities $\gamma$ and
$\hbar\omega < 1.5 g$ they do not agree. However, this is due to numerical
instabilities of the algorithm for the $W_{3}$-invariant. In order to
show that there is indeed no difference between the sum over Chern
numbers and $W_{3}$ we analyzed the sizes of the gaps at zero
quasienergy and $-\omega/2$.
\begin{figure}[!htb]
	\includegraphics[width=1.0\columnwidth]{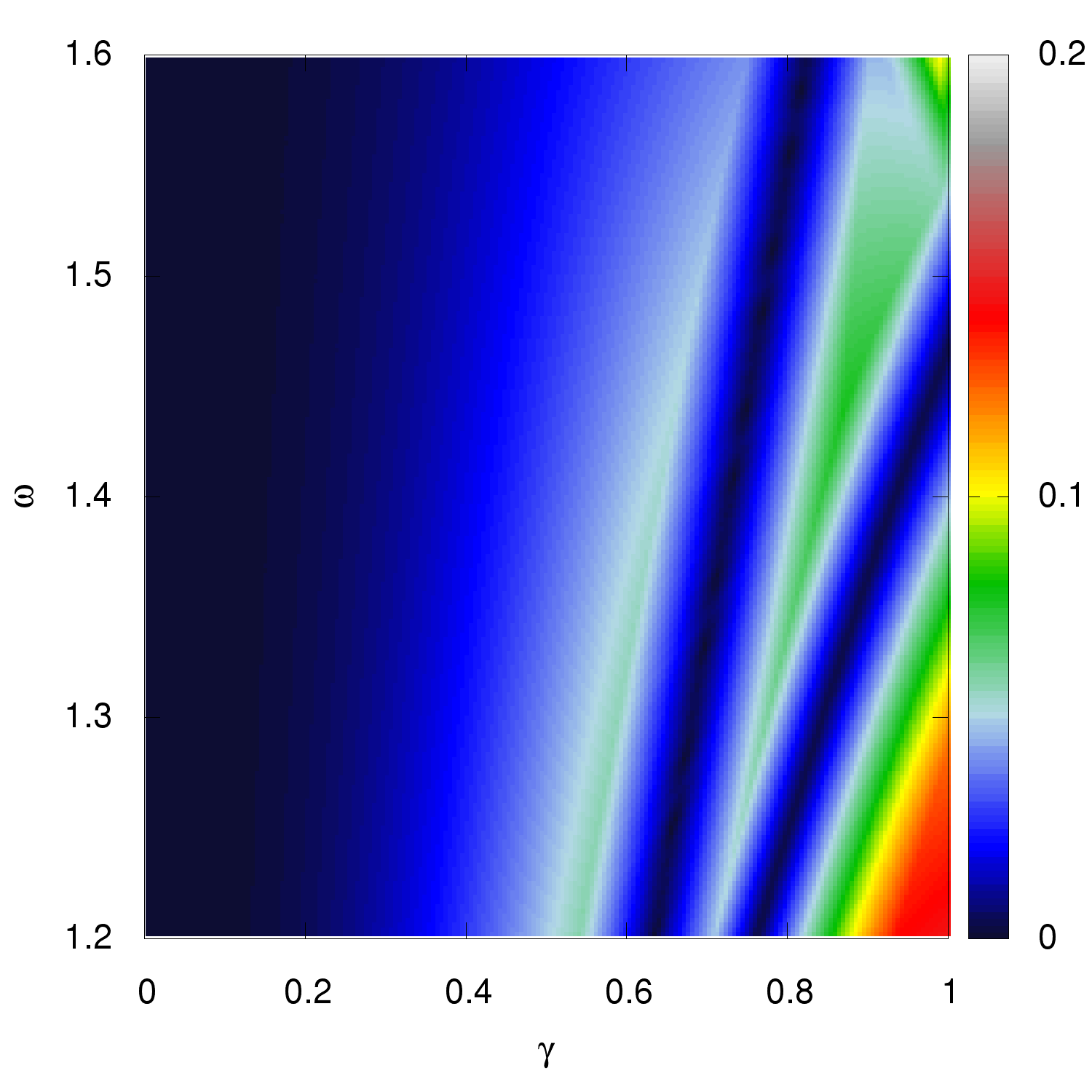}
	\caption{The size of the zone edge gap in dependence of intensity
      $\gamma$ and driving frequency $\omega$. The data were
      calculated as distance between the minimum of the lower band of
      the central Floquet zone and $-\omega/2$. The zero lines at the
      right half of the plot are also visible as topological phase
      transition int Fig.~\ref{C2_F_B_0}. }\label{Pi_Gap}
\end{figure}
Fig.~\ref{Pi_Gap} shows the difference between $-\omega/2$ and the
minimum of the lower band of the central Floquet zone. Comparing the
regions where the $-\omega/2$-gap is closed with the corresponding
regions where the Chern number changes, Fig.~\ref{C2_F_B_0}, one can
see that the zeros of the $-\omega/2$-gap are responsible for a change
of Chern numbers.
\begin{figure}[!htb]
	\includegraphics[width=1.0\columnwidth]{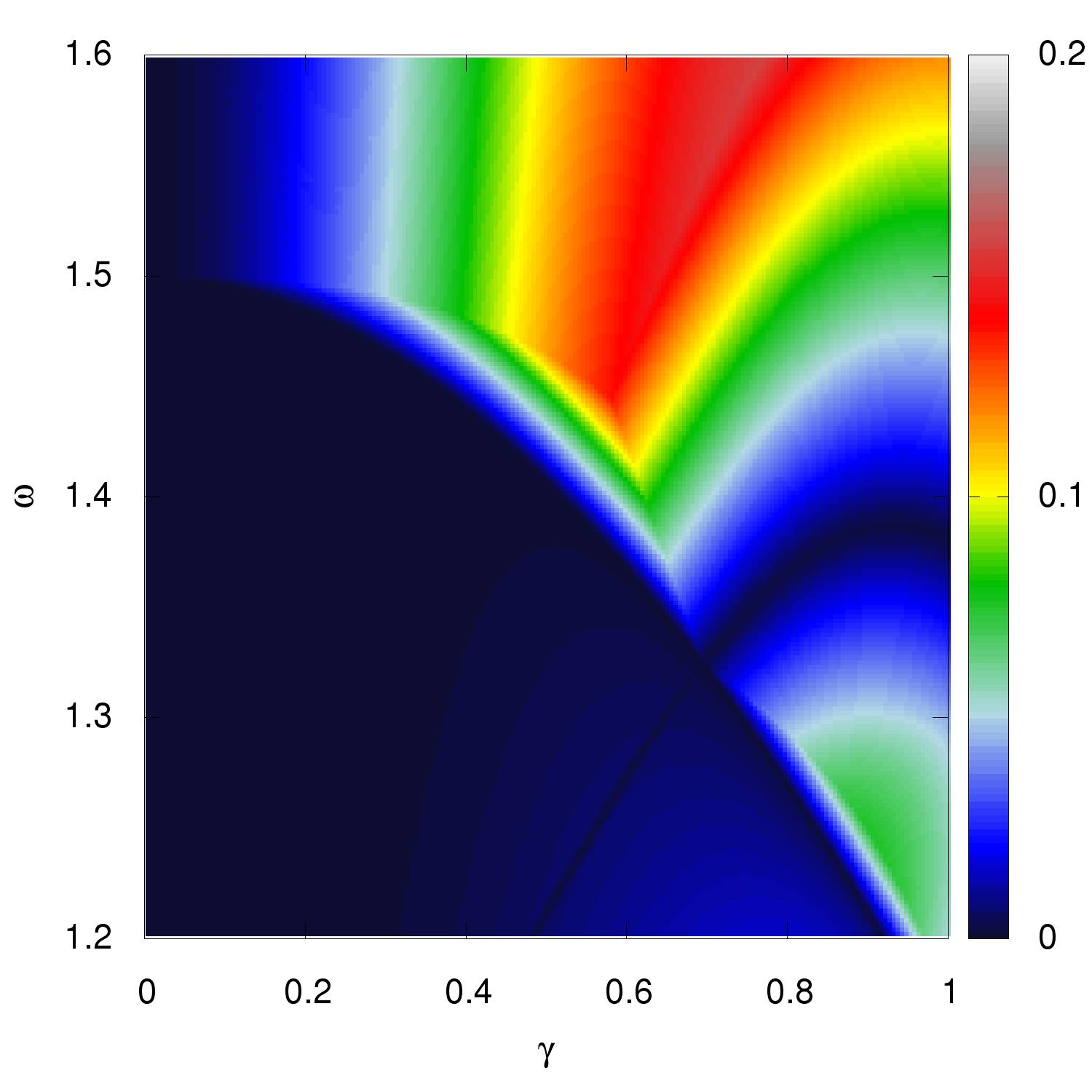}
	\caption{The minimum of the upper band of the central Floquet zone
      in dependence of intensity $\gamma$ and driving frequency
      $\omega$ is plotted. The zeros, and with that the band
      touchings, can be directly mapped to a change of the sum over
      Chern numbers, compare Fig.~\ref{C2_sumB_0}.}\label{Zero_Gap}
\end{figure}
Whereas, the arc in Fig.~\ref{Zero_Gap} starting from
$(\gamma,\omega)=(0.5\,eAa/\hbar,1.2\, g/\hbar)$ to
$(\gamma,\omega)=(1.0\, eAa/\hbar,1.36\,g/\hbar)$, where the zero gap
is closed, can be seen in Fig.~\ref{W3B_0} as well as in Fig.~\ref{C2_F_B_0}.
\begin{figure}[!htb]
	\includegraphics[width=1.0\columnwidth]{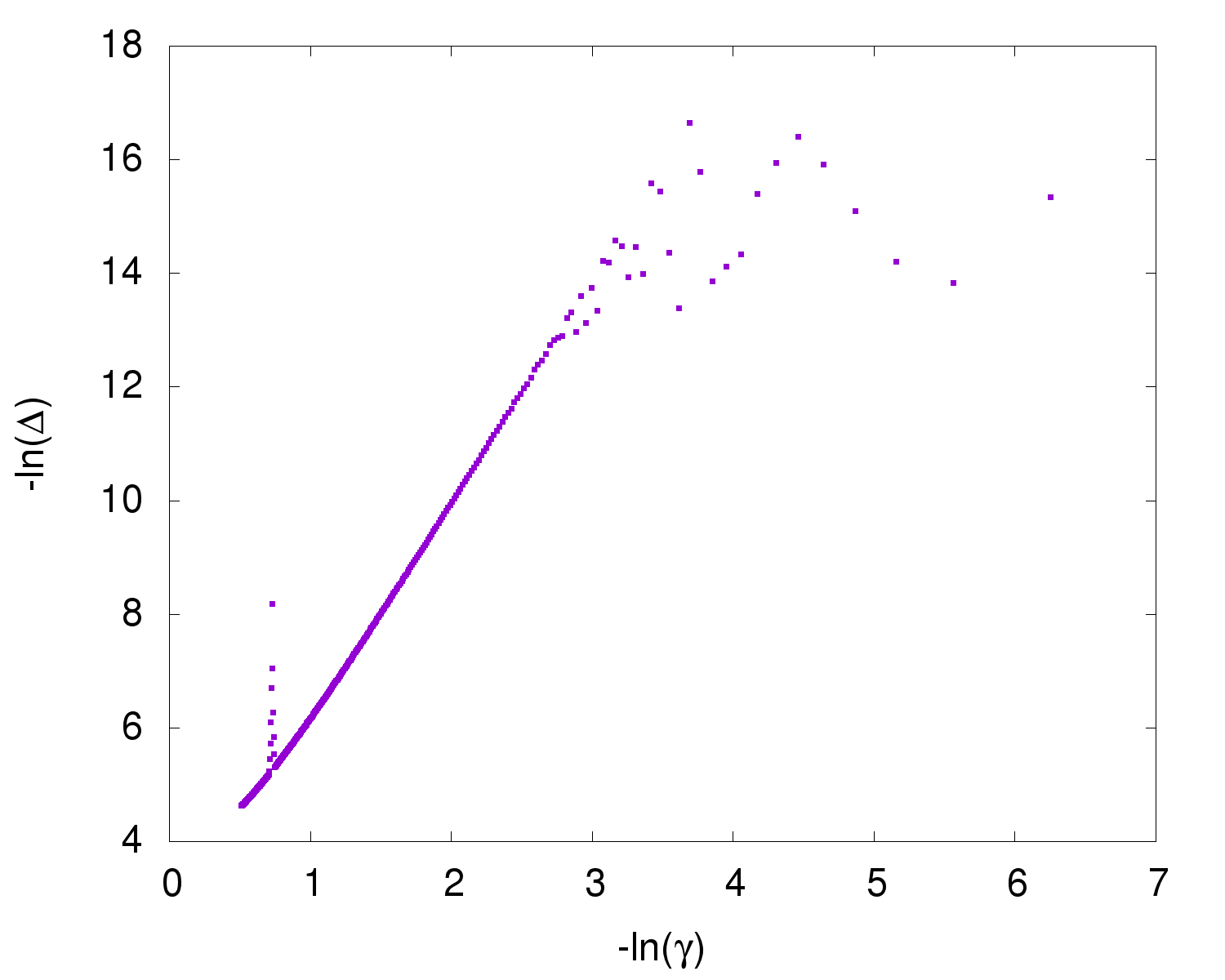}
	\caption{$\Delta$ is the minimum of the upper band of the central
      Floquet zone and $\gamma$ is here understood as dimensionless
      intensity $\gamma \to \gamma\, \hbar/eAa$. The plot shows the
      gap size, i.e., the difference between the minimum of the upper
      band and zero, for $\gamma=1/520$ to $\gamma = 3/5$ at fixed
      $\omega = 1.2\, g/\hbar$. The peak at $-\ln(\gamma)\approx 0.7$ is
      an evidence for a gap closing at $\gamma=0.5$. Whereas for
      $-\ln(0.2) \approx 1.6$ no peak is visible, giving a hint that
      there is no topological phase transition at
      $\gamma=0.2$. }\label{Zero_Gap_line}
\end{figure}
In the following we clarify if there is a difference between the sum
over Chern number of the truncated Floquet Hamiltonian and the
$W_{3}$-invariant. We calculated the gap sizes in the interval
$\gamma \in [0.0,0.6]\, eAa/\hbar$ for $\omega = 1.2\,g/\hbar$. The BZ
is discretized by using 3500$\times$3500 points. If there would be a
gap closing, e.g, at $(\gamma,\omega)=(0.2\, eAa/\hbar,1.2\,g/\hbar)$
in Fig.~\ref{W3B_0} we should see a signature of a gap closing either
in Fig.~\ref{Zero_Gap_line} or in Fig.~\ref{Pi_Gap_line}. The latter
show the gap sizes in a double logarithmically plot for the zero and
the $-\omega/2$ gap. If there would be a gap closing there should be a
signature at $-\ln(\gamma)\approx 1.6$ which is not the case. This
shows that the deviations between Chern number summation and $W_{3}$
can be traced back to numerical instabilities. Indeed, we were able to
achieve agreement between the results of the summation over Chern
numbers and the $W_{3}$-invariant when increasing the discretization
of the time-momentum BZ for some representative points. As an example
we investigated $(\gamma,\omega)=(0.1\,eAa/\hbar,1.4\,g/\hbar)$: An
increase of the number of discretization points to
$800\times 800\times 800$ is necessary in order to achieve convergence
of the $W_{3}$-algorithm and with that agreement with the summation
over Chern numbers.
\begin{figure}[!htb]
	\includegraphics[width=1.0\columnwidth]{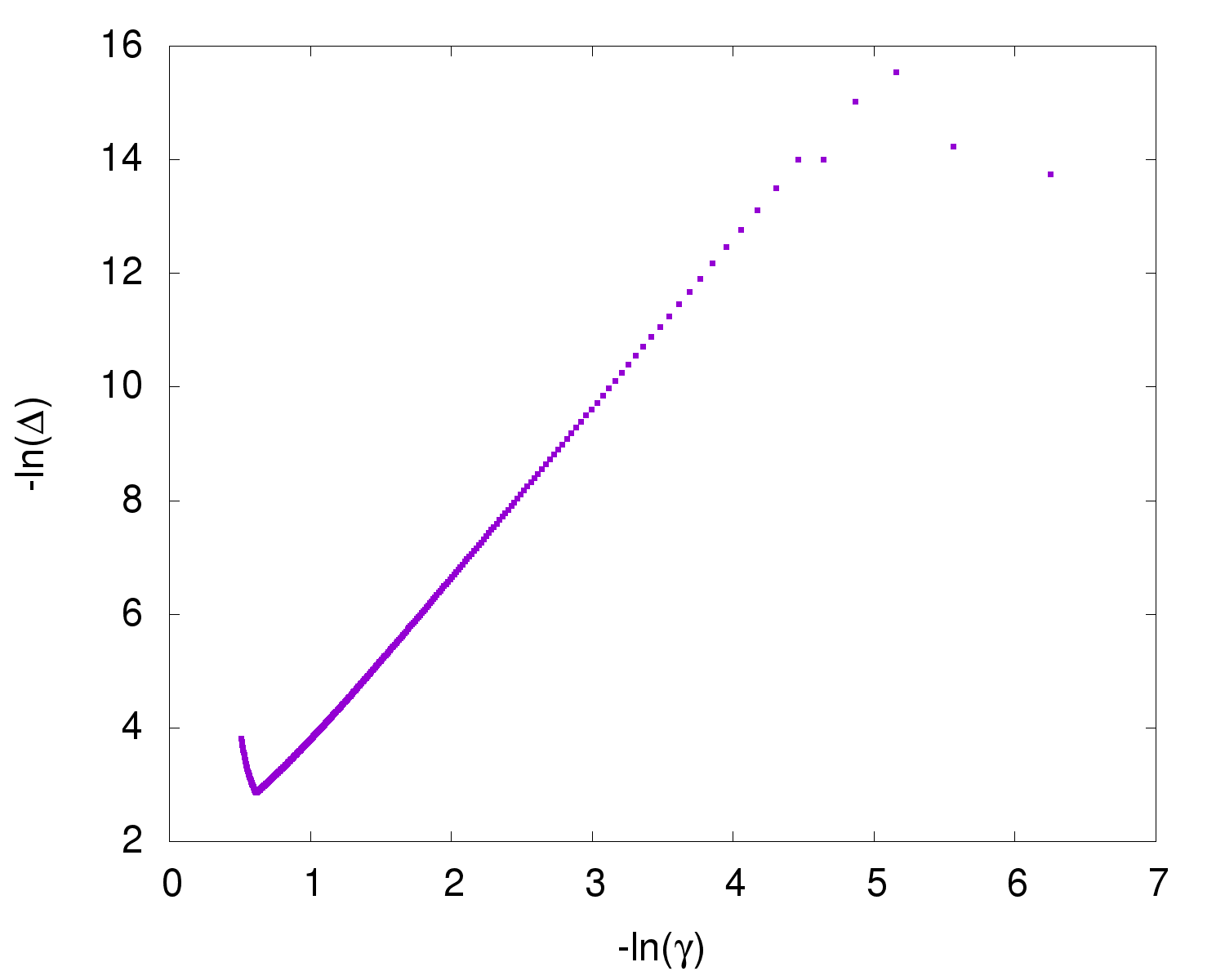}
	\caption{$\Delta$ is the distance between the minimum of the lower
      band of the central Floquet zone and $-\omega/2$ and, as in
      Fig.~\ref{Zero_Gap_line}, $\gamma$ is again dimensionless. No
      peak is visible in this plot, where $\gamma$ and $\omega$ are in
      the same parameter range as in
      Fig.~\ref{Zero_Gap_line}. }\label{Pi_Gap_line}
\end{figure}
Besides from numerical demanding regions, both topological
characterizations show a striking agreement, colored with gray in
Fig.~\ref{W3_DiffB_0}. To our knowledge, apart from the observation
that the sum over the Chern numbers of the truncated Floquet
Hamiltonian and the $W_{3}$-invariant seem to coincide for circularly
driven graphene, a proof, so far, is missing.
\begin{figure}[t]
  \includegraphics[width=1.0\columnwidth]{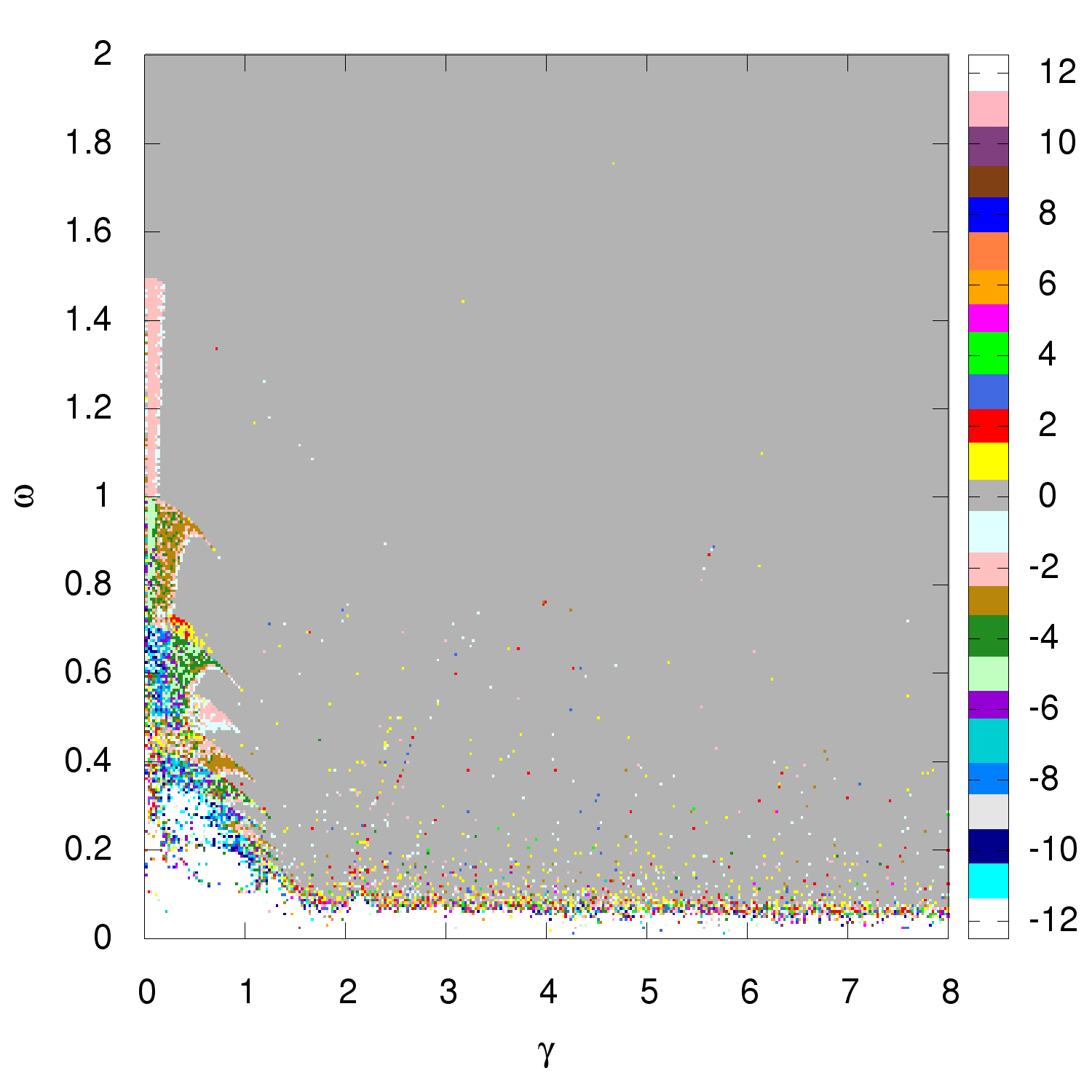}
  \caption{The Difference between the $W_{3}$-invariant and the sum
    over Chern numbers. }\label{W3_DiffB_0}
  \end{figure}
  Remarkably, even in the cases where both the Chern number and the
  $W_{3}$-invariant coincide (e.g., compare
  $(\gamma,\omega)=(4.0\,eAa/\hbar,2.0\,g/\hbar)$ Fig.~\ref{C2_F_B_0}
  and Fig.~\ref{W3B_0}) not all Floquet zones of the truncated Floquet
  Hamiltonian have the same Chern numbers as the central Floquet zone,
  as depicted in Fig.~\ref{highest_diff_mode_B_0}.
\begin{figure}[!t]
  \includegraphics[width=1.0\columnwidth]{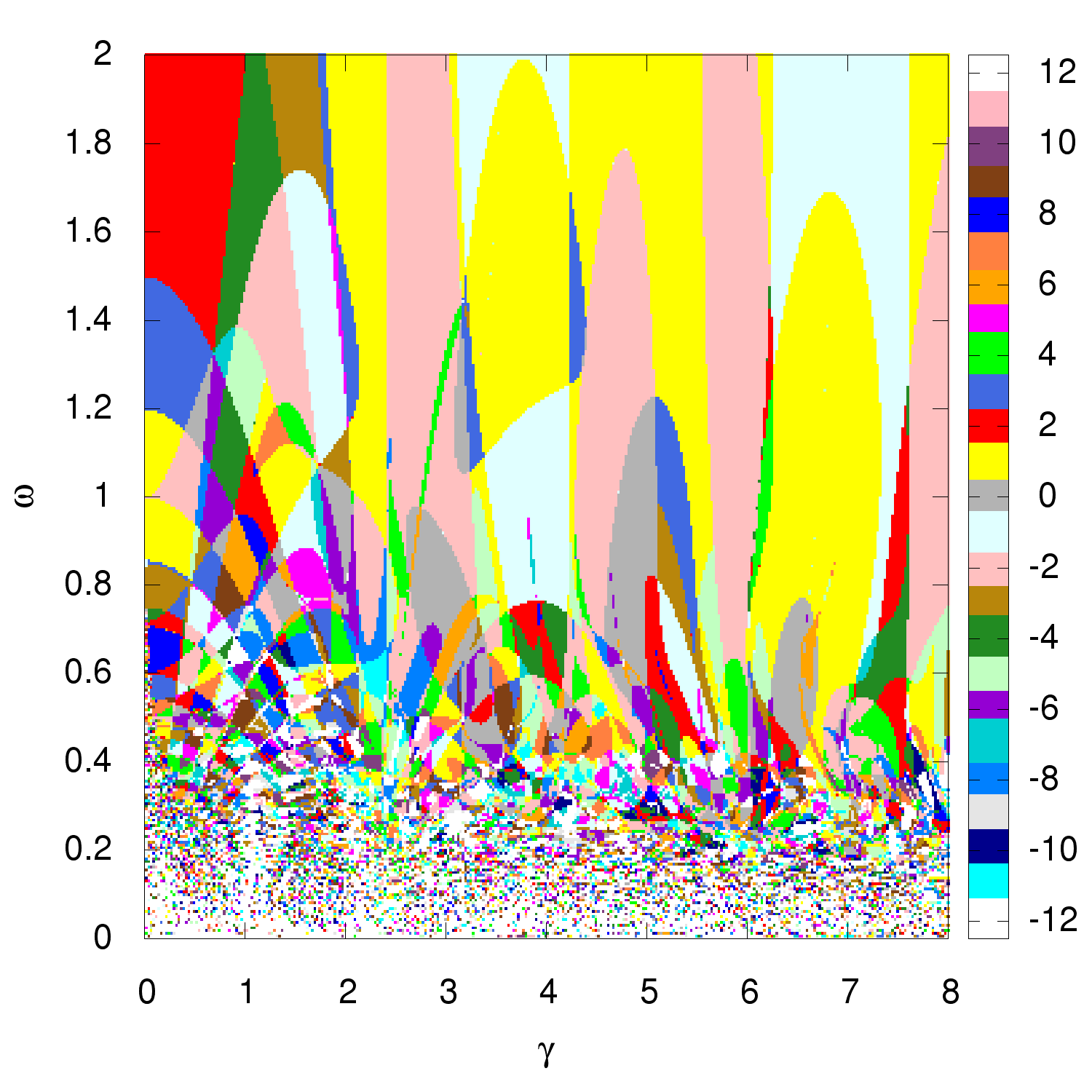}
  \caption{The Chern number of the lower band of graphene of the
    central Floquet zone. The driving is again circularly
    polarized. One can see the difference to the $W_{3}$-invariant in
    Fig.~\ref{W3B_0}. }\label{C2_F_B_0}
\end{figure}
This holds even for the off resonant
regime. Fig.~\ref{highest_diff_mode_B_0off} extends
Fig.~\ref{highest_diff_mode_B_0} to higher driving
frequencies. However, this feature survives for even higher driving
frequencies $\omega \propto 10^{6}\,g/\hbar$. Again, this can be
understood when having a closer look at the quasienergy band
structure. In the far off resonant regime the gap between the two
bands of graphene is very small. Hence, even when the Floquet zones
are far away from each other a small coupling is enough to close and
reopen the small gap of some Floquet zones.
\begin{figure}[t]
	\includegraphics[width=1.0\columnwidth]{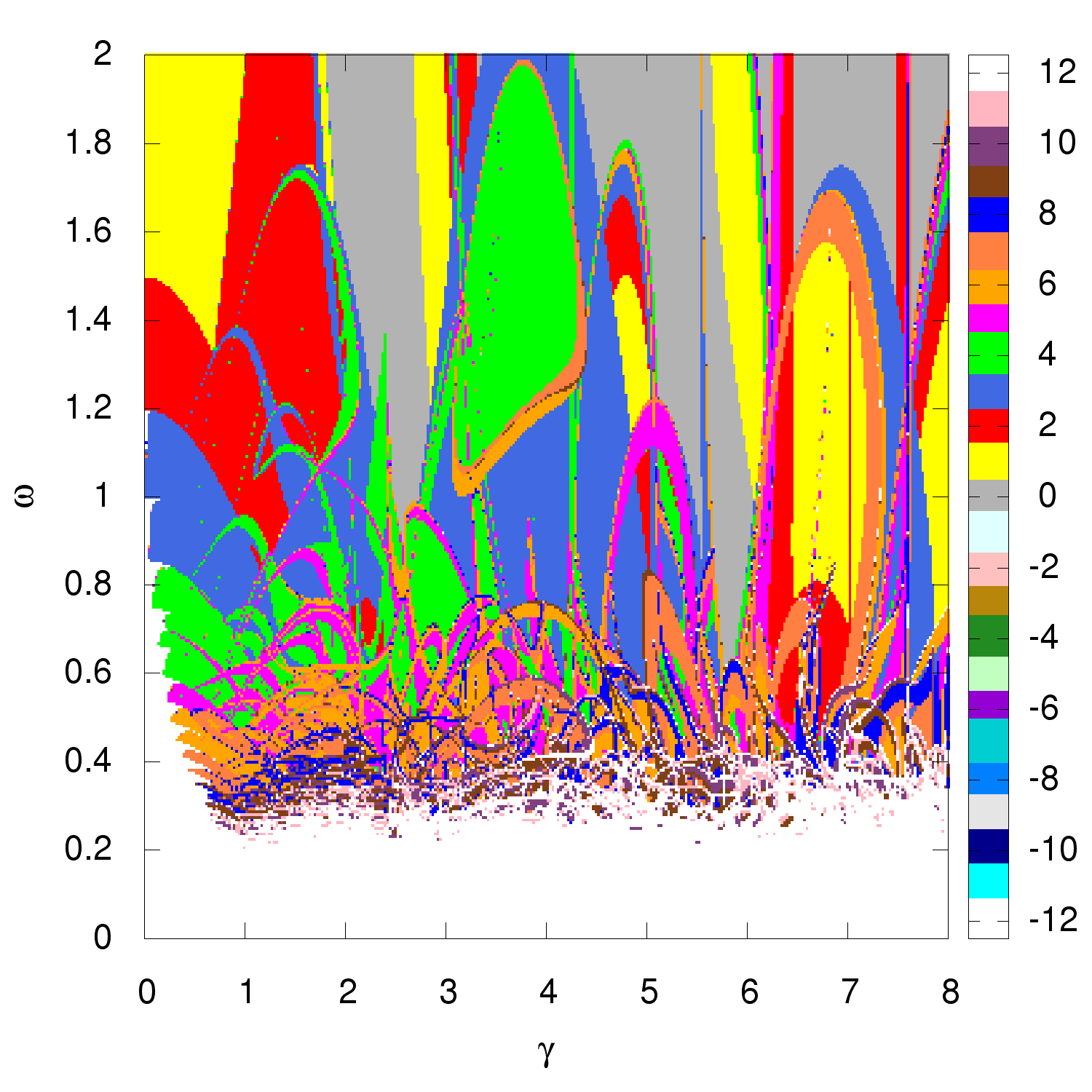}
	\caption{The innermost Floquet zone having different Chern numbers
      from that of the central Floquet zone. The counting of the
      Floquet zones starts here with the lowest mode, e.g. for
      $(\gamma,\omega)=(4.0\,eAa/\hbar,1.6\,g/\hbar)$ the (-50+4)'th
      Floquet zone has different Chern numbers from the Chern numbers
      of the central Floquet zone.}\label{highest_diff_mode_B_0}
\end{figure}
\begin{figure}[t]
	\includegraphics[width=1.0\columnwidth]{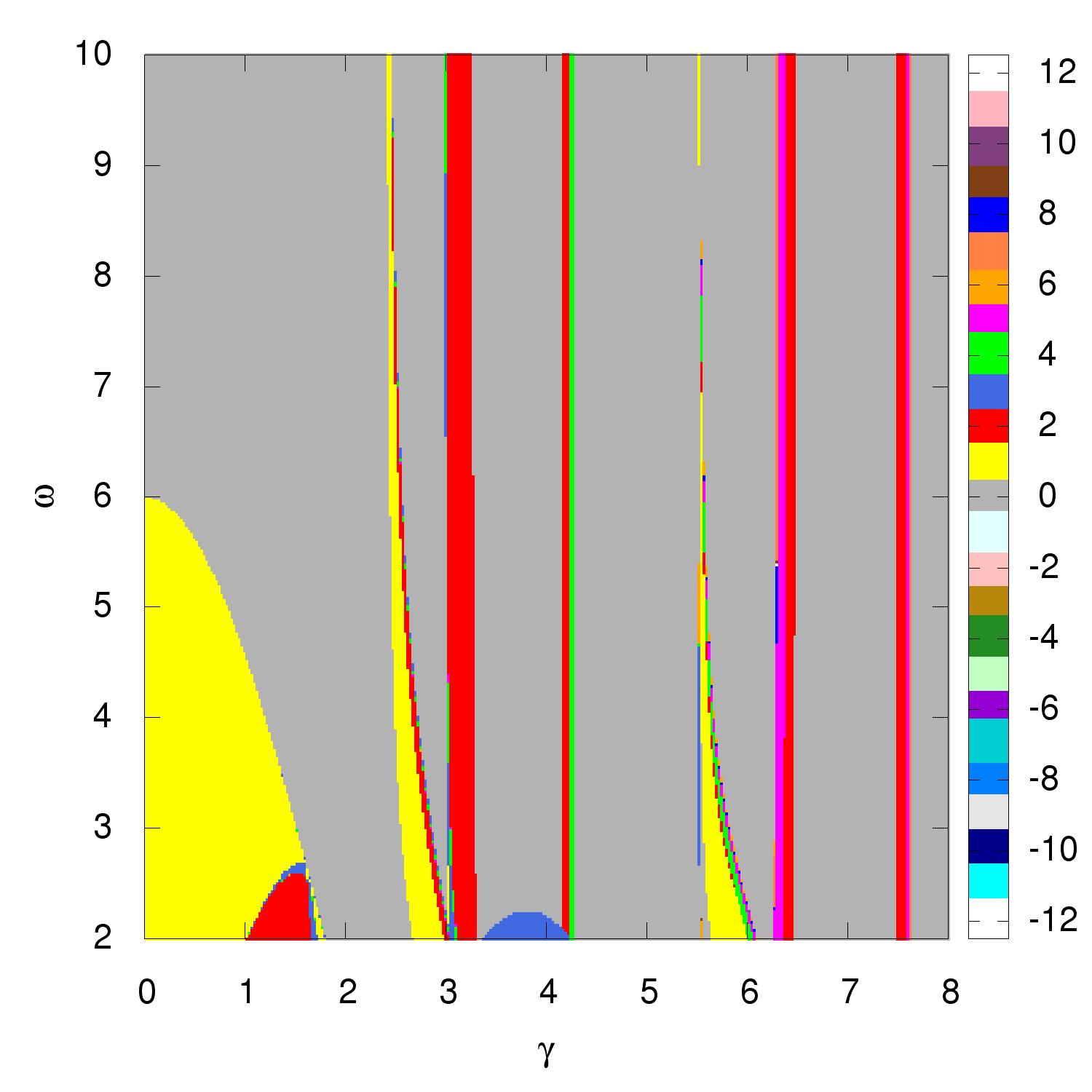}
	\caption{Even in the far off resonant regime not all Chern numbers
      of the Floquet zones of the truncated Floquet Hamiltonian agree
      with the Chern numbers of the central Floquet
      zone.}\label{highest_diff_mode_B_0off}
  \end{figure}
\section{$W_{3}$-invariant for spin-1/2 rotations} \label{SecondExample}
Besides the example given in the main text, there is a second case
given in Ref.~\onlinecite{Hockendorf17} where the summation over the
truncated Floquet Hamiltonian does not give the correct topological
invariant in the driven case. In this case the time evolution operator
reads
\begin{align}
   U(\vec{\mu}) = e^{-i2\pi w \vec{g}(\vec{\mu})\cdot\vec{\sigma}}\;,
\end{align}
where $\vec{g}(\vec{\mu})$ is a bijective map from the cube
$[0,1]^{3}$ to the unit ball $|\vec{\mu}|\leq 1$ that maps the surface
(center) of the cube to the surface (center) of the unit
ball\cite{Hockendorf17}. Let us set $w=1$ in order calculate the
$W_{3}$-invariant for one period. The mapping
$\vec{g}(\vec{\mu})$ can be constructed by applying two mappings. The first one is shifting the
unit cube and stretching it
\begin{align}
   \vec{s}(\vec{\mu}) &: [0,1]^{3} \to [-1,1]^{3}\\
   \vec{s}(\vec{\mu}) &: \begin{pmatrix} x \\ y \\ z \end{pmatrix} \mapsto \begin{pmatrix} 2x-1 \\ 2y-1 \\ 2z-1 \end{pmatrix}\;,
\end{align}
and the second one is the mapping to the unit ball
\begin{align}
  \vec{c}(\vec{\mu}) &: [-1,1]^{3} \to \{ |\vec{\mu}|\leq 1 : \vec{\mu}\in \mathbb{R}^{3} \} \\
  \vec{c}(\vec{\mu}) &: \begin{pmatrix}x\\ y\\ z \end{pmatrix} \mapsto \begin{pmatrix} x\sqrt{1-\tfrac{y^{2}}{2} - z^{2}\big( \tfrac{1}{2}-\tfrac{y^{2}}{3} \big) }\\ y\sqrt{1-\tfrac{z^{2}}{2} - x^{2}\big( \tfrac{1}{2}-\tfrac{z^{2}}{3} \big) }\\ z\sqrt{1-\tfrac{x^{2}}{2} - y^{2}\big( \tfrac{1}{2}-\tfrac{x^{2}}{3} \big) } \end{pmatrix}\ .
\end{align}
By concatenation we yield 
\begin{align}
   \vec{g}(\vec{\mu}) = \vec{c}\big( \vec{s}(\vec{\mu}) \big)\;.
\end{align}
With the explicit form given for the mapping from the cube to the ball we
can calculate the eigenvalues of the operator
$\vec{g}(\vec{\mu})\cdot\vec{\sigma}$ which are
\begin{align}
   \lambda_{\pm} = \pm \sqrt{1+64xyz(x-1)(y-1)(z-1)}\ .
\end{align}
By identifying $x \equiv k_x$, $y \equiv k_y$ and $z \equiv t/T$, the
time-dependent Hamiltonian can be reconstructed with
\begin{align}
   H(t) = i\hbar\big( \partial_{t}U(t) \big) U^{\dagger}(t)\;.
\end{align}
Having $H(t)$ we can calculate the corresponding Floquet-Hamiltonian
which has a driving period of $\omega = 2\pi$ since we have chosen
$w=1$. But we know that the quasienergies $\varepsilon_{\pm}$ of the
Floquet Hamiltonian are equal to the eigenvalues of $U(\vec{\mu})$
evaluated after one period, i.e.,
\begin{align}
   \varepsilon_{\pm} & = i\ln e^{\pm i2\pi\lambda_{\pm}}\Big|_{t/T=1} \\
                     & = \pm 2\pi\;,
\end{align}
and by shifting the quasienergies into the central Floquet zone we get
two degenerate bands with zero quasienergy
\begin{align}
   \varepsilon_{\pm} ={}& 0\;.
\end{align}
The Floquet spectrum is everywhere degenerate but the Chern numbers
are well defined. However, the summation over Chern numbers of the
truncated Floquet Hamiltonian gives not the correct topological
invariant which is in this case $W_{3}=2$.
\bibliographystyle{apsrev4-1}
\bibliography{paper}

\def\url#1{}
\begin{thebibliography}{53}%
\makeatletter
\providecommand \@ifxundefined [1]{%
 \@ifx{#1\undefined}
}%
\providecommand \@ifnum [1]{%
 \ifnum #1\expandafter \@firstoftwo
 \else \expandafter \@secondoftwo
 \fi
}%
\providecommand \@ifx [1]{%
 \ifx #1\expandafter \@firstoftwo
 \else \expandafter \@secondoftwo
 \fi
}%
\providecommand \natexlab [1]{#1}%
\providecommand \enquote  [1]{``#1''}%
\providecommand \bibnamefont  [1]{#1}%
\providecommand \bibfnamefont [1]{#1}%
\providecommand \citenamefont [1]{#1}%
\providecommand \href@noop [0]{\@secondoftwo}%
\providecommand \href [0]{\begingroup \@sanitize@url \@href}%
\providecommand \@href[1]{\@@startlink{#1}\@@href}%
\providecommand \@@href[1]{\endgroup#1\@@endlink}%
\providecommand \@sanitize@url [0]{\catcode `\\12\catcode `\$12\catcode
  `\&12\catcode `\#12\catcode `\^12\catcode `\_12\catcode `\%12\relax}%
\providecommand \@@startlink[1]{}%
\providecommand \@@endlink[0]{}%
\providecommand \url  [0]{\begingroup\@sanitize@url \@url }%
\providecommand \@url [1]{\endgroup\@href {#1}{\urlprefix }}%
\providecommand \urlprefix  [0]{URL }%
\providecommand \Eprint [0]{\href }%
\providecommand \doibase [0]{http://dx.doi.org/}%
\providecommand \selectlanguage [0]{\@gobble}%
\providecommand \bibinfo  [0]{\@secondoftwo}%
\providecommand \bibfield  [0]{\@secondoftwo}%
\providecommand \translation [1]{[#1]}%
\providecommand \BibitemOpen [0]{}%
\providecommand \bibitemStop [0]{}%
\providecommand \bibitemNoStop [0]{.\EOS\space}%
\providecommand \EOS [0]{\spacefactor3000\relax}%
\providecommand \BibitemShut  [1]{\csname bibitem#1\endcsname}%
\let\auto@bib@innerbib\@empty
\bibitem [{\citenamefont {Klitzing}\ \emph {et~al.}(1980)\citenamefont
  {Klitzing}, \citenamefont {Dorda},\ and\ \citenamefont {Pepper}}]{Klitzing}%
  \BibitemOpen
  \bibfield  {author} {\bibinfo {author} {\bibfnamefont {K.~v.}\ \bibnamefont
  {Klitzing}}, \bibinfo {author} {\bibfnamefont {G.}~\bibnamefont {Dorda}}, \
  and\ \bibinfo {author} {\bibfnamefont {M.}~\bibnamefont {Pepper}},\ }\href
  {\doibase 10.1103/PhysRevLett.45.494} {\bibfield  {journal} {\bibinfo
  {journal} {Phys. Rev. Lett.}\ }\textbf {\bibinfo {volume} {45}},\ \bibinfo
  {pages} {494} (\bibinfo {year} {1980})}\BibitemShut {NoStop}%
\bibitem [{\citenamefont {von Klitzing}(1986)}]{KlitzingQHE}%
  \BibitemOpen
  \bibfield  {author} {\bibinfo {author} {\bibfnamefont {K.}~\bibnamefont {von
  Klitzing}},\ }\href {\doibase 10.1103/RevModPhys.58.519} {\bibfield
  {journal} {\bibinfo  {journal} {Rev. Mod. Phys.}\ }\textbf {\bibinfo {volume}
  {58}},\ \bibinfo {pages} {519} (\bibinfo {year} {1986})}\BibitemShut
  {NoStop}%
\bibitem [{\citenamefont {Hasan}\ and\ \citenamefont {Kane}(2010)}]{Hasan10}%
  \BibitemOpen
  \bibfield  {author} {\bibinfo {author} {\bibfnamefont {M.~Z.}\ \bibnamefont
  {Hasan}}\ and\ \bibinfo {author} {\bibfnamefont {C.~L.}\ \bibnamefont
  {Kane}},\ }\href {\doibase 10.1103/RevModPhys.82.3045} {\bibfield  {journal}
  {\bibinfo  {journal} {Rev. Mod. Phys.}\ }\textbf {\bibinfo {volume} {82}},\
  \bibinfo {pages} {3045} (\bibinfo {year} {2010})}\BibitemShut {NoStop}%
\bibitem [{\citenamefont {Qi}\ and\ \citenamefont {Zhang}(2011)}]{Qi11}%
  \BibitemOpen
  \bibfield  {author} {\bibinfo {author} {\bibfnamefont {X.-L.}\ \bibnamefont
  {Qi}}\ and\ \bibinfo {author} {\bibfnamefont {S.-C.}\ \bibnamefont {Zhang}},\
  }\href {\doibase 10.1103/RevModPhys.83.1057} {\bibfield  {journal} {\bibinfo
  {journal} {Rev. Mod. Phys.}\ }\textbf {\bibinfo {volume} {83}},\ \bibinfo
  {pages} {1057} (\bibinfo {year} {2011})}\BibitemShut {NoStop}%
\bibitem [{\citenamefont {Hofstadter}(1976)}]{Hofstadter76}%
  \BibitemOpen
  \bibfield  {author} {\bibinfo {author} {\bibfnamefont {D.~R.}\ \bibnamefont
  {Hofstadter}},\ }\href {\doibase 10.1103/PhysRevB.14.2239} {\bibfield
  {journal} {\bibinfo  {journal} {Phys. Rev. B}\ }\textbf {\bibinfo {volume}
  {14}},\ \bibinfo {pages} {2239} (\bibinfo {year} {1976})}\BibitemShut
  {NoStop}%
\bibitem [{\citenamefont {Thouless}\ \emph {et~al.}(1982)\citenamefont
  {Thouless}, \citenamefont {Kohmoto}, \citenamefont {Nightingale},\ and\
  \citenamefont {den Nijs}}]{TKNN}%
  \BibitemOpen
  \bibfield  {author} {\bibinfo {author} {\bibfnamefont {D.~J.}\ \bibnamefont
  {Thouless}}, \bibinfo {author} {\bibfnamefont {M.}~\bibnamefont {Kohmoto}},
  \bibinfo {author} {\bibfnamefont {M.~P.}\ \bibnamefont {Nightingale}}, \ and\
  \bibinfo {author} {\bibfnamefont {M.}~\bibnamefont {den Nijs}},\ }\href
  {\doibase 10.1103/PhysRevLett.49.405} {\bibfield  {journal} {\bibinfo
  {journal} {Phys. Rev. Lett.}\ }\textbf {\bibinfo {volume} {49}},\ \bibinfo
  {pages} {405} (\bibinfo {year} {1982})}\BibitemShut {NoStop}%
\bibitem [{\citenamefont {Fukui}\ \emph {et~al.}(2005)\citenamefont {Fukui},
  \citenamefont {Hatsugai},\ and\ \citenamefont {Suzuki}}]{FukuiChern}%
  \BibitemOpen
  \bibfield  {author} {\bibinfo {author} {\bibfnamefont {T.}~\bibnamefont
  {Fukui}}, \bibinfo {author} {\bibfnamefont {Y.}~\bibnamefont {Hatsugai}}, \
  and\ \bibinfo {author} {\bibfnamefont {H.}~\bibnamefont {Suzuki}},\ }\href
  {\doibase 10.1143/JPSJ.74.1674} {\bibfield  {journal} {\bibinfo  {journal}
  {J. Phys. Soc. Jpn.}\ }\textbf {\bibinfo {volume} {74}},\ \bibinfo {pages}
  {1674} (\bibinfo {year} {2005})}\BibitemShut {NoStop}%
\bibitem [{\citenamefont {Oka}\ and\ \citenamefont {Aoki}(2009)}]{Oka09}%
  \BibitemOpen
  \bibfield  {author} {\bibinfo {author} {\bibfnamefont {T.}~\bibnamefont
  {Oka}}\ and\ \bibinfo {author} {\bibfnamefont {H.}~\bibnamefont {Aoki}},\
  }\href {\doibase 10.1103/PhysRevB.79.081406} {\bibfield  {journal} {\bibinfo
  {journal} {Phys. Rev. B}\ }\textbf {\bibinfo {volume} {79}},\ \bibinfo
  {pages} {081406} (\bibinfo {year} {2009})}\BibitemShut {NoStop}%
\bibitem [{\citenamefont {Kitagawa}\ \emph {et~al.}(2010)\citenamefont
  {Kitagawa}, \citenamefont {Berg}, \citenamefont {Rudner},\ and\ \citenamefont
  {Demler}}]{Kitagawa10}%
  \BibitemOpen
  \bibfield  {author} {\bibinfo {author} {\bibfnamefont {T.}~\bibnamefont
  {Kitagawa}}, \bibinfo {author} {\bibfnamefont {E.}~\bibnamefont {Berg}},
  \bibinfo {author} {\bibfnamefont {M.}~\bibnamefont {Rudner}}, \ and\ \bibinfo
  {author} {\bibfnamefont {E.}~\bibnamefont {Demler}},\ }\href {\doibase
  10.1103/PhysRevB.82.235114} {\bibfield  {journal} {\bibinfo  {journal} {Phys.
  Rev. B}\ }\textbf {\bibinfo {volume} {82}},\ \bibinfo {pages} {235114}
  (\bibinfo {year} {2010})}\BibitemShut {NoStop}%
\bibitem [{\citenamefont {Lindner}\ \emph {et~al.}(2011)\citenamefont
  {Lindner}, \citenamefont {Refael},\ and\ \citenamefont
  {Galitski}}]{Lindner11}%
  \BibitemOpen
  \bibfield  {author} {\bibinfo {author} {\bibfnamefont {N.~H.}\ \bibnamefont
  {Lindner}}, \bibinfo {author} {\bibfnamefont {G.}~\bibnamefont {Refael}}, \
  and\ \bibinfo {author} {\bibfnamefont {V.}~\bibnamefont {Galitski}},\ }\href
  {\doibase 10.1038/nphys1926} {\bibfield  {journal} {\bibinfo  {journal}
  {Nature Physics}\ }\textbf {\bibinfo {volume} {7}},\ \bibinfo {pages} {490}
  (\bibinfo {year} {2011})}\BibitemShut {NoStop}%
\bibitem [{\citenamefont {Gu}\ \emph {et~al.}(2011)\citenamefont {Gu},
  \citenamefont {Fertig}, \citenamefont {Arovas},\ and\ \citenamefont
  {Auerbach}}]{Gu11}%
  \BibitemOpen
  \bibfield  {author} {\bibinfo {author} {\bibfnamefont {Z.}~\bibnamefont
  {Gu}}, \bibinfo {author} {\bibfnamefont {H.~A.}\ \bibnamefont {Fertig}},
  \bibinfo {author} {\bibfnamefont {D.~P.}\ \bibnamefont {Arovas}}, \ and\
  \bibinfo {author} {\bibfnamefont {A.}~\bibnamefont {Auerbach}},\ }\href
  {\doibase 10.1103/PhysRevLett.107.216601} {\bibfield  {journal} {\bibinfo
  {journal} {Phys. Rev. Lett.}\ }\textbf {\bibinfo {volume} {107}},\ \bibinfo
  {pages} {216601} (\bibinfo {year} {2011})}\BibitemShut {NoStop}%
\bibitem [{\citenamefont {Cayssol}\ \emph {et~al.}(2013)\citenamefont
  {Cayssol}, \citenamefont {D\'ora}, \citenamefont {Simon},\ and\ \citenamefont
  {Moessner}}]{Cayssol13}%
  \BibitemOpen
  \bibfield  {author} {\bibinfo {author} {\bibfnamefont {J.}~\bibnamefont
  {Cayssol}}, \bibinfo {author} {\bibfnamefont {B.}~\bibnamefont {D\'ora}},
  \bibinfo {author} {\bibfnamefont {F.}~\bibnamefont {Simon}}, \ and\ \bibinfo
  {author} {\bibfnamefont {R.}~\bibnamefont {Moessner}},\ }\href {\doibase
  10.1002/pssr.201206451} {\bibfield  {journal} {\bibinfo  {journal} {phys.
  stat. sol. (RRL)}\ }\textbf {\bibinfo {volume} {7}},\ \bibinfo {pages} {101}
  (\bibinfo {year} {2013})}\BibitemShut {NoStop}%
\bibitem [{\citenamefont {Rudner}\ \emph {et~al.}(2013)\citenamefont {Rudner},
  \citenamefont {Lindner}, \citenamefont {Berg},\ and\ \citenamefont
  {Levin}}]{Rudner13}%
  \BibitemOpen
  \bibfield  {author} {\bibinfo {author} {\bibfnamefont {M.~S.}\ \bibnamefont
  {Rudner}}, \bibinfo {author} {\bibfnamefont {N.~H.}\ \bibnamefont {Lindner}},
  \bibinfo {author} {\bibfnamefont {E.}~\bibnamefont {Berg}}, \ and\ \bibinfo
  {author} {\bibfnamefont {M.}~\bibnamefont {Levin}},\ }\href {\doibase
  10.1103/PhysRevX.3.031005} {\bibfield  {journal} {\bibinfo  {journal} {Phys.
  Rev. X}\ }\textbf {\bibinfo {volume} {3}},\ \bibinfo {pages} {031005}
  (\bibinfo {year} {2013})}\BibitemShut {NoStop}%
\bibitem [{\citenamefont {Mikami}\ \emph {et~al.}(2016)\citenamefont {Mikami},
  \citenamefont {Kitamura}, \citenamefont {Yasuda}, \citenamefont {Tsuji},
  \citenamefont {Oka},\ and\ \citenamefont {Aoki}}]{Mikami16}%
  \BibitemOpen
  \bibfield  {author} {\bibinfo {author} {\bibfnamefont {T.}~\bibnamefont
  {Mikami}}, \bibinfo {author} {\bibfnamefont {S.}~\bibnamefont {Kitamura}},
  \bibinfo {author} {\bibfnamefont {K.}~\bibnamefont {Yasuda}}, \bibinfo
  {author} {\bibfnamefont {N.}~\bibnamefont {Tsuji}}, \bibinfo {author}
  {\bibfnamefont {T.}~\bibnamefont {Oka}}, \ and\ \bibinfo {author}
  {\bibfnamefont {H.}~\bibnamefont {Aoki}},\ }\href {\doibase
  10.1103/PhysRevB.93.144307} {\bibfield  {journal} {\bibinfo  {journal} {Phys.
  Rev. B}\ }\textbf {\bibinfo {volume} {93}},\ \bibinfo {pages} {144307}
  (\bibinfo {year} {2016})}\BibitemShut {NoStop}%
\bibitem [{\citenamefont {Holthaus}(2016)}]{Holthaus16}%
  \BibitemOpen
  \bibfield  {author} {\bibinfo {author} {\bibfnamefont {M.}~\bibnamefont
  {Holthaus}},\ }\href {\doibase 10.1088/0953-4075/49/1/013001} {\bibfield
  {journal} {\bibinfo  {journal} {J. Phys. B: Atm. Mol. Opt.}\ }\textbf
  {\bibinfo {volume} {49}},\ \bibinfo {pages} {013001} (\bibinfo {year}
  {2016})}\BibitemShut {NoStop}%
\bibitem [{\citenamefont {Klinovaja}\ \emph {et~al.}(2016)\citenamefont
  {Klinovaja}, \citenamefont {Stano},\ and\ \citenamefont
  {Loss}}]{Klinovaja16}%
  \BibitemOpen
  \bibfield  {author} {\bibinfo {author} {\bibfnamefont {J.}~\bibnamefont
  {Klinovaja}}, \bibinfo {author} {\bibfnamefont {P.}~\bibnamefont {Stano}}, \
  and\ \bibinfo {author} {\bibfnamefont {D.}~\bibnamefont {Loss}},\ }\href
  {\doibase 10.1103/PhysRevLett.116.176401} {\bibfield  {journal} {\bibinfo
  {journal} {Phys. Rev. Lett.}\ }\textbf {\bibinfo {volume} {116}},\ \bibinfo
  {pages} {176401} (\bibinfo {year} {2016})}\BibitemShut {NoStop}%
\bibitem [{\citenamefont {Karch}\ \emph {et~al.}(2010)\citenamefont {Karch},
  \citenamefont {Olbrich}, \citenamefont {Schmalzbauer}, \citenamefont {Zoth},
  \citenamefont {Brinsteiner}, \citenamefont {Fehrenbacher}, \citenamefont
  {Wurstbauer}, \citenamefont {Glazov}, \citenamefont {Tarasenko},
  \citenamefont {Ivchenko}, \citenamefont {Weiss}, \citenamefont {Eroms},
  \citenamefont {Yakimova}, \citenamefont {Lara-Avila}, \citenamefont
  {Kubatkin},\ and\ \citenamefont {Ganichev}}]{Karch10}%
  \BibitemOpen
  \bibfield  {author} {\bibinfo {author} {\bibfnamefont {J.}~\bibnamefont
  {Karch}}, \bibinfo {author} {\bibfnamefont {P.}~\bibnamefont {Olbrich}},
  \bibinfo {author} {\bibfnamefont {M.}~\bibnamefont {Schmalzbauer}}, \bibinfo
  {author} {\bibfnamefont {C.}~\bibnamefont {Zoth}}, \bibinfo {author}
  {\bibfnamefont {C.}~\bibnamefont {Brinsteiner}}, \bibinfo {author}
  {\bibfnamefont {M.}~\bibnamefont {Fehrenbacher}}, \bibinfo {author}
  {\bibfnamefont {U.}~\bibnamefont {Wurstbauer}}, \bibinfo {author}
  {\bibfnamefont {M.~M.}\ \bibnamefont {Glazov}}, \bibinfo {author}
  {\bibfnamefont {S.~A.}\ \bibnamefont {Tarasenko}}, \bibinfo {author}
  {\bibfnamefont {E.~L.}\ \bibnamefont {Ivchenko}}, \bibinfo {author}
  {\bibfnamefont {D.}~\bibnamefont {Weiss}}, \bibinfo {author} {\bibfnamefont
  {J.}~\bibnamefont {Eroms}}, \bibinfo {author} {\bibfnamefont
  {R.}~\bibnamefont {Yakimova}}, \bibinfo {author} {\bibfnamefont
  {S.}~\bibnamefont {Lara-Avila}}, \bibinfo {author} {\bibfnamefont
  {S.}~\bibnamefont {Kubatkin}}, \ and\ \bibinfo {author} {\bibfnamefont
  {S.~D.}\ \bibnamefont {Ganichev}},\ }\href {\doibase
  10.1103/PhysRevLett.105.227402} {\bibfield  {journal} {\bibinfo  {journal}
  {Phys. Rev. Lett.}\ }\textbf {\bibinfo {volume} {105}},\ \bibinfo {pages}
  {227402} (\bibinfo {year} {2010})}\BibitemShut {NoStop}%
\bibitem [{\citenamefont {Calvo}\ \emph {et~al.}(2011)\citenamefont {Calvo},
  \citenamefont {Pastawski}, \citenamefont {Roche},\ and\ \citenamefont
  {Torres}}]{Calvo11}%
  \BibitemOpen
  \bibfield  {author} {\bibinfo {author} {\bibfnamefont {H.~L.}\ \bibnamefont
  {Calvo}}, \bibinfo {author} {\bibfnamefont {H.~M.}\ \bibnamefont
  {Pastawski}}, \bibinfo {author} {\bibfnamefont {S.}~\bibnamefont {Roche}}, \
  and\ \bibinfo {author} {\bibfnamefont {L.~E. F.~F.}\ \bibnamefont {Torres}},\
  }\href {\doibase 10.1063/1.3597412} {\bibfield  {journal} {\bibinfo
  {journal} {Appl. Phys. Lett.}\ }\textbf {\bibinfo {volume} {98}},\ \bibinfo
  {pages} {232103} (\bibinfo {year} {2011})}\BibitemShut {NoStop}%
\bibitem [{\citenamefont {Zhou}\ and\ \citenamefont {Wu}(2011)}]{Zhou11}%
  \BibitemOpen
  \bibfield  {author} {\bibinfo {author} {\bibfnamefont {Y.}~\bibnamefont
  {Zhou}}\ and\ \bibinfo {author} {\bibfnamefont {M.~W.}\ \bibnamefont {Wu}},\
  }\href {\doibase 10.1103/PhysRevB.83.245436} {\bibfield  {journal} {\bibinfo
  {journal} {Phys. Rev. B}\ }\textbf {\bibinfo {volume} {83}},\ \bibinfo
  {pages} {245436} (\bibinfo {year} {2011})}\BibitemShut {NoStop}%
\bibitem [{\citenamefont {Scholz}\ \emph {et~al.}(2013)\citenamefont {Scholz},
  \citenamefont {L\'opez},\ and\ \citenamefont {Schliemann}}]{Scholz13}%
  \BibitemOpen
  \bibfield  {author} {\bibinfo {author} {\bibfnamefont {A.}~\bibnamefont
  {Scholz}}, \bibinfo {author} {\bibfnamefont {A.}~\bibnamefont {L\'opez}}, \
  and\ \bibinfo {author} {\bibfnamefont {J.}~\bibnamefont {Schliemann}},\
  }\href {\doibase 10.1103/PhysRevB.88.045118} {\bibfield  {journal} {\bibinfo
  {journal} {Phys. Rev. B}\ }\textbf {\bibinfo {volume} {88}},\ \bibinfo
  {pages} {045118} (\bibinfo {year} {2013})}\BibitemShut {NoStop}%
\bibitem [{\citenamefont {Usaj}\ \emph {et~al.}(2014)\citenamefont {Usaj},
  \citenamefont {Perez-Piskunow}, \citenamefont {Foa~Torres},\ and\
  \citenamefont {Balseiro}}]{Usaj14}%
  \BibitemOpen
  \bibfield  {author} {\bibinfo {author} {\bibfnamefont {G.}~\bibnamefont
  {Usaj}}, \bibinfo {author} {\bibfnamefont {P.~M.}\ \bibnamefont
  {Perez-Piskunow}}, \bibinfo {author} {\bibfnamefont {L.~E.~F.}\ \bibnamefont
  {Foa~Torres}}, \ and\ \bibinfo {author} {\bibfnamefont {C.~A.}\ \bibnamefont
  {Balseiro}},\ }\href {\doibase 10.1103/PhysRevB.90.115423} {\bibfield
  {journal} {\bibinfo  {journal} {Phys. Rev. B}\ }\textbf {\bibinfo {volume}
  {90}},\ \bibinfo {pages} {115423} (\bibinfo {year} {2014})}\BibitemShut
  {NoStop}%
\bibitem [{\citenamefont {{M.A. Sentef}}\ \emph {et~al.}(2015)\citenamefont
  {{M.A. Sentef}}, \citenamefont {{M. Claassen}}, \citenamefont {{A. F.
  Kemper}}, \citenamefont {{B. Moritz}}, \citenamefont {{T. Oka}},
  \citenamefont {{J. K. Freericks}},\ and\ \citenamefont {{T. P.
  Devereaux}}}]{Sentef15}%
  \BibitemOpen
  \bibfield  {author} {\bibinfo {author} {\bibnamefont {{M.A. Sentef}}},
  \bibinfo {author} {\bibnamefont {{M. Claassen}}}, \bibinfo {author}
  {\bibnamefont {{A. F. Kemper}}}, \bibinfo {author} {\bibnamefont {{B.
  Moritz}}}, \bibinfo {author} {\bibnamefont {{T. Oka}}}, \bibinfo {author}
  {\bibnamefont {{J. K. Freericks}}}, \ and\ \bibinfo {author} {\bibnamefont
  {{T. P. Devereaux}}},\ }\href {\doibase 10.1038/ncomms8047} {\bibfield
  {journal} {\bibinfo  {journal} {Nat. Comm.}\ }\textbf {\bibinfo {volume}
  {6}},\ \bibinfo {pages} {7047} (\bibinfo {year} {2015})}\BibitemShut
  {NoStop}%
\bibitem [{\citenamefont {L\'opez}\ \emph
  {et~al.}(2015{\natexlab{a}})\citenamefont {L\'opez}, \citenamefont
  {Di~Teodoro}, \citenamefont {Schliemann}, \citenamefont {Berche},\ and\
  \citenamefont {Santos}}]{Lopez15a}%
  \BibitemOpen
  \bibfield  {author} {\bibinfo {author} {\bibfnamefont {A.}~\bibnamefont
  {L\'opez}}, \bibinfo {author} {\bibfnamefont {A.}~\bibnamefont {Di~Teodoro}},
  \bibinfo {author} {\bibfnamefont {J.}~\bibnamefont {Schliemann}}, \bibinfo
  {author} {\bibfnamefont {B.}~\bibnamefont {Berche}}, \ and\ \bibinfo {author}
  {\bibfnamefont {B.}~\bibnamefont {Santos}},\ }\href {\doibase
  10.1103/PhysRevB.92.235411} {\bibfield  {journal} {\bibinfo  {journal} {Phys.
  Rev. B}\ }\textbf {\bibinfo {volume} {92}},\ \bibinfo {pages} {235411}
  (\bibinfo {year} {2015}{\natexlab{a}})}\BibitemShut {NoStop}%
\bibitem [{\citenamefont {Wang}\ and\ \citenamefont {Li}(2016)}]{Wang16}%
  \BibitemOpen
  \bibfield  {author} {\bibinfo {author} {\bibfnamefont {Y.-X.}\ \bibnamefont
  {Wang}}\ and\ \bibinfo {author} {\bibfnamefont {F.}~\bibnamefont {Li}},\
  }\href {\doibase 10.1016/j.physb.2016.03.029} {\bibfield  {journal} {\bibinfo
   {journal} {Physica B: Condensed Matter}\ }\textbf {\bibinfo {volume}
  {492}},\ \bibinfo {pages} {1 } (\bibinfo {year} {2016})}\BibitemShut
  {NoStop}%
\bibitem [{\citenamefont {L\'opez}\ \emph
  {et~al.}(2015{\natexlab{b}})\citenamefont {L\'opez}, \citenamefont {Scholz},
  \citenamefont {Santos},\ and\ \citenamefont {Schliemann}}]{Lopez15b}%
  \BibitemOpen
  \bibfield  {author} {\bibinfo {author} {\bibfnamefont {A.}~\bibnamefont
  {L\'opez}}, \bibinfo {author} {\bibfnamefont {A.}~\bibnamefont {Scholz}},
  \bibinfo {author} {\bibfnamefont {B.}~\bibnamefont {Santos}}, \ and\ \bibinfo
  {author} {\bibfnamefont {J.}~\bibnamefont {Schliemann}},\ }\href {\doibase
  10.1103/PhysRevB.91.125105} {\bibfield  {journal} {\bibinfo  {journal} {Phys.
  Rev. B}\ }\textbf {\bibinfo {volume} {91}},\ \bibinfo {pages} {125105}
  (\bibinfo {year} {2015}{\natexlab{b}})}\BibitemShut {NoStop}%
\bibitem [{\citenamefont {Mohan}\ \emph {et~al.}(2016)\citenamefont {Mohan},
  \citenamefont {Saxena}, \citenamefont {Kundu},\ and\ \citenamefont
  {Rao}}]{Mohan16}%
  \BibitemOpen
  \bibfield  {author} {\bibinfo {author} {\bibfnamefont {P.}~\bibnamefont
  {Mohan}}, \bibinfo {author} {\bibfnamefont {R.}~\bibnamefont {Saxena}},
  \bibinfo {author} {\bibfnamefont {A.}~\bibnamefont {Kundu}}, \ and\ \bibinfo
  {author} {\bibfnamefont {S.}~\bibnamefont {Rao}},\ }\href {\doibase
  10.1103/PhysRevB.94.235419} {\bibfield  {journal} {\bibinfo  {journal} {Phys.
  Rev. B}\ }\textbf {\bibinfo {volume} {94}},\ \bibinfo {pages} {235419}
  (\bibinfo {year} {2016})}\BibitemShut {NoStop}%
\bibitem [{\citenamefont {{M. Tahir}}\ \emph {et~al.}(2016)\citenamefont {{M.
  Tahir}}, \citenamefont {{Q. Y. Zhang}},\ and\ \citenamefont {{U.
  Schwingenschl\"ogl}}}]{Tahir16}%
  \BibitemOpen
  \bibfield  {author} {\bibinfo {author} {\bibnamefont {{M. Tahir}}}, \bibinfo
  {author} {\bibnamefont {{Q. Y. Zhang}}}, \ and\ \bibinfo {author}
  {\bibnamefont {{U. Schwingenschl\"ogl}}},\ }\href {\doibase
  10.1038/srep31821} {\bibfield  {journal} {\bibinfo  {journal} {Sci. Rep.}\
  }\textbf {\bibinfo {volume} {6}},\ \bibinfo {pages} {31821} (\bibinfo {year}
  {2016})}\BibitemShut {NoStop}%
\bibitem [{\citenamefont {{M. Claassen}}\ \emph {et~al.}(2016)\citenamefont
  {{M. Claassen}}, \citenamefont {{C. Jia}}, \citenamefont {{B. Moritz}},\ and\
  \citenamefont {{T. P. Devereaux}}}]{Claasen16}%
  \BibitemOpen
  \bibfield  {author} {\bibinfo {author} {\bibnamefont {{M. Claassen}}},
  \bibinfo {author} {\bibnamefont {{C. Jia}}}, \bibinfo {author} {\bibnamefont
  {{B. Moritz}}}, \ and\ \bibinfo {author} {\bibnamefont {{T. P. Devereaux}}},\
  }\href {\doibase 10.1038/ncomms13074} {\bibfield  {journal} {\bibinfo
  {journal} {Nat. Comm.}\ }\textbf {\bibinfo {volume} {7}},\ \bibinfo {pages}
  {13074} (\bibinfo {year} {2016})}\BibitemShut {NoStop}%
\bibitem [{\citenamefont {L\'opez}\ \emph {et~al.}(2012)\citenamefont
  {L\'opez}, \citenamefont {Sun},\ and\ \citenamefont {Schliemann}}]{Lopez12}%
  \BibitemOpen
  \bibfield  {author} {\bibinfo {author} {\bibfnamefont {A.}~\bibnamefont
  {L\'opez}}, \bibinfo {author} {\bibfnamefont {Z.~Z.}\ \bibnamefont {Sun}}, \
  and\ \bibinfo {author} {\bibfnamefont {J.}~\bibnamefont {Schliemann}},\
  }\href {\doibase 10.1103/PhysRevB.85.205428} {\bibfield  {journal} {\bibinfo
  {journal} {Phys. Rev. B}\ }\textbf {\bibinfo {volume} {85}},\ \bibinfo
  {pages} {205428} (\bibinfo {year} {2012})}\BibitemShut {NoStop}%
\bibitem [{\citenamefont {{A. L\'opez}}\ \emph {et~al.}(2013)\citenamefont {{A.
  L\'opez}}, \citenamefont {{A. Scholz}}, \citenamefont {{Z. Z. Sun}},\ and\
  \citenamefont {{J. Schliemann}}}]{Lopez13}%
  \BibitemOpen
  \bibfield  {author} {\bibinfo {author} {\bibnamefont {{A. L\'opez}}},
  \bibinfo {author} {\bibnamefont {{A. Scholz}}}, \bibinfo {author}
  {\bibnamefont {{Z. Z. Sun}}}, \ and\ \bibinfo {author} {\bibnamefont {{J.
  Schliemann}}},\ }\href {\doibase 10.1140/epjb/e2013-40488-1} {\bibfield
  {journal} {\bibinfo  {journal} {Eur. Phys. J. B}\ }\textbf {\bibinfo {volume}
  {86}},\ \bibinfo {pages} {366} (\bibinfo {year} {2013})}\BibitemShut
  {NoStop}%
\bibitem [{\citenamefont {{R. Rammal}}(1985)}]{Rammal85}%
  \BibitemOpen
  \bibfield  {author} {\bibinfo {author} {\bibnamefont {{R. Rammal}}},\ }\href
  {\doibase 10.1051/jphys:019850046080134500} {\bibfield  {journal} {\bibinfo
  {journal} {J. Phys. France}\ }\textbf {\bibinfo {volume} {46}},\ \bibinfo
  {pages} {1345} (\bibinfo {year} {1985})}\BibitemShut {NoStop}%
\bibitem [{\citenamefont {Hasegawa}\ and\ \citenamefont
  {Kohmoto}(2006)}]{Hasegawa06}%
  \BibitemOpen
  \bibfield  {author} {\bibinfo {author} {\bibfnamefont {Y.}~\bibnamefont
  {Hasegawa}}\ and\ \bibinfo {author} {\bibfnamefont {M.}~\bibnamefont
  {Kohmoto}},\ }\href {\doibase 10.1103/PhysRevB.74.155415} {\bibfield
  {journal} {\bibinfo  {journal} {Phys. Rev. B}\ }\textbf {\bibinfo {volume}
  {74}},\ \bibinfo {pages} {155415} (\bibinfo {year} {2006})}\BibitemShut
  {NoStop}%
\bibitem [{\citenamefont {Wang}\ and\ \citenamefont {Gong}(2009)}]{Wang09}%
  \BibitemOpen
  \bibfield  {author} {\bibinfo {author} {\bibfnamefont {J.}~\bibnamefont
  {Wang}}\ and\ \bibinfo {author} {\bibfnamefont {J.}~\bibnamefont {Gong}},\
  }\href {\doibase 10.1103/PhysRevLett.102.244102} {\bibfield  {journal}
  {\bibinfo  {journal} {Phys. Rev. Lett.}\ }\textbf {\bibinfo {volume} {102}},\
  \bibinfo {pages} {244102} (\bibinfo {year} {2009})}\BibitemShut {NoStop}%
\bibitem [{\citenamefont {Rhim}\ and\ \citenamefont {Park}(2012)}]{Rhim12}%
  \BibitemOpen
  \bibfield  {author} {\bibinfo {author} {\bibfnamefont {J.-W.}\ \bibnamefont
  {Rhim}}\ and\ \bibinfo {author} {\bibfnamefont {K.}~\bibnamefont {Park}},\
  }\href {\doibase 10.1103/PhysRevB.86.235411} {\bibfield  {journal} {\bibinfo
  {journal} {Phys. Rev. B}\ }\textbf {\bibinfo {volume} {86}},\ \bibinfo
  {pages} {235411} (\bibinfo {year} {2012})}\BibitemShut {NoStop}%
\bibitem [{\citenamefont {Yilmaz}\ \emph {et~al.}(2015)\citenamefont {Yilmaz},
  \citenamefont {\"Unal},\ and\ \citenamefont {Oktel}}]{Yilmaz15}%
  \BibitemOpen
  \bibfield  {author} {\bibinfo {author} {\bibfnamefont {F.}~\bibnamefont
  {Yilmaz}}, \bibinfo {author} {\bibfnamefont {F.~N.}\ \bibnamefont {\"Unal}},
  \ and\ \bibinfo {author} {\bibfnamefont {M.~O.}\ \bibnamefont {Oktel}},\
  }\href {\doibase 10.1103/PhysRevA.91.063628} {\bibfield  {journal} {\bibinfo
  {journal} {Phys. Rev. A}\ }\textbf {\bibinfo {volume} {91}},\ \bibinfo
  {pages} {063628} (\bibinfo {year} {2015})}\BibitemShut {NoStop}%
\bibitem [{\citenamefont {Yilmaz}\ and\ \citenamefont
  {Oktel}(2017)}]{Yilmaz17}%
  \BibitemOpen
  \bibfield  {author} {\bibinfo {author} {\bibfnamefont {F.}~\bibnamefont
  {Yilmaz}}\ and\ \bibinfo {author} {\bibfnamefont {M.~O.}\ \bibnamefont
  {Oktel}},\ }\href {\doibase 10.1103/PhysRevA.95.063628} {\bibfield  {journal}
  {\bibinfo  {journal} {Phys. Rev. A}\ }\textbf {\bibinfo {volume} {95}},\
  \bibinfo {pages} {063628} (\bibinfo {year} {2017})}\BibitemShut {NoStop}%
\bibitem [{\citenamefont {Asb\'oth}\ and\ \citenamefont
  {Alberti}(2017)}]{Asboth17}%
  \BibitemOpen
  \bibfield  {author} {\bibinfo {author} {\bibfnamefont {J.~K.}\ \bibnamefont
  {Asb\'oth}}\ and\ \bibinfo {author} {\bibfnamefont {A.}~\bibnamefont
  {Alberti}},\ }\href {\doibase 10.1103/PhysRevLett.118.216801} {\bibfield
  {journal} {\bibinfo  {journal} {Phys. Rev. Lett.}\ }\textbf {\bibinfo
  {volume} {118}},\ \bibinfo {pages} {216801} (\bibinfo {year}
  {2017})}\BibitemShut {NoStop}%
\bibitem [{\citenamefont {Dean}\ \emph {et~al.}(2013)\citenamefont {Dean},
  \citenamefont {Wang}, \citenamefont {Maher}, \citenamefont {Forsythe},
  \citenamefont {Ghahari}, \citenamefont {Gao}, \citenamefont {Katoch},
  \citenamefont {Ishigami}, \citenamefont {Moon}, \citenamefont {Koshino},
  \citenamefont {Taniguchi}, \citenamefont {Watanabe}, \citenamefont {Shepard},
  \citenamefont {Hone},\ and\ \citenamefont {Kim}}]{Dean13}%
  \BibitemOpen
  \bibfield  {author} {\bibinfo {author} {\bibfnamefont {C.~R.}\ \bibnamefont
  {Dean}}, \bibinfo {author} {\bibfnamefont {L.}~\bibnamefont {Wang}}, \bibinfo
  {author} {\bibfnamefont {P.}~\bibnamefont {Maher}}, \bibinfo {author}
  {\bibfnamefont {C.}~\bibnamefont {Forsythe}}, \bibinfo {author}
  {\bibfnamefont {F.}~\bibnamefont {Ghahari}}, \bibinfo {author} {\bibfnamefont
  {Y.}~\bibnamefont {Gao}}, \bibinfo {author} {\bibfnamefont {J.}~\bibnamefont
  {Katoch}}, \bibinfo {author} {\bibfnamefont {M.}~\bibnamefont {Ishigami}},
  \bibinfo {author} {\bibfnamefont {P.}~\bibnamefont {Moon}}, \bibinfo {author}
  {\bibfnamefont {M.}~\bibnamefont {Koshino}}, \bibinfo {author} {\bibfnamefont
  {T.}~\bibnamefont {Taniguchi}}, \bibinfo {author} {\bibfnamefont
  {K.}~\bibnamefont {Watanabe}}, \bibinfo {author} {\bibfnamefont {K.~L.}\
  \bibnamefont {Shepard}}, \bibinfo {author} {\bibfnamefont {J.}~\bibnamefont
  {Hone}}, \ and\ \bibinfo {author} {\bibfnamefont {P.}~\bibnamefont {Kim}},\
  }\href {\doibase https://doi.org/10.1038/nature12186} {\bibfield  {journal}
  {\bibinfo  {journal} {Nature}\ }\textbf {\bibinfo {volume} {497}},\ \bibinfo
  {pages} {598} (\bibinfo {year} {2013})}\BibitemShut {NoStop}%
\bibitem [{\citenamefont {Wang}\ \emph {et~al.}(2013)\citenamefont {Wang},
  \citenamefont {Steinberg}, \citenamefont {Jarillo-Herrero},\ and\
  \citenamefont {Gedik}}]{Wang13}%
  \BibitemOpen
  \bibfield  {author} {\bibinfo {author} {\bibfnamefont {Y.~H.}\ \bibnamefont
  {Wang}}, \bibinfo {author} {\bibfnamefont {H.}~\bibnamefont {Steinberg}},
  \bibinfo {author} {\bibfnamefont {P.}~\bibnamefont {Jarillo-Herrero}}, \ and\
  \bibinfo {author} {\bibfnamefont {N.}~\bibnamefont {Gedik}},\ }\href
  {\doibase 10.1126/science.1239834} {\bibfield  {journal} {\bibinfo  {journal}
  {Science}\ }\textbf {\bibinfo {volume} {342}},\ \bibinfo {pages} {453}
  (\bibinfo {year} {2013})}\BibitemShut {NoStop}%
\bibitem [{\citenamefont {Owerre}(2018)}]{Owerre2018}%
  \BibitemOpen
  \bibfield  {author} {\bibinfo {author} {\bibfnamefont {S.}~\bibnamefont
  {Owerre}},\ }\href {\doibase 10.1016/j.aop.2018.10.005} {\bibfield  {journal}
  {\bibinfo  {journal} {Annals of Physics}\ }\textbf {\bibinfo {volume}
  {399}},\ \bibinfo {pages} {93} (\bibinfo {year} {2018})}\BibitemShut
  {NoStop}%
\bibitem [{\citenamefont {Kooi}\ \emph {et~al.}(2018)\citenamefont {Kooi},
  \citenamefont {Quelle}, \citenamefont {Beugeling},\ and\ \citenamefont
  {Smith}}]{Kooi18}%
  \BibitemOpen
  \bibfield  {author} {\bibinfo {author} {\bibfnamefont {S.~H.}\ \bibnamefont
  {Kooi}}, \bibinfo {author} {\bibfnamefont {A.}~\bibnamefont {Quelle}},
  \bibinfo {author} {\bibfnamefont {W.}~\bibnamefont {Beugeling}}, \ and\
  \bibinfo {author} {\bibfnamefont {C.~M.}\ \bibnamefont {Smith}},\ }\href
  {\doibase 10.1103/PhysRevB.98.115124} {\bibfield  {journal} {\bibinfo
  {journal} {Phys. Rev. B}\ }\textbf {\bibinfo {volume} {98}},\ \bibinfo
  {pages} {115124} (\bibinfo {year} {2018})}\BibitemShut {NoStop}%
\bibitem [{\citenamefont {Du}\ \emph {et~al.}(2018)\citenamefont {Du},
  \citenamefont {Chen}, \citenamefont {Barr}, \citenamefont {Barr},\ and\
  \citenamefont {Fiete}}]{Du18}%
  \BibitemOpen
  \bibfield  {author} {\bibinfo {author} {\bibfnamefont {L.}~\bibnamefont
  {Du}}, \bibinfo {author} {\bibfnamefont {Q.}~\bibnamefont {Chen}}, \bibinfo
  {author} {\bibfnamefont {A.~D.}\ \bibnamefont {Barr}}, \bibinfo {author}
  {\bibfnamefont {A.~R.}\ \bibnamefont {Barr}}, \ and\ \bibinfo {author}
  {\bibfnamefont {G.~A.}\ \bibnamefont {Fiete}},\ }\href {\doibase
  10.1103/physrevb.98.245145} {\bibfield  {journal} {\bibinfo  {journal}
  {Physical Review B}\ }\textbf {\bibinfo {volume} {98}},\ \bibinfo {pages}
  {245145} (\bibinfo {year} {2018})}\BibitemShut {NoStop}%
\bibitem [{\citenamefont {H\"ockendorf}\ \emph {et~al.}(2018)\citenamefont
  {H\"ockendorf}, \citenamefont {Alvermann},\ and\ \citenamefont
  {Fehske}}]{Hockendorf18}%
  \BibitemOpen
  \bibfield  {author} {\bibinfo {author} {\bibfnamefont {B.}~\bibnamefont
  {H\"ockendorf}}, \bibinfo {author} {\bibfnamefont {A.}~\bibnamefont
  {Alvermann}}, \ and\ \bibinfo {author} {\bibfnamefont {H.}~\bibnamefont
  {Fehske}},\ }\href {\doibase 10.1103/PhysRevB.97.045140} {\bibfield
  {journal} {\bibinfo  {journal} {Phys. Rev. B}\ }\textbf {\bibinfo {volume}
  {97}},\ \bibinfo {pages} {045140} (\bibinfo {year} {2018})}\BibitemShut
  {NoStop}%
\bibitem [{\citenamefont {H\"ockendorf}\ \emph {et~al.}(2017)\citenamefont
  {H\"ockendorf}, \citenamefont {Alvermann},\ and\ \citenamefont
  {Fehske}}]{Hockendorf17}%
  \BibitemOpen
  \bibfield  {author} {\bibinfo {author} {\bibfnamefont {B.}~\bibnamefont
  {H\"ockendorf}}, \bibinfo {author} {\bibfnamefont {A.}~\bibnamefont
  {Alvermann}}, \ and\ \bibinfo {author} {\bibfnamefont {H.}~\bibnamefont
  {Fehske}},\ }\href {\doibase 10.1088/1751-8121/aa7591} {\bibfield  {journal}
  {\bibinfo  {journal} {Journal of Physics A: Mathematical and Theoretical}\
  }\textbf {\bibinfo {volume} {50}},\ \bibinfo {pages} {295301} (\bibinfo
  {year} {2017})}\BibitemShut {NoStop}%
\bibitem [{\citenamefont {Nathan}\ and\ \citenamefont
  {Rudner}(2015)}]{Nathan15}%
  \BibitemOpen
  \bibfield  {author} {\bibinfo {author} {\bibfnamefont {F.}~\bibnamefont
  {Nathan}}\ and\ \bibinfo {author} {\bibfnamefont {M.~S.}\ \bibnamefont
  {Rudner}},\ }\href {\doibase 10.1088/1367-2630/17/12/125014} {\bibfield
  {journal} {\bibinfo  {journal} {New Journal of Physics}\ }\textbf {\bibinfo
  {volume} {17}},\ \bibinfo {pages} {125014} (\bibinfo {year}
  {2015})}\BibitemShut {NoStop}%
\bibitem [{\citenamefont {Berry}(1984)}]{Berry1984}%
  \BibitemOpen
  \bibfield  {author} {\bibinfo {author} {\bibfnamefont {M.~V.}\ \bibnamefont
  {Berry}},\ }\href {\doibase 10.1098/rspa.1984.0023} {\bibfield  {journal}
  {\bibinfo  {journal} {Proceedings of the Royal Society A: Mathematical,
  Physical and Engineering Sciences}\ }\textbf {\bibinfo {volume} {392}},\
  \bibinfo {pages} {45} (\bibinfo {year} {1984})}\BibitemShut {NoStop}%
\bibitem [{\citenamefont {Simon}(1983)}]{Simon1983}%
  \BibitemOpen
  \bibfield  {author} {\bibinfo {author} {\bibfnamefont {B.}~\bibnamefont
  {Simon}},\ }\href {\doibase 10.1103/physrevlett.51.2167} {\bibfield
  {journal} {\bibinfo  {journal} {Physical Review Letters}\ }\textbf {\bibinfo
  {volume} {51}},\ \bibinfo {pages} {2167} (\bibinfo {year}
  {1983})}\BibitemShut {NoStop}%
\bibitem [{\citenamefont {Goldman}(2009)}]{Goldman}%
  \BibitemOpen
  \bibfield  {author} {\bibinfo {author} {\bibfnamefont {N.}~\bibnamefont
  {Goldman}},\ }\href {\doibase https://doi.org/10.1088/0953-4075/42/5/055302}
  {\bibfield  {journal} {\bibinfo  {journal} {J. Phys. B: Atm. Mol. Opt.}\
  }\textbf {\bibinfo {volume} {42}},\ \bibinfo {pages} {055302} (\bibinfo
  {year} {2009})}\BibitemShut {NoStop}%
\bibitem [{\citenamefont {Seetharam}\ \emph {et~al.}(2015)\citenamefont
  {Seetharam}, \citenamefont {Bardyn}, \citenamefont {Lindner}, \citenamefont
  {Rudner},\ and\ \citenamefont {Refael}}]{Seetharam15}%
  \BibitemOpen
  \bibfield  {author} {\bibinfo {author} {\bibfnamefont {K.~I.}\ \bibnamefont
  {Seetharam}}, \bibinfo {author} {\bibfnamefont {C.-E.}\ \bibnamefont
  {Bardyn}}, \bibinfo {author} {\bibfnamefont {N.~H.}\ \bibnamefont {Lindner}},
  \bibinfo {author} {\bibfnamefont {M.~S.}\ \bibnamefont {Rudner}}, \ and\
  \bibinfo {author} {\bibfnamefont {G.}~\bibnamefont {Refael}},\ }\href
  {\doibase 10.1103/PhysRevX.5.041050} {\bibfield  {journal} {\bibinfo
  {journal} {Phys. Rev. X}\ }\textbf {\bibinfo {volume} {5}},\ \bibinfo {pages}
  {041050} (\bibinfo {year} {2015})}\BibitemShut {NoStop}%
\bibitem [{\citenamefont {Desbuquois}\ \emph {et~al.}(2017)\citenamefont
  {Desbuquois}, \citenamefont {Messer}, \citenamefont {G\"org}, \citenamefont
  {Sandholzer}, \citenamefont {Jotzu},\ and\ \citenamefont
  {Esslinger}}]{Desbuquois17}%
  \BibitemOpen
  \bibfield  {author} {\bibinfo {author} {\bibfnamefont {R.}~\bibnamefont
  {Desbuquois}}, \bibinfo {author} {\bibfnamefont {M.}~\bibnamefont {Messer}},
  \bibinfo {author} {\bibfnamefont {F.}~\bibnamefont {G\"org}}, \bibinfo
  {author} {\bibfnamefont {K.}~\bibnamefont {Sandholzer}}, \bibinfo {author}
  {\bibfnamefont {G.}~\bibnamefont {Jotzu}}, \ and\ \bibinfo {author}
  {\bibfnamefont {T.}~\bibnamefont {Esslinger}},\ }\href {\doibase
  10.1103/PhysRevA.96.053602} {\bibfield  {journal} {\bibinfo  {journal} {Phys.
  Rev. A}\ }\textbf {\bibinfo {volume} {96}},\ \bibinfo {pages} {053602}
  (\bibinfo {year} {2017})}\BibitemShut {NoStop}%
\bibitem [{Note1()}]{Note1}%
  \BibitemOpen
  \bibinfo {note} {More precisely, the following conditions have to be
  fulfilled\cite {Rudner13}: 1) ${\protect \mathaccentV {tilde}07EU}(\protect
  \mathaccentV {vec}17E{k},T) = \protect \mathbb {I}$. 2) There should exist a
  one-parameter family of evolution operators $\protect \{ U_s:s\in [0,1]
  \protect \}$ which interpolates between $U$ and $\protect \mathaccentV
  {tilde}07EU$ as follows: $U_{s=0}(\protect \mathaccentV {vec}17E{k},t) =
  U(\protect \mathaccentV {vec}17E{k},t)$ and $U_{s=1}(\protect \mathaccentV
  {vec}17E{k},t) = \protect \mathaccentV {tilde}07EU(\protect \mathaccentV
  {vec}17E{k},t)$. 3) $\protect \mathaccentV {tilde}07EU(\protect \mathaccentV
  {vec}17E{k}, T)$ has to maintain a gap around $\varepsilon _s$ with
  $\varepsilon _{s=0} = \varepsilon $, $\varepsilon _{s=1} = \pi /T$ and a
  smooth interpolation from $s=0$ to $s=1$.}\BibitemShut {Stop}%
\bibitem [{\citenamefont {Shirley}(1963)}]{Shirley63}%
  \BibitemOpen
  \bibfield  {author} {\bibinfo {author} {\bibfnamefont {J.~H.}\ \bibnamefont
  {Shirley}},\ }\emph {\bibinfo {title} {Interaction of a quantum system with a
  strong oscillating field}},\ \href
  {http://resolver.caltech.edu/CaltechETD:etd-05142008-103758} {Ph.D. thesis},\
  \bibinfo  {school} {California Institute of Technology} (\bibinfo {year}
  {1963})\BibitemShut {NoStop}%
\bibitem [{\citenamefont {Shirley}(1965)}]{Shirley65}%
  \BibitemOpen
  \bibfield  {author} {\bibinfo {author} {\bibfnamefont {J.~H.}\ \bibnamefont
  {Shirley}},\ }\href {\doibase 10.1103/PhysRev.138.B979} {\bibfield  {journal}
  {\bibinfo  {journal} {Phys. Rev.}\ }\textbf {\bibinfo {volume} {138}},\
  \bibinfo {pages} {B979} (\bibinfo {year} {1965})}\BibitemShut {NoStop}%
\end{thebibliography}%
\end{document}